\documentclass[aps,prd,onecolumn,showpacs,amsmath,amssymb,floatfix,nofootinbib,11pt]{revtex4-1}
\usepackage[colorlinks=true, citecolor=blue, linkcolor=blue, urlcolor=blue]{hyperref}
\usepackage{color,xcolor,graphicx,amsfonts,multirow,amsmath,multirow}

\newcommand{\DS}[1]{/\!\!\!#1}
\newcommand{\DSS}[1]{/\!\!\!\!#1}
\allowdisplaybreaks

\begin{document}
\title{$\eta^{(\prime)}$ -meson twist-2 distribution amplitude within QCD sum rule approach and its application to the semi-leptonic decay $ D_s^+ \to\eta^{(\prime)}\ell^+ \nu_\ell$}
\author{Dan-Dan Hu$^1$}
\author{Hai-Bing Fu$^{1,3}$}
\email{fuhb@cqu.edu.cn (Corresponding author)}
\author{Tao Zhong$^1$}
\author{Long Zeng$^{2,3}$}
\author{Wei Cheng$^{4}$}
\author{Xing-Gang Wu$^{2,3}$}
\email{wuxg@cqu.edu.cn}
\address{$^1$Department of Physics, Guizhou Minzu University, Guiyang 550025, P.R. China}
\address{$^2$Department of Physics, Chongqing University, Chongqing 401331, P.R. China}
\address{$^3$Chongqing Key Laboratory for Strongly Coupled Physics, Chongqing 401331, P.R. China}
\address{$^4$School of Science, Chongqing University of Posts and Telecommunications, Chongqing 400065, P.R. China}

\date{\today}

\begin{abstract}
In this paper, we make a detailed discussion on the $\eta$ and $\eta^\prime$-meson leading-twist light-cone distribution amplitude $\phi_{2;\eta^{(\prime)}}(u,\mu)$ by using the QCD sum rules approach under the background field theory. Taking both the non-perturbative condensates up to dimension-six and the next-to-leading order (NLO) QCD corrections to the perturbative part, its first three moments $\langle\xi^n_{2;\eta^{(\prime)}}\rangle|_{\mu_0} $ with $n = (2, 4, 6)$ can be determined, where the initial scale $\mu_0$ is set as the usual choice of $1$ GeV. Numerically, we obtain $\langle\xi_{2;\eta}^2\rangle|_{\mu_0} =0.231_{-0.013}^{+0.010}$, $\langle\xi_{2;\eta}^4 \rangle|_{\mu_0} =0.109_{ - 0.007}^{ + 0.007}$, and $\langle\xi_{2;\eta}^6 \rangle|_{\mu_0} =0.066_{-0.006}^{+0.006}$ for $\eta$-meson, $\langle\xi_{2;\eta'}^2\rangle|_{\mu_0} =0.211_{-0.017}^{+0.015}$, $\langle\xi_{2;\eta'}^4 \rangle|_{\mu_0} =0.093_{ - 0.009}^{ + 0.009}$, and $\langle\xi_{2;\eta'}^6 \rangle|_{\mu_0} =0.054_{-0.008}^{+0.008}$ for $\eta'$-meson. Next, we calculate the $D_s\to\eta^{(\prime)}$ transition form factors (TFFs) $f^{\eta^{(\prime)}}_{+}(q^2)$ within QCD light-cone sum rules approach up to NLO level. The values at large recoil region are $f^{\eta}_+(0) = 0.476_{-0.036}^{+0.040}$ and $f^{\eta'}_+(0) = 0.544_{-0.042}^{+0.046}$. After extrapolating TFFs to the allowable physical regions within the series expansion, we obtain the branching fractions of the semi-leptonic decay, i.e. $D_s^+\to\eta^{(\prime)}\ell^+ \nu_\ell$, i.e. ${\cal B}(D_s^+ \to\eta^{(\prime)}  e^+\nu_e)=2.346_{-0.331}^{+0.418}(0.792_{-0.118}^{+0.141})\times10^{-2}$ and ${\cal B}(D_s^+ \to\eta^{(\prime)} \mu^+\nu_\mu)=2.320_{-0.327}^{+0.413}(0.773_{-0.115}^{+0.138})\times10^{-2}$ for $\ell = (e, \mu)$ channels respectively. And in addition to that, the mixing angle for $\eta-\eta'$ with $\varphi$ and ratio for the different decay channels ${\cal R}_{\eta'/\eta}^\ell$ are given, which show good agreement with the recent BESIII measurements.
\end{abstract}

\maketitle
	
\section{Introduction}

The $D^+_s$-meson, which is composed of a charm quark and a strange antiquark, has been discovered in year 1993. There are rich physics contents in $D^+_s$-meson decays. The $D_s^+$-meson semileptonic or exclusive decay processes provide important heavy-to-light theoretical bases for studying heavy quark decays, investigating light meson spectroscopy and supplying a bridge between weak and strong interaction couplings of quarks. More and more experimental results have been reported from the BABAR, the CLEO, the BESIII collaborations, and etc., such as the $D_s^+ \to (\eta, \eta', K^0, a_0(980), f_0(980), \phi, K^{*0})\ell^+\nu_\ell$ decays' branching fractions are within the range of $[0.12, 2.61]\%$~\cite{BaBar:2008gpr, CLEO:2009dyb, CLEO:2009ugx, BESIII:2017ikf, BESIII:2018xre, BESIII:2021duu}. Total semileptonic branching fractions provide useful discrimination on the different theoretical evaluations of hadronic matrix elements, which sizably affect the charm quark semileptonic decays. The $D_s \to \eta^{(\prime)}\ell^+\nu_\ell$ is different from the usually considered channels with final state containing light-quark composition only, and it has attracted much attention from both theoretical and experimental groups. Moreover, the $\eta^{(\prime)}$-mesons, composed by $s\bar s$ quark pair, are especially intriguing, since the $s$-quark plays an important role for the flavor physics. In deep leaning of those two processes, one can obtain useful information on the CKM matrix element $|V_{cs}|$ and the heavy-to-light transition form factors (TFFs).

Experimentally, the semileptonic decay processes for $D_s^+ \to \eta^{(\prime)} \ell^+\nu_\ell$ have been found by the CLEO collaboration early in year 1995, and their measured value of the ratio of branching fractions ${\cal B}(D_s^+\to \eta'e^+ \nu_e)/{\cal B}(D_s^+\to \eta e^+\nu_e)$ is $0.35\pm0.09\pm0.07$~\cite{Brandenburg:1995qq}. Then, the CLEO collaboration issued the measured value of the branching fractions in years 2009 and 2015, i.e. ${\cal B}(D_s^+\to \eta e^+ \nu_e)=(2.48\pm0.29\pm0.13)\%$, ${\cal B}(D_s^+\to \eta e^+ \nu_e )=(2.28 \pm0.14 \pm0.20)\%$, ${\cal B}(D_s^+\to \eta' e^+ \nu_e) = (0.91 \pm0.33 \pm0.05)\%$, and ${\cal B}(D_s^+\to \eta' e^+ \nu_e )=(0.68\pm0.15\pm0.06)\%$~\cite{Yelton:2009aa, Hietala:2015jqa}. In year 2017, the BESIII collaboration measured the branching fractions by using the same channels based on the integrated luminosity of $482~{\rm pb^{-1}}$ of the $e^+e^-$ collision at the center-of-mass energy $\sqrt s=4.009~{\rm GeV}$, and they issued ${\cal B} (D_s^+ \to \eta \mu^+\nu_\mu)=(2.42\pm 0.46\pm 0.11)\%$, ${\cal B} (D_s^+ \to \eta' \mu^+\nu_\mu)=(1.06\pm 0.54\pm 0.07)\%$, ${\cal B}(D_s^+\to \eta e^+\nu_e)=(2.30\pm 0.31\pm 0.08)\%$, and ${\cal B}(D_s^+\to \eta' e^+\nu_e)=(0.93\pm 0.30\pm 0.05)\%$~\cite{Ablikim:2017omq, Ablikim:2016rqq}. In year 2019, the BESIII collaboration finished the improved measurements on the branching fractions by using $e^+e^-$ annihilation data corresponding to an integrated luminosity of $3.19~{\rm fb}^{-1}$ collected at a center-of-mass energy of $4.178~{\rm GeV}$ and then gave the first determination of $D_s\to \eta^{(\prime)}$ TFF, e.g. ${\cal B}(D_s^+\to \eta e^+\nu_e)=(2.323\pm 0.063\pm 0.063)\%$, ${\cal B}(D_s^+\to \eta' e^+\nu_e)=(0.824\pm 0.073\pm 0.027)\%$, $f^{\eta}_+(0) |V_{cs}| = 0.4455\pm0.0053\pm0.0044$, and $f^{\eta'}_+(0) |V_{cs}| = 0.477\pm0.049\pm0.011$ ~\cite{Ablikim:2019rjz}. There are large discrepancies for $D_s^+\to \eta^{(\prime)} \ell^+\nu_\ell$ among different experimental collaborations. With more and more data accumulated in the near future, the experimental precision shall be greatly improved and the gap among different measurements could be shrunk.

Theoretically, the decay widths or branching fractions for the semileptonic decay $D_s^+\to\eta^{(\prime)}\ell^+\nu_\ell$ depends heavily on the precision of the $D_s\to\eta^{(\prime)}$ TFFs. At present, the $D_s\to\eta^{(\prime)}$ TFFs have been studies under various approaches, such as the lattice QCD (LQCD)~\cite{Bali:2014pva}, the traditional and covariant light-front quark model (LFQM)~\cite{Verma:2011yw, Cheng:2017pcq, Wei:2009nc}, the constituent quark model (CQM)~\cite{Melikhov:2000yu}, the covariant confined quark model (CCQM)~\cite{Soni:2018adu, Ivanov:2019nqd}, the light-cone sum rules (LCSR)~\cite{Duplancic:2015zna, Offen:2013nma}, the QCD sum rules (QCD SR)~\cite{Colangelo:2001cv}. The LCSR approach is based on the operator product expansion (OPE) near the light-cone $x^2\rightsquigarrow 0$ and parameterizes all the non-perturbative dynamics into the light-cone distribution amplitudes (LCDAs), which have been applied for dealing with many semileptonic decay processes~\cite{Duplancic:2008ix, Momeni:2019uag, Descotes-Genon:2019bud, Cheng:2019tgh, Momeni:2020zrb, Emmerich:2018rug, Momeni:2018tjf, Shen:2016hyv, Straub:2015ica, YanJun:2011rn, Wang:2010tz, Wang:2008bw, Beneke:2018wjp,Du:2003ja, Yang:2005bv, Singh:2018yvt, Fu:2018yin}. One may observe that the predicted values of $D_s\to\eta^{(\prime)}$ TFFs behave differently from various groups. Those discrepancies indicate that it is important to improve the accuracy of theoretical calculation. In this paper, we will calculate the $D_s\to\eta^{(\prime)}$ TFFs by using the LCSR approach up to next-to-leading order (NLO) QCD corrections.

It is well known that the states $\eta$ and $\eta^{\prime}$ are considered as candidates for mixing. The mixing among pseudoscalar mesons is of great theoretical interests and significance for understanding the dynamics and hadronic structures, which is caused by the QCD anomaly and related to the chiral symmetry breaking. Thus, one can gain a better insight into the dynamics if the mixing parameters are more accurately determined. The semileptonic decays $D_s^+ \to \eta^{(\prime)} \ell^+\nu_\ell$ probe the $s \bar s$ components of $\eta$ and $\eta^{\prime}$, which can well separate the strong and weak effects in theory, are expected to be sensitive to $\eta-\eta^{\prime}$ mixing angle~\cite{Anisovich:1997dz}. Many measurements on the processes, where $\eta$ and $\eta'$ are involved, have been carried out to fix the mixing parameters. The $\eta-\eta'$ mixing can be described in different forms. To investigate the $\eta-\eta'$ mixing, two schemes have been suggested in the literature~\cite{Ball:2007hb, Ball:1995zv, Feldmann:1999uf, Huang:2006as, Ke:2009mn, DeFazio:2000my, Choi:2010zb}, i.e. the singlet-octet (SO) mixing scheme and the quark-flavor (QF) mixing scheme. The SO mixing angle $\theta$ between $\eta$ and $\eta'$ is known to be in the range of $\theta\in[-10^\circ, -23^\circ]$. In this scheme, The $\eta$ and $\eta'$ are the mixtures of the flavor SU(3) octet $\eta_8$ and single $\eta_0$ states:
\begin{align}
\left( \begin{array}{l}
\eta \\
{\eta'}
\end{array} \right) = \left( \begin{array}{l}
\cos \theta ~~~ - \sin \theta \\
\sin \theta ~~~~~ \cos \theta
\end{array} \right)\left( \begin{array}{l}
{\eta _8}\\
{\eta _0}
\end{array} \right),
\end{align}
where $\eta_8=(u\bar u+d\bar d-2s\bar s)/\sqrt6$ and $\eta_0=(u\bar u+d\bar d+s\bar s)/\sqrt3$. Analogously, information could be gathered on the mixing scheme in the QF basis, which is consists with the form $\eta$ and $\eta'$ states as combinations of $|\eta_q\rangle=|\bar uu + \bar dd\rangle/\sqrt 2$ and $|\eta_s\rangle=|\bar ss\rangle$:
\begin{align}
&|\eta \rangle  = \cos \varphi |{\eta _q}\rangle  - \sin \varphi |{\eta _s}\rangle,
\nonumber\\
&|\eta '\rangle  = \sin \varphi |{\eta _q}\rangle  + \cos \varphi |{\eta _s}\rangle.
\end{align}
It has been shown that in this scheme a single angle is essentially required. In year 2007, the KLOE Collaboration provides the value $\varphi =(41.5\pm0.3_{\rm stat}\pm 0.7_{\rm syst}\pm 0.6_{\rm th})^\circ $ by extracting the pseudoscalar mixing angle $\varphi$ in the QF basis by measuring the radio ${\cal B}(\phi\to\eta'\gamma) / {\cal B}(\phi \to \eta \gamma)$~\cite{Ambrosino:2006gk}. Some theoretical groups have calculated the single mixing angle $\varphi$~\cite{Ball:2007hb, Duplancic:2015zna, Colangelo:2001cv, Azizi:2010zj}, their predicted values are within the range of $\varphi\in[39^ \circ, 41.8^\circ]$. One can put forward the ratio ${\cal R}_{\eta'/\eta} = {\cal B}(D_s\to\eta '\ell^+\nu_\ell)/{\cal B} (D_s\to\eta\ell^+\nu_\ell)$ to access the $\eta-\eta'$ mixing angle through the ratio of the TFFs $f_+^\eta(q^2)/f_+^{\eta'}(q^2)$~\cite{Colangelo:2001cv}, which are related to the $\eta-\eta'$ mixing scheme. In particular, information could be gathered on the mixing scheme in the QF basis. In this paper, we will use the QF basis with the single mixing angle $\varphi$ to analyze the $D_s\to\eta^{(\prime)}$ decay modes, and the corresponding TFFs satisfy the relation
\begin{align}
\tan \varphi = \frac{|f_+^\eta (q^2)|}{|f_+^{\eta'}(q^2)|}. \label{bt}
\end{align}
One usually takes the large recoil point of the squared momentum transfer $q^2 = 0$ to do the calculation, i.e. $\tan \varphi = {|f_+^\eta (0)|}/{|f_+^{\eta'}(0)|}$. A more accurate $D_s^+\to\eta^{(\prime)}\ell^+\nu_\ell$ LCSR analysis is important.

The $\eta$ and $\eta'$ mesons full of rich phenomenology, which are predominantly flavor-singlet states. This means that their wave functions are approximately symmetric in the three lightest quark types (up, down, and strange), which build up the light-hadron spectroscopy. Thus, the LCDAs for $\eta^{(\prime)}$-meson, as one of the most important parameters, composed by $s\bar s$ are significant to the $D_s\to\eta^{(\prime)}$ TFFs, which can be expanded as a Gegenbauer polynomial series:
\begin{eqnarray}
\phi_{2;\eta^{(\prime)}}(u,\mu) = 6 u \bar u \left[1 + \sum_{n=1}^\infty a^n_{2;\eta^{(\prime)}}(\mu) C^{3/2}_n(\xi) \right],  \label{HPDA_CZ}
\end{eqnarray}
where $a^n_{2;\eta^{(\prime)}}(\mu)$ stands for the $n{\rm th}$-order Gegenbauer moment, $\bar{u}=(1-u)$ and $\xi=(2u-1)$. When the factorization scale $\mu$ is large enough, the twist-2 LCDAs $\phi_{2;\eta^{(\prime)}}(u,\mu)$ tends to the asymptotic form $\phi_{2;\eta^{(\prime)}}(u,\infty) = 6u\bar u$. There are some theoretical and experimental predictions for the Gegenbauer moments $a^n_{2;\eta}(\mu)$, such as the fitting results coming from CLEO collaboration $a^2_{2;\eta}(\mu_0) = -0.07\pm0.03$~\cite{Gronberg:1997fj}, the fitting results from BABAR result $a^2_{2;\eta}(\mu_0) = -0.05\pm0.02$~\cite{BABAR:2011ad}, the results predicted by Kroll and Passek-Kumericki $a^2_{2;\eta}(\mu_0) = -0.05\pm0.02$~\cite{Kroll:2013iwa}, and the fitting results from a sum rule analysis $a_{2;\eta}^2 (\mu_0)= 0.25\pm0.15$~\cite{Offen:2013nma}. By taking the approximation with $\pi, K$-meson, Ball and Zwicky predicted $a^2_{2;\eta}(\mu_0) = 0.115$ and $a^4_{2;\eta}(\mu_0) = -0.015$~\cite{Ball:2004ye}. At present, few works have been done to calculate the second and higher order moment. Particularly, the twist-2 LCDA of $\eta'$-meson is rarely studied. So it is important and meaningful to make a more accurate calculation on the second order moment and the higher order moments for $\eta^{(\prime)}$-meson LCDAs within the QCD sum rule approach.

An effective way to calculate the $n$th-order moments of the $\eta^{(\prime)}$-meson LCDAs is to use the following definition,
\begin{eqnarray}
\langle\xi_{2;\eta^{(\prime)}}^n\rangle|_\mu = \int_0^1 du ~\xi^n~ \phi_{2;\eta^{(\prime)}}(u,\mu). \label{Eq:xin}
\end{eqnarray}
$\langle\xi_{2;\eta^{(\prime)}}^n\rangle|_\mu$ can be calculated by using the Shifman-Vainshtein-Zakharov (SVZ) sum rules~\cite{Shifman:1978by, Huang:1986wm, Govaerts:1983ka, Huang:1989gv, Novikov:1983gd, Hubschmid:1982pa}. In which, the perturbative QCD is established on the assumption that the perturbative vacuum and the short-distance interaction are not affected by the long-distance structure of the non-Abelian gauge field. The QCD physical vacuum contains a series of vacuum condensates, such as the quark condensate $\langle q\bar q\rangle$, the gluon condensate $\langle G^2 \rangle$, and etc.. These vacuum condensates reflect the non-perturbation characteristics of QCD. The QCD sum rules based on the background field theory (BFTSR) method gives a possible way to consider the non-perturbation effect, which also provides a systematic description of these vacuum condensates from the field theory point of view. At present, the BFTSR has been used in calculating the twist-2 or twist-3 LCDAs for $\pi$, $K$, $D$, $D_s$, $a_0$, $K_0^*$, $f_0$, $\rho$, $J/\psi$, $a_1(1260)$-mesons~\cite{Zhong:2014jla, Fu:2016yzx, Fu:2018vap, Zhong:2014fma, Zhong:2016kuv,Zhong:2011jf, Han:2013zg, Huang:2004tp, Huang:2005av, Zhong:2011rg, Zhang:2017rwz, Zhang:2021wnv,Hu:2021lkl}. Besides, other methods in studying the LCDAs can be found in Refs.~\cite{Cloet:2013tta, Polyakov:2020cnc, Cheng:2020vwr, Wang:2019msf, Zhang:2017bzy, Zuo:2011sk, Wu:2010zc, Khodjamirian:2006st, Zuo:2006re}. In this paper, we will calculate the $\eta^{(\prime)}$-meson twist-2 LCDAs within the BFTSR approach for the first time.

The rest of the paper are organized as follows. In Sec.~\ref{sec:2}, we present the basic idea of the QCD background field theory, the detailed BFTSR procedures for calculating the moments of $\phi_{2;\eta^{(\prime)}}(u,\mu)$, the branching fractions and the transition form factors involved in the semileptonic decay $D_s^+\to \eta^{(\prime)} \ell^+ \nu_\ell$. In Sec.~\ref{sec:3}, we present our numerical results and make a detailed comparison with other experimental and theoretical predictions. Section~\ref{sec:summary} is reserved for a summery. The intermediate processes for calculating the moments of $\eta^{(\prime)}$-meson LCDA within the BFTSR and the basic definition of $\eta^{(\prime)}$-meson twist-2, 3, 4 LCDAs are given in the Appendixes~\ref{sec:appendixA}, \ref{sec:appendixB} and \ref{sec:appendixC}.

\section{Calculation Technology}\label{sec:2}

\subsection{Basic idea and formulas for background field theory}

The important aspect for the SVZ sum rules approach is the OPE, which has been introduced by using the QCD physical vacuum $ \langle 0|{\cal O}_N|0\rangle_{\rm phys.}$. Its special property is the non-perturbation effect, which can be described by the classical background field satisfying the equations of motion. The main idea of the background field theory method is to describe the non-perturbation effect with the classical background field satisfying the equation of motion and to describe the quantum fluctuation, namely the perturbation effect, on this basis within the frame of the quantum field theory. More specifically, one can use the following substitution in the theoretical Lagrangian and Green's functions~\cite{Makeenko:1979pb, Shifman:1978zq, Huang:1986wm}
\begin{align}
&{\cal A}_\mu^a (x)\to {\cal A}_\mu^a (x) + \phi_\mu^a(x),
\nonumber\\
&\psi(x)\to\psi(x)+\eta(x), \label{AF}
\end{align}
where ${\cal A}_\mu^a (x)$ and $\psi(x)$ stand for the gluon and quark background fields, $\phi_\mu^a(x)$ and $\eta(x)$ represent their quantum fluctuations, respectively. In the presence of background fields, the quantization of $\phi_\mu^a(x)$ and $\eta(x)$ has been completed in Ref.~\cite{Ambjorn:1982bp}. Among them, the gluon quantum field satisfies the background field gauge,
\begin{align}
& D_\mu ^{ab}({\cal A})\phi _b^\mu  = 0,
\nonumber\\
& D_\mu ^{ab}({\cal A}) = \delta^{ab}\partial_\mu - g_sf^{abc}{\cal A}_\mu ^c , \label{ca}
\end{align}
where the color indices $a,b,c = (1,2,...,8)$, $g_s$ is the coupling constant of strong interactions, $f^{abc}$ is the structure constant of the SU$_f$(3) group. The advantage of choosing the background field gauge is that it makes the theory be invariant under the background field gauge, making the calculated physical quantity independent of the gauge. With the help of Eq.~\eqref{AF}, one can obtain the following effective Lagrangian~\cite{Huang:1989gv}
\begin{align}
{\cal L}_{\rm eff} & = {\cal L}_{\rm QCD}({\cal A},\psi ) + {\cal L}({\rm ghosts}) + \bar \eta (i\DSS D - m)\eta  + \frac12\phi _\mu ^a \bigg\{g^{\mu \nu}D_{ac}^2-\Big(1-\frac1\alpha\Big) [D^\mu {D^\nu }]_{ac}
\nonumber\\
&+ 2{g_s}f^{abc} G_b^{\mu \nu }\bigg\}\phi_\nu^c ~+~ g_s(\bar \psi \DS\phi^a T^a\eta  +\bar\eta \DS\phi^aT^a\eta) ~-~ g_s^2f^{adf}f_{abc}{\cal A}_d^\mu \phi _f^\nu \phi _\mu ^b\phi _\nu ^c - g_s f^{abc}
\nonumber\\
& \times (\partial_\mu \phi _\nu ^a)\phi _b^\mu \phi _c^\nu-\frac{1}{4}g_s^2 f^{abc} f_{acf}\phi _b^\mu \phi _c^\nu \phi _\mu ^d\phi _\nu ^f + {g_s}\bar \eta {\DS\phi ^a}{T^a}\eta, ~\label{la}
\end{align}
where the ${\cal L}_{\rm QCD}({\cal A},\psi )$ with ${\cal A}_\mu^a (x)$, $\psi(x)$ has the usual form of QCD Lagrangian and can be minimized to zero when the classical fields ${\cal A}_\mu^a (x)$ and $\psi(x)$ are the solutions of the equation of motion, and $\alpha$ is the gauge-fixing parameter. The ${\cal L}({\rm ghosts})$ is the contribution of the ghost particle term. Especially, in order to describe the various quark-antiquark pairs and gluons in a vacuum, one can follow the classical QCD Lagrangian~\cite{Huang:1989gv}
\begin{align}
{\cal L}_{\rm QCD} ({\cal A},\psi) = -\frac14 G_{\mu\nu}^a G^{a\mu\nu} + \bar\psi(i\DSS D - m)\psi,
\end{align}
with $ G_{\mu\nu}^a = \partial_\mu {\cal A}_\nu^a - \partial_\nu {\cal A}_\mu^a + g_s f^{abc} {\cal A}_\nu^b {\cal A}_\mu^c$ stands for the gluon field strength tensor. The gluon field ${\cal A}_\mu^a(x)$ and quark field $\psi(x)$ satisfy the QCD equations of motion,
\begin{align}
&(i/\!\!\!\! D - m)\psi (x)=0 , \nonumber\\
&\widetilde D_\mu^{ab} G^{\nu\mu}_b(x)= g_s \bar\psi(x)\gamma^\nu T^a \psi(x) , \label{eq:mot}
\end{align}
where $D_\mu = \partial_\mu - i g_s T^a {\cal A}_\mu^a(x)$ with $a,b,c = (1,2,...,8)$ and $\widetilde D_\mu^{ab} = \delta^{ab}\partial_\mu - g_s f^{abc} {\cal A}_\mu^c(x)$ are fundamental and adjoint representations of the gauge covariant derivative, respectively. As an advantage of using the background field theory, one can take different gauges for dealing with the quantum fluctuations and background fields. More specifically, one can adopt the background gauge, i.e. $\widetilde{D}^{AB}_\mu \phi^{B \mu}(x) = 0$ for the gluon quantum field~\cite{BG3,BG4}, the Schwinger gauge or the fixed-point gauge, i.e. $x^\mu \mathcal{A}^A_\mu(x) = 0$ for the background field~\cite{Shifman:1980ui}.

As the background field satisfies the motion equation~\eqref{eq:mot}, at least one background field is included in the coupling between the quantum fluctuation field and the background field. In the effective Lagrangian, there is no contribution from the vertex ${g_s}\bar \psi {\gamma ^\mu }\phi_\mu ^a{T^a}\psi$, and it only has the vertex ${g_s}\bar \psi \DS\phi _\mu ^a{T^a}\eta$ and its conjugate contributions. According to the effective Lagrangian, the quantum quark and gluon propagators in the background field are~\cite{Huang:2004tp}
\begin{align}
& S_F(x) = i[i\gamma^\mu D_\mu-m]^{-1}, \label{qu}
\\
& S_{\mu \nu }^{ab}(x) = i[g_{\mu \nu}({D^2})^{ab} + 2{g_s} f^{abc}G_{\mu \nu }^c]^{ - 1}   \label{glu},
\end{align}
where the gauge-fixing parameter is taken as $\alpha=1$. Within the framework of BFT, the quark propagator will be affected by the background quark and/or gluon fields, which satisfies the equation
\begin{eqnarray}
(i \DSS D - m) S_F(x,0) = \delta^4 (x).
\label{quaprofun1}
\end{eqnarray}
If taking
\begin{eqnarray}
S_F(x,0) = (i\DSS D + m) \mathcal{D}(x,0),
\label{quaprofun2}
\end{eqnarray}
Eq.~(\ref{quaprofun1}) can be changed as
\begin{eqnarray}
(\Box - \mathcal{P}_\mu \partial^\mu - \mathcal{Q} + m^2) \mathcal{D}(x,0) = \delta^4(x),
\label{quaprofun3}
\end{eqnarray}
where $\Box = \partial^2$, and
\begin{align}
&\mathcal{P}_\mu = 2i\mathcal{A}_\mu(x), \nonumber\\
&\mathcal{Q} = \gamma^\nu \gamma^\mu \mathcal{A}_\nu(x) \mathcal{A}_\mu(x) + i \gamma^\nu \gamma^\mu \partial_\nu \mathcal{A}_\mu(x).
\end{align}
Moreover, after applying the fixed-point gauge, the gluon background field can be expressed by using the gauge invariant $G_{\mu\nu;\alpha_1 \cdots \alpha_n}$ as
\begin{align}
\mathcal{A}_\mu(x) &= \frac{1}{2} x^\nu G_{\nu\mu} + \frac{1}{3} x^\nu x^\alpha G_{\nu\mu;\alpha} + \frac{1}{8} x^\nu x^\alpha x^\beta G_{\nu\mu;\alpha\beta}
+ \frac{1}{30} x^\nu x^\alpha x^\beta x^\gamma G_{\nu\mu;\alpha\beta\gamma}
\nonumber\\
& + \frac{1}{144} x^\nu x^\alpha x^\beta x^\gamma x^\delta G_{\nu\mu;\alpha\beta\gamma\delta} + \cdots, \label{gluexp}
\end{align}
where $G_{\mu\nu;\alpha_1 \cdots \alpha_n}$ is the notation for $(D_{\alpha_1}\cdots D_{\alpha_n}G_{\mu\nu})(0)$, where the indexes $\alpha_1 \cdots \alpha_n$ indicates the covariant derivative up to $n$-th order. Substituting Eq.~(\ref{gluexp}) into Eq.~(\ref{quaprofun3}), we obtain the expressions for $\mathcal{D}(x,0)$. By further using Eq.~(\ref{quaprofun2}), we obtain the required quark propagators in the background field, i.e.,
\begin{equation}
S_F(x,0)=S_F^0(x,0)+S_F^2(x,0)+S_F^3(x,0)+ \sum_{i=1}^{2} S_F^{4(i)}(x,0) + \sum_{i=1}^{3} S_F^{5(i)}(x,0) + \sum_{i=1}^{5} S_F^{6(i)}(x,0). \label{prop4}
\end{equation}
We present the quark propagators with various gauge invariant tensors $G_{\mu\nu;\alpha_1 \cdots \alpha_n}$ that shall result in up to dimension-six operators in the Appendix~\ref{sec:appendix_I1}. Because the fixed-point gauge  violates the translation invariance, the quark propagator from $x$ to $0$, $S_F(0,x)$, can not be directly obtained by applying the replacement $x \to -x$ in Eq.~(\ref{prop4}). However, it can be related with $S_F(x,0)$ via the relation~\cite{Hubschmid:1982pa}
\begin{eqnarray}
S_F(0,x|\mathcal{A}) = C S^{\rm T}_F(x,0|-\mathcal A^{\rm T}) C^{-1},
\label{revpro}
\end{eqnarray}
where $C$ stands for the charge conjugation matrix and the symbol $T$ indicates transposition of both the Dirac and the color matrices.

Furthermore, one will encounter the vertex operator $\Gamma(z\cdot \tensor{D})^n$ with $\Gamma$ indicates all kinds of Dirac matrices for heavy/light-meson twist-2, 3 LCDAs. Generally, we have the following expansion~\cite{Zhong:2014jla}
\begin{eqnarray}
(z\cdot \tensor{D})^n = (z\cdot \overrightarrow{D} - z\cdot \overleftarrow{D})^n = (z\cdot \tensor{\partial} + z\cdot B)^n + \cdots,
\end{eqnarray}
where the ellipsis denotes for the higher-order terms, which are irrelevant for our present analysis and
\begin{eqnarray}
z\cdot B &=& -2 i z\cdot \mathcal{A} \nonumber\\
         &=& -i x^\mu z^\nu G_{\mu\nu} - \frac{2i}{3} x^\mu x^\rho z^\nu G_{\mu\nu;\rho}  - \frac{i}{4} x^\mu x^\rho x^\sigma z^\nu G_{\mu\nu;\rho\sigma} - \frac{i}{15} x^\mu x^\rho x^\sigma x^\lambda z^\nu G_{\mu\nu;\rho\sigma\lambda} \nonumber\\
&-& \frac{i}{72} x^\mu x^\rho x^\sigma x^\lambda x^\tau z^\nu G_{\mu\nu;\rho\sigma\lambda\tau} + \cdots. \label{zB}
\end{eqnarray}
We can expand the operator $(z\cdot \tensor{D})^n$ into series of the operators $(z\cdot \tensor{\partial})^n$ and $G_{\mu\nu;\rho\cdots}$. For the purpose, we first expand $(z\cdot \tensor{D})^n$ as
\begin{eqnarray}
(z\cdot \tensor{D})^0 &=& 1, \nonumber\\
(z\cdot \tensor{D})^1 &=& z\cdot \tensor{\partial} + z\cdot B, \nonumber\\
(z\cdot \tensor{D})^2 &=& (z\cdot \tensor{\partial})^2 + 2(z\cdot \tensor{\partial}) (z\cdot \underline{B}) + (z\cdot B)^2, \nonumber\\
(z\cdot \tensor{D})^3 &=& (z\cdot \tensor{\partial})^3 + 3(z\cdot \tensor{\partial})^2 (z\cdot \underline{B}) + \left[ (z\cdot \partial)^2 (z\cdot B) \right]  + 3 (z\cdot \tensor{\partial}) (z\cdot \underline{B})^2 + (z\cdot B)^3,  \nonumber\\
&& \cdots \cdots ,
\end{eqnarray}
where, the operator ``underline'' below symbol ``$B$'' (or ``$x$'' in the Appendix~\ref{sec:appendix_I2}) indicates that the operation $\tensor{\partial}$ does not act on it. In deriving those equations, the following equation has been adopted,
\begin{align}
(z\cdot \tensor{\partial})^n (z\cdot B) = \sum^n_{k=0} \frac{n!}{k! (n-k!)} (z\cdot \tensor{\partial})^{n-k} \left[ (z\cdot \partial)^k (z\cdot \underline{B}) \right].
\end{align}
By keeping only those terms that shall leads to operators up to dimension-six, we can obtain the full expression for the vertex operator, which are listed in the Appendix~\ref{sec:appendix_I2}.

\subsection{SVZ sum rules for the moments of $\phi _{2;\eta^{(\prime)}}(x,\mu)$}

Following the traditional way of constructing the light pseudoscalar meson, we take the following definition
\begin{align}
\langle 0| \bar s (0){\cal C}_s\DS z\gamma_5[z,-z] {(iz\cdot\tensor D)^{n}}s(0)|\eta^{(\prime)}(q)\rangle =i{(z\cdot q)}^{n+1}  f_{\eta^{(\prime)}} \langle\xi^n_{2;\eta^{(\prime)}}\rangle|_\mu, \label{Eq:xi_defination}
\end{align}
where $z_\mu$ stand for the light-like vector and $[z,-z]$ is the path-ordered gauge connection, $f_{\eta^{(\prime)}}$ are the $\eta$ and $\eta^{\prime}$-meson decay constant, $(iz \cdot \tensor D)^n=(iz\cdot \overrightarrow D - iz\cdot \overleftarrow D)^n$. Basis on the QF scheme, the flavour content is ${\cal C}_s = ({\cal C}_1 - \sqrt{2}{\cal C}_8)/ \sqrt{3}$ with SO basis ${\cal C}_1 = \mathbf{1}/\sqrt{n_f}$ and ${\cal C}_8 = \lambda_8/\sqrt{2}$~\cite{Duplancic:2015zna,Ball:2007hb}. In which the $\lambda_i$ is the standard SU$_f$(3) Gell-Mann matrix and $\mathbf{1}$ is the $3\times 3$ unit matrix. As a special case, the 0th-order LCDA's moment for Eq.~\eqref{Eq:xin} satisfies the normalization condition
\begin{align}
\langle \xi^0_{2;\eta^{(\prime)}}\rangle|_\mu  =\int_0^1 du \phi_{2;\eta^{(\prime)}}(u,\mu)=1.
\end{align}
To derive the SVZ sum rules for the moments $\langle \xi^n_{2;\eta^{(\prime)}}\rangle|_\mu$, we introduce the following correlation function,
\begin{align}
\Pi_{2;\eta^{(\prime)}}^{(n,0)}(z,q) &=i \int d^4x e^{iq\cdot x}\langle 0 |T\{ J_n(x), J_0^\dagger (0) \}|0\rangle
\nonumber\\
&=(z\cdot q)^{n+2} I_{2;\eta^{(\prime)}}^{(n,0)}(q^2),  \label{gp}
\end{align}
where $J_n(x)= \bar s(x){\cal C}_s\DS z\gamma_5 (iz \cdot\tensor D)^ns(x)$, $J_0^{\dagger}(0)= \bar s(0){\cal C}_s\DS z\gamma_5 s(0)$ and $ z^2 \rightsquigarrow 0$. Only even moments are non-zero and the odd moments of the LCDA are zero because of G-parity, i.e. $n=(0,2,4,6,...)$ will contribute to the final results~\footnote{This point can also been seen in Eq. (16) of Ref.~\cite{Ball:2007hb}.}. At one hand, in deep Euclidean region $q^2 \ll 0$, one can apply the OPE for the correlation function Eq.~\eqref{gp}.
\begin{align}
\Pi_{2;\eta^{(\prime)}}^{(n,0)}(z,q) =i\int d^4x e^{iq\cdot x}{\rm Tr}[{\cal C}_s{\cal C}_s]
\Big\{\!&-{\rm Tr}\langle 0|S_F^s(0,x)\DS z{{\gamma }_{5}}(iz\cdot \tensor D)^n S_F^s(x,0)\DS z\gamma_5|0\rangle
\nonumber\\
& +{\rm Tr}\langle 0|\bar{s}(x)s(0)\DS z{{\gamma }_{5}}{(iz\cdot \tensor{D})^{n}}S_F^s(x,0)\DS z{{\gamma }_{5}}|0\rangle
\nonumber\\
&+{\rm Tr}\langle 0|S_F^s(0,x)\DS z{{\gamma }_{5}}{(iz\cdot \tensor{D})^{n}}\bar{s}(0)s(x)\DS z{{\gamma }_{5}}|0\rangle
\nonumber\\
& +\cdots \Big\},
\end{align}
where ${\rm Tr}[{\cal C}_s{\cal C}_s] =1$. In the detailed OPE calculation, we adopt the $\overline{\rm MS}$-scheme to deal with the infrared divergences. Lorentz invariant scalar function $I_{2;\eta^{(\prime)}}^{(n,0)}(q^2)$ in Eq.~\eqref{gp} depends on the condensation parameter and will encounter the vacuum matrix elements of the following form,
\begin{align}
&\langle 0|G_{\mu\nu}^a G_{\mu\nu}^b|0\rangle,
&&\langle 0|G_{\mu \nu }^{a}G_{\rho\sigma }^{b}G_{\lambda \tau }^{c}|0\rangle ,
\nonumber\\
&\langle 0|G_{\mu \nu ;\lambda }^{a}G_{\rho \sigma ;\tau }^{b}|0\rangle, && \langle 0|G_{\mu \nu }^{a}G_{\rho \sigma ;\lambda \tau }^{b}|0\rangle ,
\nonumber\\
&\langle 0|G_{\mu \nu ;\lambda \tau }^{a}G_{\rho \sigma }^{c}|0\rangle, && \langle 0|\bar q_\alpha^a(x) q_\beta^b(y)|0\rangle,
\nonumber\\
&\langle 0|\bar q_\alpha^a(x)q_\beta^b(y)G_{\mu\nu}^A|0\rangle,
&& \langle 0|\bar q_\alpha^a(0) q_\beta^b(0) G_{\mu\nu;\rho}^A|0\rangle.
\end{align}
The full expression for the vacuum condensates which one may encounter in the following calculations are listed in the Appendix~\ref{sec:appendix_I3}, which can also be found in our previous work~\cite{Zhong:2014jla}. On the other side, the correlation function ~\eqref{gp} can be treated by inserting a complete set of intermediate hadronic states in physical region to obtain its hadronic representation
\begin{align}
{\rm Im}I_{2;\eta^{(\prime)},{\rm Had}}^{(n,0)}( q^2 )&=\pi \delta(q^2 - m_{\eta^{(\prime)}}^2) f_{\eta^{(\prime)}}^2 \langle \xi^n_{2;\eta^{(\prime)}} \rangle|_\mu  +\pi \frac{3}{4\pi^2 (n+1)(n+3)}\theta ( q^2 -s_{\eta^{(\prime)}}),  \label{rm}
\end{align}
in which $s_{\eta^{(\prime)}}$ stand for the continuum threshold for the lowest continuum state. The first term is the contribution of $\eta^{(\prime)}$-meson poles, and the second term is the contribution of continuum states above poles. Then both of the OPE part and the hadronic representation of the invariant function $I_{2;\eta^{(\prime)}}^{(n,0)}( q^2)$ can be marched with the following dispersion relation
\begin{align}
\frac{1}{\pi}\int_{4m_s^2}^{\infty }{ds}\frac{{\rm Im}I_{2;\eta^{(\prime)} ,{\rm Had}}^{(n,0)}(s)}{s-q^2}=I_{2;\eta^{(\prime)} ,{\rm QCD}}^{(n,0)}({{q}^{2}}).  \label{Ds}
\end{align}
The Borel transform helps to reduce the contribution from the continuum states on the left of Eq.~\eqref{Ds} and the contribution of the higher-dimension condensates on the right, and finally the sum rules can be obtained,
\begin{align}
\frac1{\pi}\frac1{M^2}\int ds e^{-s/M^2}{\rm Im}I_{2;\eta^{(\prime)}  , {\rm had}}^{(n,0)}(s)=\hat{\cal B}_{M^2} I_{2;\eta^{(\prime)} ,{\rm QCD}}^{(n,0)}(q^2),
\end{align}
where $M^2$ is Borel parameter, $\hat{\cal B}_{M^2}$ is Borel transformation operator,
\begin{align}
\hat{\cal B}_{M^2} = \mathop {\lim }\limits_{\scriptstyle - q^2, n\to\infty \hfill\atop
\scriptstyle{-q^2/n} = {M^2}\hfill} \frac{1}{(n - 1)!}(-q^2)^n \Bigg( - \frac{d}{d( -q^2)}\Bigg)^n.
\end{align}
In order to deal with the $s$-quark mass ($m_s$) contribution to $\langle\xi_{2;\eta^{(\prime)}}^{n}\rangle|_\mu$, we take the expansion according to different orders of $m_s^{k}$ with $k=(0,2,4,...)$, i.e. $I_{m_s^{k}}(n, M^2)$, since $m_s$ is closer to $\Lambda_{\rm QCD}$, which is different from our previous treatment for the heavy quark such as $q=(c,b)$ in Refs.~\cite{Fu:2018vap, Zhong:2014fma}. Here, we take the first two orders of the squared $s$-quark mass, i.e. $I_{m_s^0}(n, M^2)$ and $I_{m_s^2}(n, M^2)$. The reason lies in the suppression of $m_s^4 < 0.1\%$ for the third-order, which are quite small and can be safely neglected. Recently, we have suggested a new method for renormalization of various moments, i.e. the 0th-order moment $\langle\xi_{2;\eta^{(\prime)}}^0\rangle|_\mu$ should be considered to the total results. The reason lies in the accuracy is often up to dimension-six condensates and the NLO QCD corrections for the perturbative part instead of the infinite dimension or infinite-order perturbative parts~\cite{Zhong:2021epq}. The final expression for $\langle\xi_{2;\eta^{(\prime)}}^{n}\rangle|_\mu$ can then be written as
\begin{align}
&\frac{f_{\eta^{(\prime)}}^2 \langle\xi^n_{2;\eta^{(\prime)}}\rangle|_\mu \langle \xi^0_{2;\eta^{(\prime)}}\rangle|_\mu}{M^2 e^{m_{\eta^{(\prime)}}^2/M^2}}  = \frac1\pi \frac1{M^2} \int_{4m_s^2}^{s_{\eta^{(\prime)}}} ds e^{-s/M^2} \frac{3 v^{n+1}}{8\pi(n+1)(n+3)} \left(1+ \frac{\alpha_s}{\pi }A'_n\right) \bigg\{[1+(-1)^n]
\nonumber
\\
&\hspace{1.8cm}\times(n+1)\frac{1-v^2}2 + [1+(-1)^n]\bigg\}+\frac{2m_s\langle\bar ss\rangle}{M^4}+\frac{\langle \alpha_s G^2\rangle}{12\pi M^4}~\frac{1+n\theta(n-2)}{n+1} \,-\, \frac{8n+1}9
\nonumber
\\
&\hspace{1.8cm}\times\frac{m_s\langle g_s\bar s\sigma TGs\rangle}{M^6} + \frac{\langle g_s \bar ss\rangle }{81M^6}4(2n + 1) -\frac{\langle g_s^3fG^3\rangle}{48\pi^2M^6}n\theta(n-2)+\frac{\sum\langle g_s^2\bar qq\rangle ^2}{486\pi^2M^6}\bigg\{\!\!-2(51n
\nonumber
\\
&\hspace{1.8cm}
+25)\bigg(-\ln\frac{M^2}{\mu^2}\bigg)+3\,(17n+35)+\theta(n-2)\bigg[\,2n\bigg(-\ln\frac{M^2}{\mu^2}\bigg)-25\,(2n+ 1)\,\tilde \psi (n)
\nonumber
\\
&\hspace{1.8cm} + \frac1n(49n^2+100n+56)\bigg]\bigg\}+ I_{m_s^2}(n,M^2).\label{xi2}
\end{align}
Due to the mass of $s$ quark is heavier than $u$, $d$-quark, the $I_{m_s^2}(n, M^2)$-terms should be considered in this paper, which are
\begin{align}
&I_{m_s^2}(n,M^2)= m_s^2\bigg\{-\frac{\langle\alpha_s G^2\rangle}{6\pi M^6}\left[\theta(n-2)(n\tilde\psi (n)-2) + 2n\left(-\ln\frac{M^2}{\mu^2}\right)-n-2\right]+\frac{\langle g_s^3fG^3\rangle}{288\pi^2M^8}
\nonumber\\
&\hspace{0.6cm} \times \bigg\{\!-10 \delta^{n0} + \theta (n - 2) \left[ 4n(2n-1)\left( -\ln\frac{M^2}{\mu^2}\right) - 4n\tilde \psi (n) + 8({n^2} - n + 1) \right] + \theta(n-4)
\nonumber\\
&\hspace{0.6cm}\times [2n(8n-1)\tilde\psi(n) - (19{n^2} + 19n + 6)]+ 8n\,(3n - 1)\left( -\ln\frac{M^2}{\mu^2} \right) - (21n^2+53n-6)\bigg\}
\nonumber\\
&\hspace{0.6cm}- \frac{\sum \langle g_s^2q\bar q\rangle^2 }{972\pi^2M^8}~\bigg\{ 6{\delta ^{n0}}\left[ {16\left( -\ln\frac{M^2}{\mu^2} \right) - 3} \right] + \theta (n - 2) \bigg[8(n^2+ 12n -12)\left( -\ln\frac{M^2}{\mu^2} \right) - 2
\nonumber\\
&\hspace{0.6cm} \times (29n + 22)\, \tilde \psi (n)~ +~ 4\left(5{n^2} - 2n - 33 + \frac{46}{n}\right)\bigg] + \theta (n - 4)~\bigg[2\left(56{n^2} - 25n + 24\right) \tilde \psi(n)
\nonumber\\
&\hspace{0.6cm}  -\,\left(139 n^2 + 91n + 54\right)\bigg] ~+~ 8\left(27n^2-15n-11\right)~ \left(-\ln\frac{M^2}{\mu^2} \right) - 3(63n^2+159n-50) \bigg\}
\nonumber\\
&\hspace{0.6cm} +\frac{4(n-1)}3 \frac{m_s\langle\bar ss\rangle }{M^6} + \frac{8n-3}9 \frac{m_s\langle g_s\bar s\sigma TGs\rangle}{M^8} - \frac{4(2n+1)}{81}\frac{\langle g_s \bar ss\rangle^2}{M^8}\bigg\}.    \label{ms}
\end{align}
For convenience, we put the detailed terms contribute to the BFTSR $\langle\xi_{2;\eta^{(\prime)}}^n\rangle|_\mu$ in the Appendix~\ref{sec:appendixB}. By taking the index $n$ to zero, we get the sum rule of the 0th-order moment, which takes the following form,
\begin{align}
&\frac{(\langle\xi^0_{2;\eta^{(\prime)}}\rangle|_\mu)^2 f_{\eta^{(\prime)}}^2}{M^2e^{m_{\eta^{(\prime)}}^2/M^2}}= \frac1\pi \frac1{M^2}\int_{4m_s^2}^{s_{\eta^{(\prime)}}}ds e^{-s/M^2} \frac{v(3-v^2)}{8\pi} + \frac{2m_s\langle\bar ss\rangle}{M^4}-\frac{m_s\langle g_s\bar s\sigma TGs\rangle}{9M^6}+ \frac{\langle \alpha_s G^2\rangle}{12\pi M^4}
\nonumber\\
&\hspace{1.05cm} + \frac{4\langle g_ss \bar s\rangle^2}{81M^6}+\frac{\sum\langle g_s^2\bar qq\rangle^2}{M^6} \frac1{486{\pi^2}}\bigg[-50\bigg(-\ln\frac{M^2}{\mu^2}\bigg) + 105\bigg] + m_s^2\bigg\{\frac{\langle \alpha_s G^2\rangle }{3\pi M^6}- \frac{\langle g_s^3fG^3\rangle}{72\pi^2M^8}
\nonumber\\
&\hspace{1.05cm}  - \frac{\sum\langle g_s^2\bar qq\rangle^2}{972\pi ^2M^8} ~\Bigg[8\left( -\ln\frac{M^2}{\mu^2} \right)-132 \Bigg] - \frac43 \frac{m_s\langle \bar ss\rangle}{M^6} \,-\, \frac13\frac{m_s\langle g_s\bar s\sigma TGs\rangle}{M^8} \,-\, \frac4{81}\frac{\langle g_s\bar ss\rangle^2}{M^8}\bigg\},
\end{align}
where
\begin{align}
\widetilde\psi(n)=\psi\bigg(\frac{n+1}2\bigg)-\psi \bigg(\frac n2\bigg)+\ln4 ,
\end{align}
with $v^2 = 1 - 4m_s^2/s$ and $A'_0=0$, $A'_2=5/3$, $A'_4=59/27$, $A'_6=353/135$ are the NLO correction to the perturbative part~\cite{Ball:1996tb}. The 0th-order derivative of the digamma function $\psi(n+1) = \Sigma_{k=1}^n 1/k -\gamma_E$, where the Euler's constant $\gamma_E = 0.557216$. Furthermore, in order to get the Gegenbauer moment $a_{2;\eta^{(\prime)}}^n$, one can expand $\phi_{2;\eta^{(\prime)}}^n(u, \mu)$ into a Gegenbauer polynomial series by using Eq.~\eqref{HPDA_CZ} and the basic definition of $\langle\xi^n_{2;\eta^{(\prime)}}\rangle|_\mu$~\eqref{Eq:xin}. Then, one can get the following relations up to $6$th-order,
\begin{align}
&\langle\xi^2_{2;\eta^{(\prime)}}\rangle|_\mu=\frac15 + \frac{12}{35}a_{2;\eta^{(\prime)}}^2(\mu),
\nonumber\\
&\langle\xi^4_{2;\eta^{(\prime)}}\rangle|_\mu=\frac3{35} + \frac8{35}a_{2;\eta^{(\prime)}}^2(\mu) + \frac8{77}a_{2;\eta^{(\prime)}}^4(\mu),
\nonumber\\
&\langle\xi^6_{2;\eta^{(\prime)}}\rangle|_\mu=\frac1{21} + \frac{12}{77} a_{2;\eta^{(\prime)}}^2(\mu) + \frac{120}{1001}a_{2;\eta^{(\prime)}}^4(\mu) +\frac{64}{2145} a_{2;\eta^{(\prime)}}^6(\mu).  \label{xi}
\end{align}
Following this method, one can get higher-order Gegenbauer moments.

\subsection{The semileptonic decay $D_s^+\to \eta^{(\prime)}\ell^+\nu_\ell $}

All the following calculations are performed under the Standard Model (SM). In order to derive the full analytical LCSR expressions for the TFFs, we use the traditional current method to calculate the TFFs. The correlation function for the TFFs of $D_s\to\eta^{(\prime)}$ is defined as~\cite{Duplancic:2008ix}:
\begin{align}
\Pi_\mu(p,q) & =i\int d^4 x e^{iqx} \langle \eta^{(\prime)}(p)|T\{\bar s(x)\gamma_\mu c(x),\bar c(0)i\gamma_5 s(0) \}|0 \rangle \nonumber\\
& = \Pi[q^2, (p+q)^2] p_\mu + \tilde \Pi[q^2, (p+q)^2] q_\mu.
\label{Correlation function}
\end{align}
Following the basic procedure of LCSR approach, the correlation function can be treated by inserting complete intermediate states with the same quantum numbers as the current operator $(\bar c i \gamma_5 s)$ in the time-like ${(p+q)}^2$-region. After isolating the pole term of the lowest pseudoscalar $D_s$-meson, one can reach the following expression,
\begin{align}
\Pi_\mu^{\rm had}(p,q)&=\frac{\langle\eta^{(\prime)}(p)|\bar s\gamma_\mu c|D_s(p+q)\rangle \langle D_s(p+q)|\bar ci\gamma_5q|0\rangle }{m_{D_s}^2-(p+q)^2}
\nonumber\\
&+\sum\limits_{\rm H}{\frac{\langle\eta^{(\prime)}(p)|\bar s\gamma_\mu c|D_s^{\rm H}(p+q)\rangle \langle D_s^{\rm H}(p+q)|\bar ci\gamma_5q|0\rangle}{m_{D_s^{\rm H}}^2-(p+q)^2}}\nonumber\\[1.2ex]
& = \Pi^{\rm had}[q^2,(p+q)^2]p_\mu+\widetilde\Pi^{\rm had}[q^2,(p+q)^2]q_\mu,
\label{Eq:Hadronic Expression}
\end{align}
with the superscript ``had'' and ``H'' stand for the hadronic expression of the correlation function and higher-excited state of $D_s$-meson, respectively. The decay constant of $D_s$-meson can be defined via the relation, $\langle D_s|\bar ci\gamma_5q|0\rangle = m_{D_s}^2f_{D_s}/m_c $. The definition of transition matrix element for $D_s\to \eta^{(\prime)}$ has the following form
\begin{align}
\langle\eta^{(\prime)}(p)|\bar s\gamma_\mu c| D_s(p+q)\rangle  = 2f^{\eta^{(\prime)}}_+(q^2)p_\mu
 + \tilde f^{\eta^{(\prime)}}(q^2)q_\mu.
\end{align}
with $\tilde f^{\eta^{(\prime)}} = f^{\eta^{(\prime)}}_+(q^2)+f^{\eta^{(\prime)}}_-(q^2)$. Then one can take the imaginary part of the invariant amplitude $\Pi^{\rm had}[q^2,(p+q)^2]$ and $\widetilde\Pi^{\rm had}[q^2,(p+q)^2]$ which has the following form,
\begin{eqnarray}
&&{\rm Im}\Pi^{\rm had}(q^2,s) = \pi \delta(s-m_{D_s^2}) \frac{2m_{D_s^2}f_{D_s}f_+^{\eta^{(\prime)}}}{m_c} + \pi\rho^{\rm H}(s) \theta(s-s_0)
\nonumber\\
&&{\rm Im}\widetilde\Pi^{\rm had}(q^2,s) = \pi \delta(s-m_{D_s^2}) \frac{m_{D_s^2}f_{D_s}\tilde f_+^{\eta^{(\prime)}}}{m_c} + \pi\tilde\rho^{\rm H}(s) \theta (s-\tilde s_0)
\end{eqnarray}
where $\rho^{\rm H}(s)$ and $\tilde\rho^{\rm H}(s)$ denote the spectral density of higher resonance and the continuum states $D_s^{\rm H}$, which can be approximated by invoking the so-called quark-hadron duality ansatz $\varrho^{\rm H}(s) = \varrho^{\rm QCD}(s)$ with $\varrho(s) = (\rho(s), \tilde\rho(s))$ and the perturbative spectral density $\rho^{\rm QCD}$ and $\tilde \rho^{\rm QCD}$, the usual step-function $\theta(x)$. Here, the $s_0$ ($\tilde s_0$) are effective parameters, which characterize the lower limit of continuum state, namely the continuum threshold parameters. After extracting the pole terms, the $s_0(\tilde s_0)$ can separate the ground state $D_s$ and the excited state $D_s^{\rm H}$. When taking the limit for the interval between the two adjacent excited states, the sum of excited states is transformed into the integral of continuum states and $s_0(\tilde s_0)$ will be changed into the lower limit of the integration. Traditionally, the continuum threshold are often taken as the magnitude that is close to the squared mass of the first excited state. The ground state of $D_s$-meson is calculated by the LCSR approach. The excited states' contribution can be highly suppressed when making the Borel transformation. Meanwhile, the continuum states, contribution is usually required to be less than 30\% so as to make the LCSR calculation more accurate and reliable. Then, one can use a general dispersion relation in the momentum squared $(p+q)^2$, which can establish a relationship with the QCD parts, i.e.
\begin{align}
\Pi^{\rm QCD}(q^2,s)&=\frac1{\pi} \int_{m_c^2}^\infty \frac{{\rm Im}\Pi ^{\rm had}(q^2,s)}{s-(p+q)^2}ds,
\nonumber\\
\widetilde\Pi^{\rm QCD}(q^2,s)&=\frac1{\pi} \int_{m_c^2}^\infty \frac{{\rm Im}\tilde\Pi ^{\rm had}(q^2,s)}{s-(p+q)^2}ds.
\end{align}
Here, we shall only deal with ${\rm Im}\Pi^{\rm had}[q^2,s]$ for the TFFs $f^{\eta^{(\prime)}}_+(q^2)$, which are the only TFFs contribute to the required branching fractions.

On the other hand, when the correlation function~\eqref{Correlation function} is dominated by the light-like distances, it can be expanded around the light-cone. The light-cone expansion is performed by integrating out the transverse and minus degrees of freedom and leaving only the longitudinal momenta of the partons as the relevant degrees of freedom. The integration over the transverse momenta is done up to a cutoff, $\mu_{\rm IR}$, all momenta below which are included in the $\eta^{(\prime)}$-meson LCDAs. Large transverse momenta are calculated in perturbative theory. Thus, the correlation function can be separated into perturbative and nonperturbative contributions, both of which depend on the longitudinal parton momenta and the factorization scale $\mu_{\rm IR}$~\cite{Ball:2007hb}.

In order to make our result more accurate, we consider both the leading-order (LO) for all the LCDAs' part and gluon radiative corrections to the dominant twist-2, 3 parts of the correlation function. The OPE result for the correlation function $\Pi^{\rm OPE}$ is then represented as a sum of LO and NLO parts,
\begin{align}
 \Pi^{\rm OPE}[q^2, (p+q)^2] = F_0(q^2,(p+q)^2)+\frac{\alpha_s C_F}{4\pi } F_1(q^2,(p+q)^2).
\end{align}
To calculate the invariant amplitude $F_0(q^2,(p+q)^2)$ and $F_1(q^2,(p+q)^2)$, one needs to know the expression for the $c$-quark propagator, i.e.
\begin{align}
\langle 0|T\{c(x)\bar c(0)\}|0\rangle &=i\int\frac{d^4k}{(2\pi)^4}{e^{-ik\cdot x}}\frac{ m_c + \DS k}{ k^2-m_c^2}-ig \int\frac{d^4 k}{(2\pi)^4}e^{-ik\cdot x} \int_0^1 dv\Bigg[ \frac{m_c + \DS k}{2(m_c^2-k^2)^2} \nonumber \\
& \times G^{\mu\nu}(vx)\sigma_{\mu\nu} +\frac{v}{m_c^2-k^2}x_\mu G^{\mu\nu}(vx)\gamma_\nu \Bigg].
\end{align}
To do the calculation, the expression of twist-2, 3, 4 LCDAs matrix elements are needed, which are displayed in the Appendix~\ref{sec:appendixC}. To get the final LCSR expression, we need to use the Borel transformation to transform the variable $(p+q)^2$ into Borel parameter $M^2$. Then the expression of the $D_s\to \eta^{(\prime)}$ TFFs up to NLO gluon radiation correction to the twist-2, 3 LCDAs can be obtained by equating the two types of representation of the correlation function and by subtracting the contribution from higher resonances and continuum states, i.e.,
\begin{align}
f^{\eta^{(\prime)}}_+(q^2)&=\frac{e^{m_{D_s}^2/M^2}} {2m_{D_s}^2 f_{D_s}} \bigg[F_0(q^2,M^2,s_0)+\frac{\alpha_s C_F}{4\pi } F_1(q^2,M^2,s_0)\bigg].
\end{align}
where $F_{0(1)}(q^2, M^2, s_0)$ originates from the OPE result for the LO (NLO) invariant amplitude $F_{0(1)}(q^2, (p+q)^2)$. Finally, the LCSR for $D_s\to \eta^{(\prime)}$ TFFs have the following form,
\begin{align}
&f^{\eta^{(\prime)}}_+(q^2) = \frac{m_c^2 f_{\eta^{(\prime)}}}{2m_{D_s}^2 f_{D_s}}e^{m_{D_s}^2/M^2}\!\!\int_{u_0}^1\! du e^{-s(u)/M^2}\bigg\{\frac{\phi_{2;\eta^{(\prime)}}(u)}u \!+ \frac1{2 m_s m_c}\bigg[\phi_{3;\eta^{(\prime)}}^p(u) \!+\! \frac16 \bigg(2\frac{\phi_{3;\eta^{(\prime)}}^\sigma (u)}u
\nonumber\\
&\hspace{0.28cm}
- \frac{m_c^2+q^2-u^2m_{\eta^{(\prime)}}^2}{m_c^2-q^2+u^2m_{\eta^{(\prime)}}^2} \frac d{du}\phi_{3;\eta^{(\prime)}}^\sigma (u)+ \frac{4um_{\eta^{(\prime)}} ^2m_c^2}{(m_c^2 - q^2 + u^2m_{\eta^{(\prime)}}^2)^2}\phi_{3;\eta^{(\prime)}}^\sigma (u)\bigg)\bigg]\! +\! \frac1{m_c^2-q^2+u^2 m_{\eta^{(\prime)}}^2}
\nonumber\\
&\hspace{0.28cm}
\times \bigg[u\psi_{4;\eta^{(\prime)}}(u) + \left(1 - \frac{{2{u^2}m_{\eta^{(\prime)}}^2}}{m_c^2 - q^2 + u^2 m_{\eta^{(\prime)}}^2}\right) \int_0^u dv  \psi_{4;\eta^{(\prime)}}(v)- \frac{m_c^2}{4}\frac{u}{m_c^2 - q^2 + u^2m_{\eta^{(\prime)}} ^2}\,\bigg(\frac{d^2}{du^2}
\nonumber\\
&\hspace{0.28cm}
- \frac{6um_{\eta^{(\prime)}} ^2}{m_c^2-q^2+u^2m_{\eta^{(\prime)}}^2}\frac{d}{du} \,+ \frac{12um_{\eta^{(\prime)}}^4}{(m_c^2 - q^2 + u^2m_{\eta^{(\prime)}} ^2)^2}\bigg)\phi _{4;\eta^{(\prime)}}(u)
 - \bigg(\frac d{du} - \frac{2u m_{\eta^{(\prime)}}^2}{m_c^2-q^2+u^2m_{\eta^{(\prime)}}^2}\bigg)
\nonumber\\
&\hspace{0.28cm}
\times \bigg(\frac{f_{3\eta^{(\prime)}}}{f_{\eta^{(\prime)}} m_c}~I_{3;\eta^{(\prime)} }(u) ~+ I_{4;\eta^{(\prime)}}(u)\bigg)- \frac{{2{u}m_{\eta^{(\prime)}} ^2}}{m_c^2-q^2+u^2m_{\eta^{(\prime)}}^2}~\bigg(u\frac{d}{du}  - \frac{{2{u^2}m_{\eta^{(\prime)}} ^2}}{m_c^2-q^2+u^2m_{\eta^{(\prime)}}^2} + 1\bigg)
\nonumber
\\
&\hspace{0.28cm}
\times \bar I_{4\eta^{(\prime)} }(u) ~+~ \frac{2u m_{\eta^{(\prime)}}^2(m_c^2-q^2-u^2m_{\eta^{(\prime)}} ^2)}{(m_c^2-q^2+u^2m_{\eta^{(\prime)}}^2)^2}~~\bigg(\frac{d}{du} - \frac{6um_{\eta^{(\prime)}} ^2}{m_c^2-q^2+u^2m_{\eta^{(\prime)}}^2}\bigg)~\int_u^1 {d\xi} \bar I_{4\eta^{(\prime)}}(\xi )\bigg]\bigg\}
\nonumber
\\
&\hspace{0.28cm}
+\frac{m_c^4{f_{\eta^{(\prime)}} } e^{ - m_c^2/M^2}}{4(m_c^2 - q^2 + m_{\eta^{(\prime)}} ^2)^2} \bigg(\frac{d}{du}\phi_{4;\eta}(u)\bigg)\bigg|_{u\to 1} + \frac{\alpha_s C_F e^{m_{D_s}^2/M^2}}{8\pi m_{D_s}^2 f_{D_s}} F_1(q^2,m^2,s_0),
\nonumber
\\\label{Eq:fp}
\end{align}
with $s(u) = m_c^2-(q^2-m_{\eta^{(\prime)}}^2u)\bar u/u$. The LO invariant amplitudes include twist-2, 3, 4 contributions. The NLO QCD corrections to the invariant amplitudes $F_1(q^2, (p+q)^2)$ include twist-2, 3 contributions, which can be separated into the following form,
\begin{align}
F_1(q^2,m^2,s_0) &=\frac1\pi \int_{m_c^2}^{s_0}dse^{-s(u)/M^2}{\rm Im} F_1(q^2,s)
\nonumber\\
& =\frac{f_{\eta^{(\prime)}}}\pi \int_{m_c^2}^{s_0} ds e^{-s(u)/M^2}\int_0^1{du}\Big\{ {\rm Im}{T_1}(q^2,s,u)\phi_{2;\eta^{(\prime)}}(u)
\nonumber\\[2ex]
& +\frac{\mu_{\eta^{(\prime)}}}{m_c}\big[{\rm Im}T_1^p(q^2,s,u)\phi_{3;\eta^{(\prime)}}^p(u) + {\rm Im}T_1^{\sigma}(q^2,s,u)\phi_{3;\eta^{(\prime)}}^\sigma(u)\big] \Big\},
\end{align}
where ${\mu _{\eta^{(\prime)}} } = {m_{\eta^{(\prime)}} ^2} /{2{m_s}}$. The imaginary parts of the amplitudes $T_1(q^2,s,u)$, $T_1^p(q^2,s,u)$ and $T_1^\sigma(q^2,s,u)$ are the hard-scattering amplitudes calculated by the 6 diagrams for the gluon corrections. The final detailed expressions agree with those of Refs.~\cite{Duplancic:2008ix, Duplancic:2015zna}, which are not listed here. The lower limit of integral
\begin{eqnarray}
u_0 = \Big(q^2 - s_0 + m_{\eta^{(\prime)}}^2 + \sqrt{(q^2 - s_0 + m_{\eta^{(\prime)}}^2 )^2 -4 m_{\eta^{(\prime)}}^2 (q^2 - m_c^2)}\Big)/(2m_{\eta^{(\prime)}}^2)
\end{eqnarray}
and for the final expression, we need a brief introduction to the integrals over three-particle LCDAs, i.e.
\begin{align}
I_{3;\eta^{(\prime)}}(u)& =\frac d{du}\Bigg[\int_0^u d\alpha_1 \int_\Delta^1 dv \Phi_{3;\eta^{(\prime)}}(\alpha_i)\Bigg],
\nonumber\\
I_{4;\eta^{(\prime)}}(u)&=\frac d{du}\Bigg\{\int_0^u d\alpha_1 \int_\Delta^1 \frac{dv}{v}
\Bigg[2\Psi_{4;\eta^{(\prime)}}(\alpha_i)-\Phi_{4;\eta^{(\prime)}}(\alpha_i)
  +2\widetilde \Psi_{4;\eta^{(\prime)}}(\alpha_i)-\widetilde \Phi_{4;\eta^{(\prime)}}(\alpha_i)\Bigg]\Bigg\},
\nonumber\\
\bar I_{4;\eta^{(\prime)}}(u)&=\frac{d}{du}\Bigg\{\int_0^u d\alpha_1
\int_\Delta^1 \frac{dv}{v}\Bigg[\Psi_{4;\eta^{(\prime)}}(\alpha_i) ~+~ \Phi_{4;\eta^{(\prime)}}(\alpha_i)
 +\widetilde{\Psi}_{4;\eta^{(\prime)}} (\alpha_i) \,+\, \widetilde{\Phi} _{4;\eta^{(\prime)}} (\alpha_i)\Bigg]\Bigg\},
\label{eq:fplusBpiLCSR3part}
\end{align}
where $\Delta = {(u-\alpha_1)}/{(1-\alpha_1)}$, $\alpha_2=1-\alpha_1-\alpha_3$ and $\alpha_3=(u-\alpha_1)/v$. Due to the contributions from three-particle parts are quite small, i.e. $<0.1\%$, we can safely neglect these parts in this paper. We would like to figure out that the decay branching fraction for the considered decay. Using the parametrization of the transition matrix elements in terms of TFFs, in massless lepton case, we get
\begin{align}
\frac{d\Gamma}{dq^2}(D_s^+\to \eta^{(\prime)}\ell^+\nu_\ell)&= \frac {G_F^2 |V_{cs}|^2}{192\pi^3m_{D_s}^{3}} \Big[\left( m_{D_s}^2+m_{\eta^{(\prime)}}^2-q^2 \right)^2 -4m_{D_s}^2 m_{\eta^{(\prime)}}^2 \Big]^{3/2}|f^{\eta^{(\prime)}}_+(q^2)|^2,  \label{Eq:dGamma}
\end{align}
where the fermi coupling constant $G_F=1.166\times{10}^{-5}~{\rm GeV}^{-2}$.

\section{Numerical Discussions}\label{sec:3}
\subsection{Input parameters}

We adopt the following parameters to do the numerical calculation. The current charm-quark mass is $m_c = 1.27\pm0.02~ {\rm GeV}$, the masses of $D_s$, $\eta$ and $\eta'$-meson $m_{D_s}=1.9685~{\rm GeV}$, $m_{\eta}=0.5478~{\rm GeV}$, $m_{\eta'}=0.9578~{\rm GeV}$ and $s$-quark mass $m_s =0.093~{\rm GeV}$. All of them are taken from the Particle Data Group (PDG)~\cite{Zyla:2020zbs}. The $D_s$, $\eta, \eta'$-meson decay constants are taken as $f_{D_s}=0.274\pm0.013\pm0.007~{\rm GeV}$~\cite{Azizi:2010zj}, $f_\eta=0.130\pm0.003 ~{\rm GeV}$~\cite{Ball:2004ye} and $f_{\eta'}=0.157\pm0.003 ~{\rm GeV}$~\cite{Ali:1998eb}. The values of non-perturbative vacuum condensates up to 6-dimension are taken as follows~\cite{Shifman:1980ui, Colangelo:2000dp, Narison:2014ska},
\begin{align}
&\langle \alpha_s G^2 \rangle = 0.038\pm0.011~{\rm GeV}^4,   \nonumber\\
&\langle g_s^3fG^3\rangle  = 0.045\pm0.007~{\rm GeV}^6,   \nonumber\\
&\langle g_s\bar qq\rangle ^2 = (2.082_{-0.697}^{+0.734})\times 10^{-3} ~{\rm GeV}^6,   \nonumber\\
&\langle g_s^2\bar qq\rangle ^2 = (7.420_{-2.483}^{+2.614})\times 10^{-3}~{\rm GeV}^6, \nonumber\\
&\langle q\bar q\rangle  = (-2.417_{-0.114}^{+0.227})\times 10^{-2}~{\rm GeV}^3, \nonumber\\
&\langle \bar ss\rangle  = (-1.789_{-0.084}^{+0.168})\times 10^{-2}~{\rm GeV}^3, \nonumber\\
&\langle g_ss \bar s\rangle ^2  = (1.541_{ - 0.516}^{ + 0.543})\times{10^{ - 3}}~{\rm GeV}^6, \nonumber\\
&\sum {\langle g_s^2\bar qq\rangle ^2}   = (1.891_{ - 0.633}^{ + 0.665})\times{10^{ - 2}}~{\rm GeV}^6.
\end{align}
The quark-gluon mixture condensate $\langle g_s\bar q\sigma TGq\rangle  = m_0^2\langle \bar qq\rangle$ with $m_0^2 = 0.80 \pm 0.02{\rm GeV}^2$, which leads to
\begin{align}
\langle g_s\bar s\sigma TGs\rangle  = ( - 1.431_{ - 0.076}^{ + 0.139})\times{10^{ - 2}}~{\rm GeV}^5.
\end{align}
Here, the ratio $\kappa = \langle\bar ss\rangle/\langle\bar qq\rangle  = 0.74\pm0.03$ has been used. Meanwhile, the typical scale in this paper is $\mu_{\rm IR} = (m_{D_s}^2 - m_c^2)^{1/2} \approx 1.5~{\rm GeV}$. So the renormalization group equations (RGEs) should be used for running the quark masses and each vacuum condensates appearing in the BFTSR from the initial scale $\mu_0 = 1~{\rm GeV}$ to the typical scale $\mu_{\rm IR}$. The RGE can be found in Refs.~\cite{Yang:1993bp, Hwang:1994vp, Lu:2006fr}, which are not listed here.

\begin{table}[t]
\renewcommand\arraystretch{1.2}
\centering
\caption{The criteria for determining Borel windows, and the resultant Borel windows and the corresponding values of the $\eta$ and $\eta'$-meson leading twist LCDA moments $\langle \xi_{2;\eta^{(\prime)}}^n\rangle|_\mu $. ``Con.'' represents the continuum contribution and ``Six.'' represents the dimension-six condensates' contribution.}\label{tab:m2}
\begin{tabular}{c l c c c c c}
\hline
~~~~~~~~~~~~~~~~~~~~~& $n$~~~~~~      &~~~~~~~~$0$~~~~~~~~          & ~~~~~~~~$2$~~~~~~~~         & ~~~~~~~~$4$~~~~~~~~          & $6$ \\  \hline
&Con.     & $ < 20\%$    & $ < 25\%$   & $ < 30\%$    & $ < 35\%$  \\
&Six.     & $ < 5\%$    & $ < 5\%$     & $ < 5\%$       & $ < 5\%$     \\
\raisebox {2.0ex}[0pt]{$\eta$-meson}&${M^2}$  & $[0.535, 1.188]$    & $[1.026,1.402]$    & $[1.368,1.759]$    & $[1.677,2.194]$    \\
&$\langle {\xi^n_{2;\eta }}\rangle|_\mu$  & $[0.952,1.168]$  & $[0.231,0.230]$  & $[0.110, 0.102]$    & $[0.067, 0.059]$
\\  \hline
&$n$      &~~~~~~~~$0$~~~~~~~~         & ~~~~~~~~$2$~~~~~~~~         & ~~~~~~~~$4$~~~~~~~~          & $6$ \\  \hline
&Con.     & $ < 30\%$    & $ < 35\%$   & $ < 40\%$    & $ < 45\%$  \\
&Six.     & $ < 5\%$    & $ < 5\%$     & $ < 5\%$       & $ < 5\%$     \\
\raisebox {2.0ex}[0pt]{$\eta'$-meson}&${M^2}$  & $[1.049, 1.137]$    & $[1.026,1.233]$    & $[1.368,1.627]$    & $[1.677,2.082]$    \\
&$\langle {\xi^n_{2;\eta' }}\rangle|_\mu$  & $[1.076,1.061]$  & $[0.221,0.201]$  & $[0.099, 0.086]$    & $[0.059, 0.047]$  \\  \hline
\end{tabular}
\end{table}

\subsection{Determination for the Gegenbauer moments of $\eta^{(\prime)}$-meson twist-2 LCDA}

One of the significant parameters in BFTSR is the continuum threshold $s_{\eta^{(\prime)}}$ for the moments of $\eta^{(\prime)}$-meson twist-2 LCDAs. Following our previous works, we can determine $s_{\eta}= 1.3\pm0.1 ~{\rm GeV}$ and $s_{\eta'}= 0.8\pm0.1 ~{\rm GeV}$ by setting the 0th-order of Gegenbauer moment into 1, i.e. $\langle\xi_{2;\eta}^0\rangle|_\mu=\langle\xi_{2;\eta^{\prime}}^0\rangle|_\mu = 1$. Meanwhile, in order to determine the allowable range of the Borel parameter $M^2$ (i.e. the Borel Window), we adopt the following three criteria
\begin{itemize}
  \item The continuum contributions are less than $45\%$ of the total results;
  \item The contributions from the dimension-six condensates do not exceed $5\%$;
  \item We require the variations of $\langle\xi_{2;\eta^{(\prime)}}^n\rangle|_{\mu}$ within the Borel window to be less than $10\%$.
\end{itemize}

\begin{figure}[t]
\begin{center}
\includegraphics[width=0.43\textwidth]{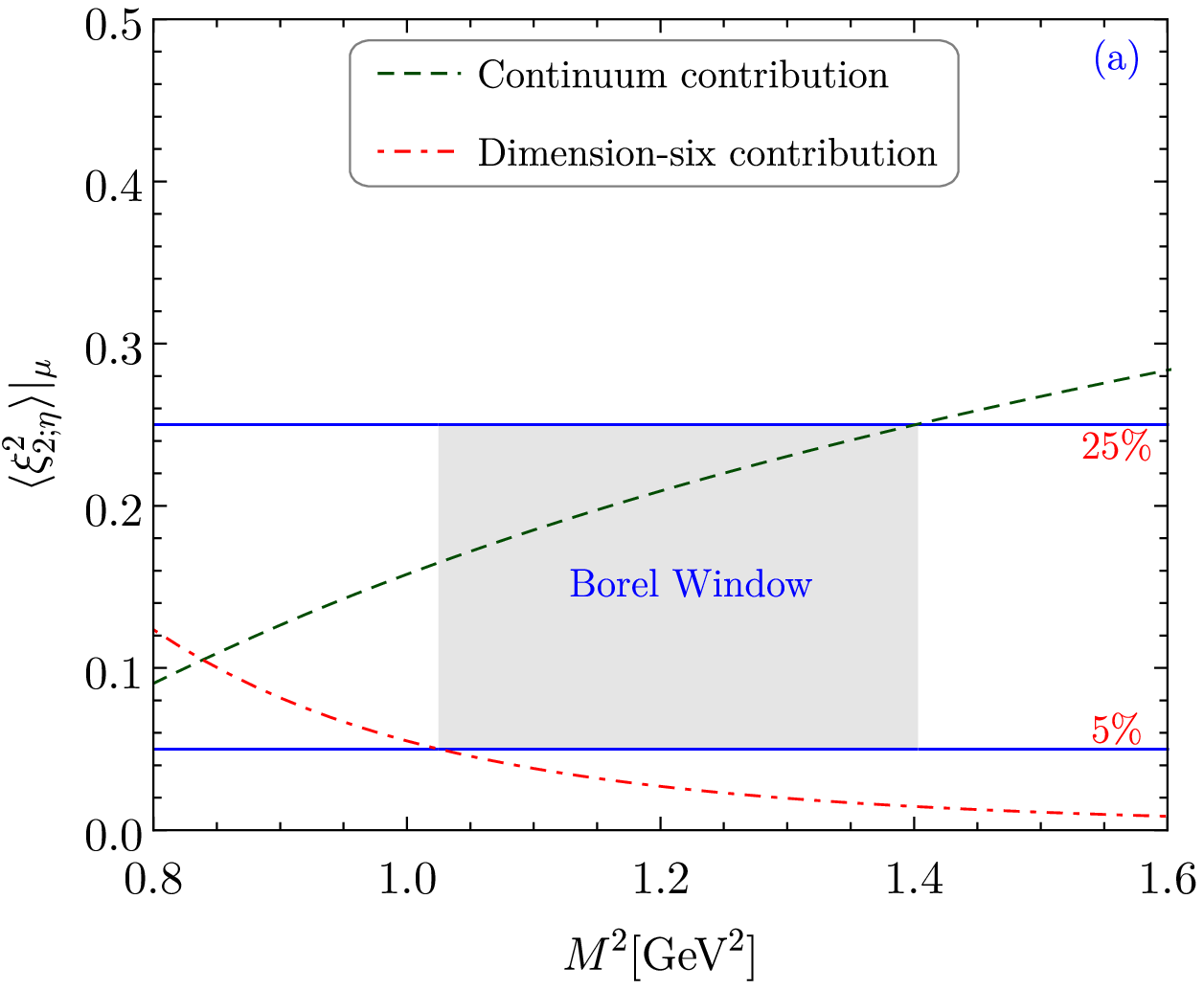}\includegraphics[width=0.43\textwidth]{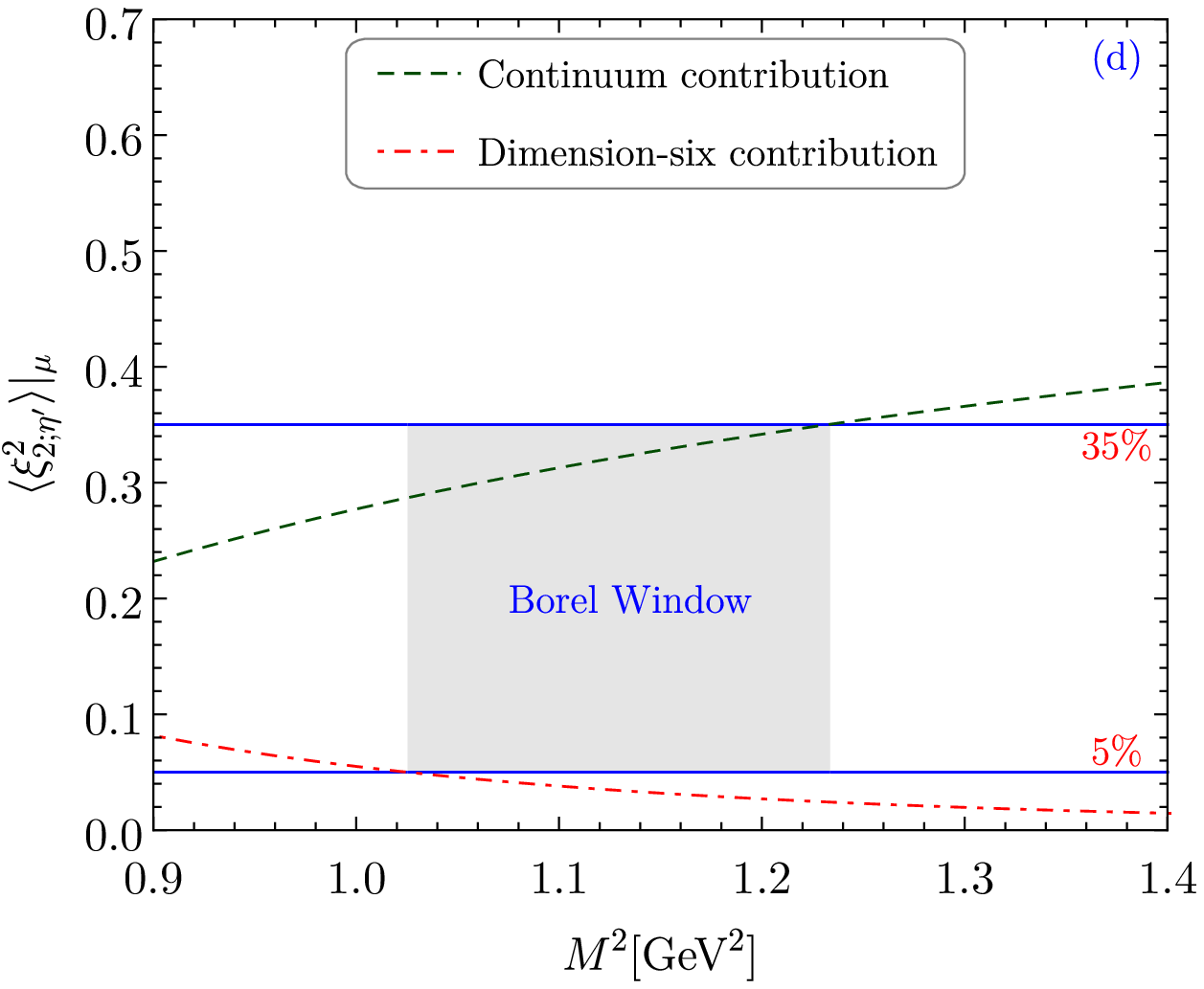}\\
\includegraphics[width=0.43\textwidth]{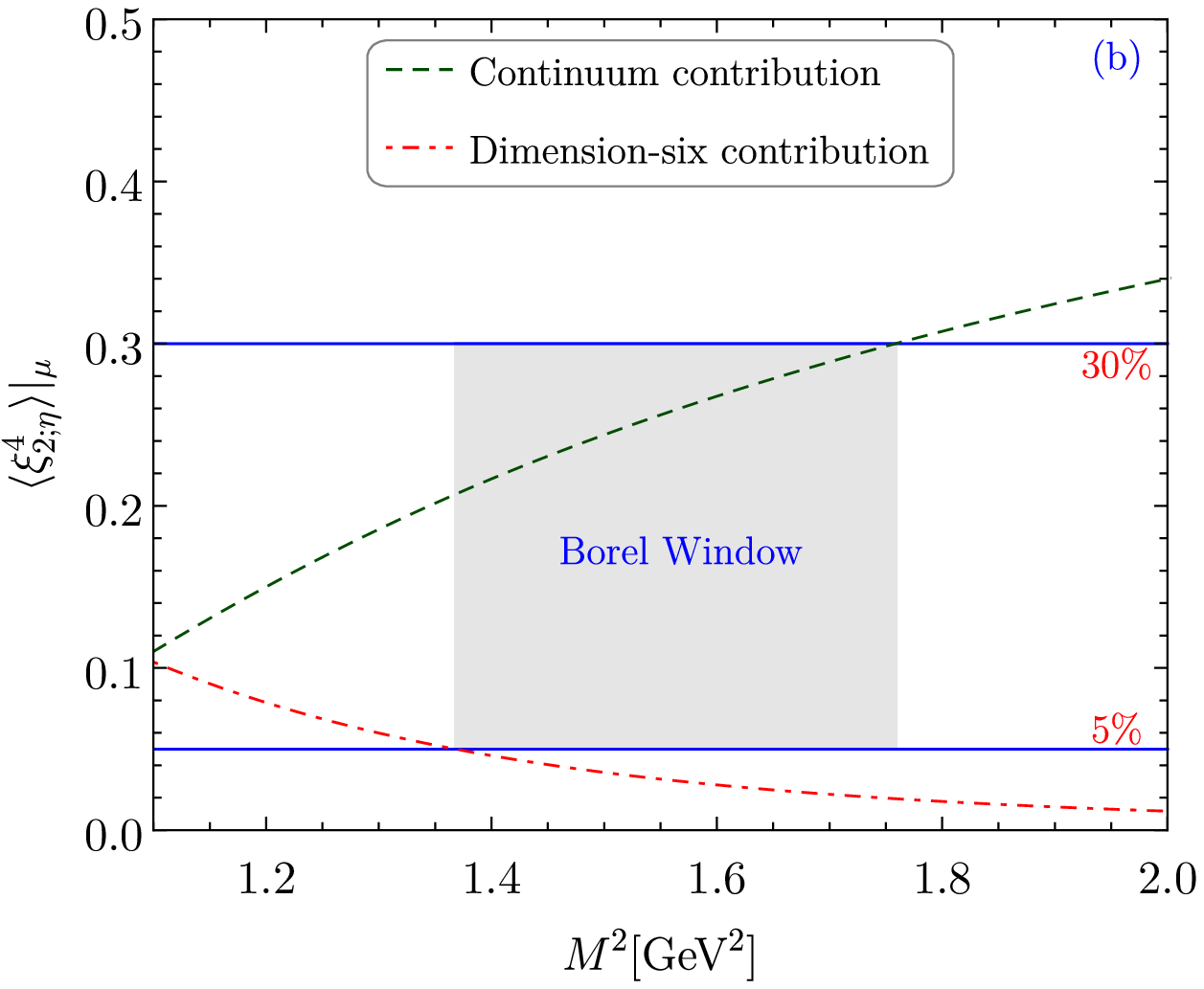}\includegraphics[width=0.43\textwidth]{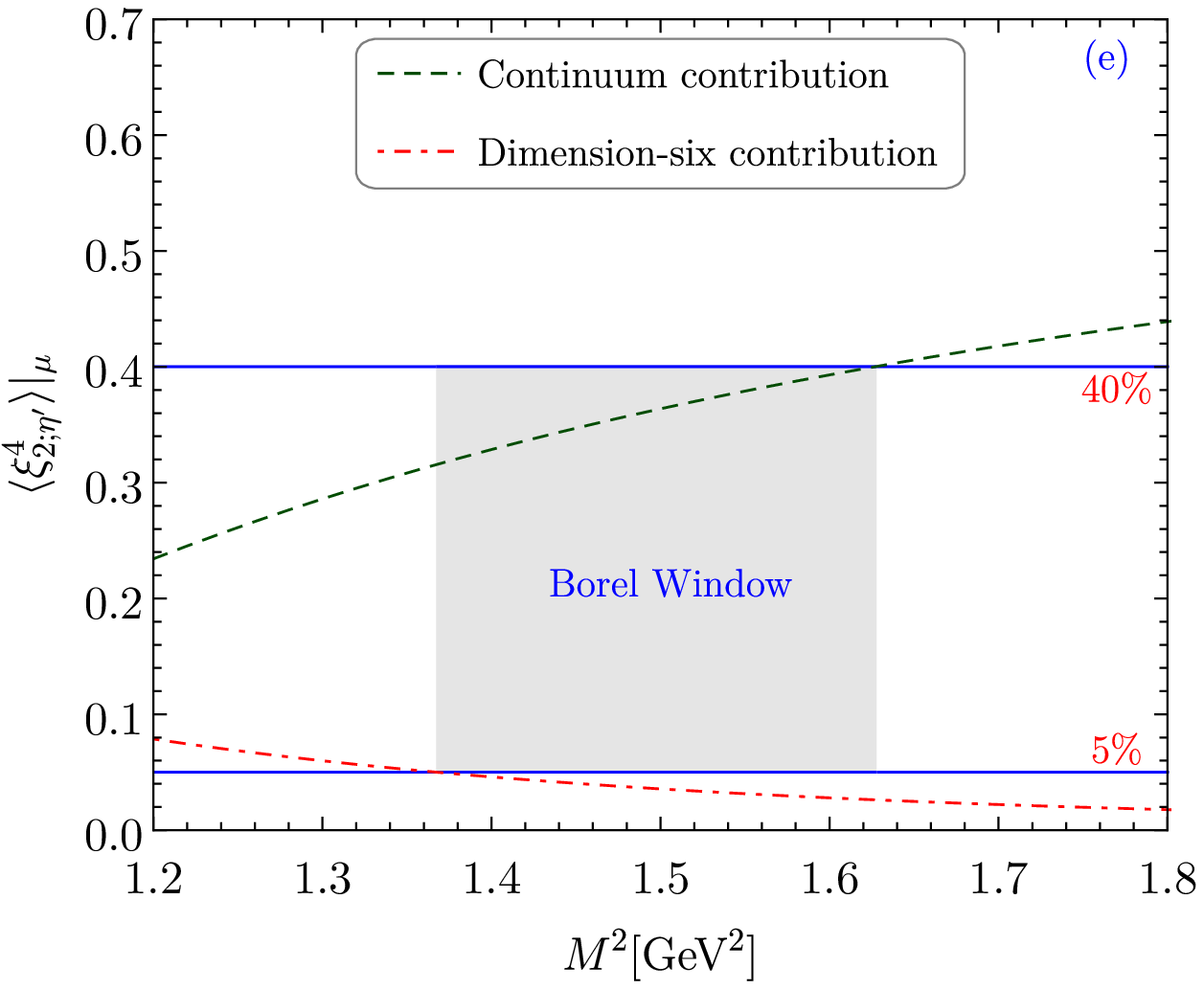}\\
\includegraphics[width=0.43\textwidth]{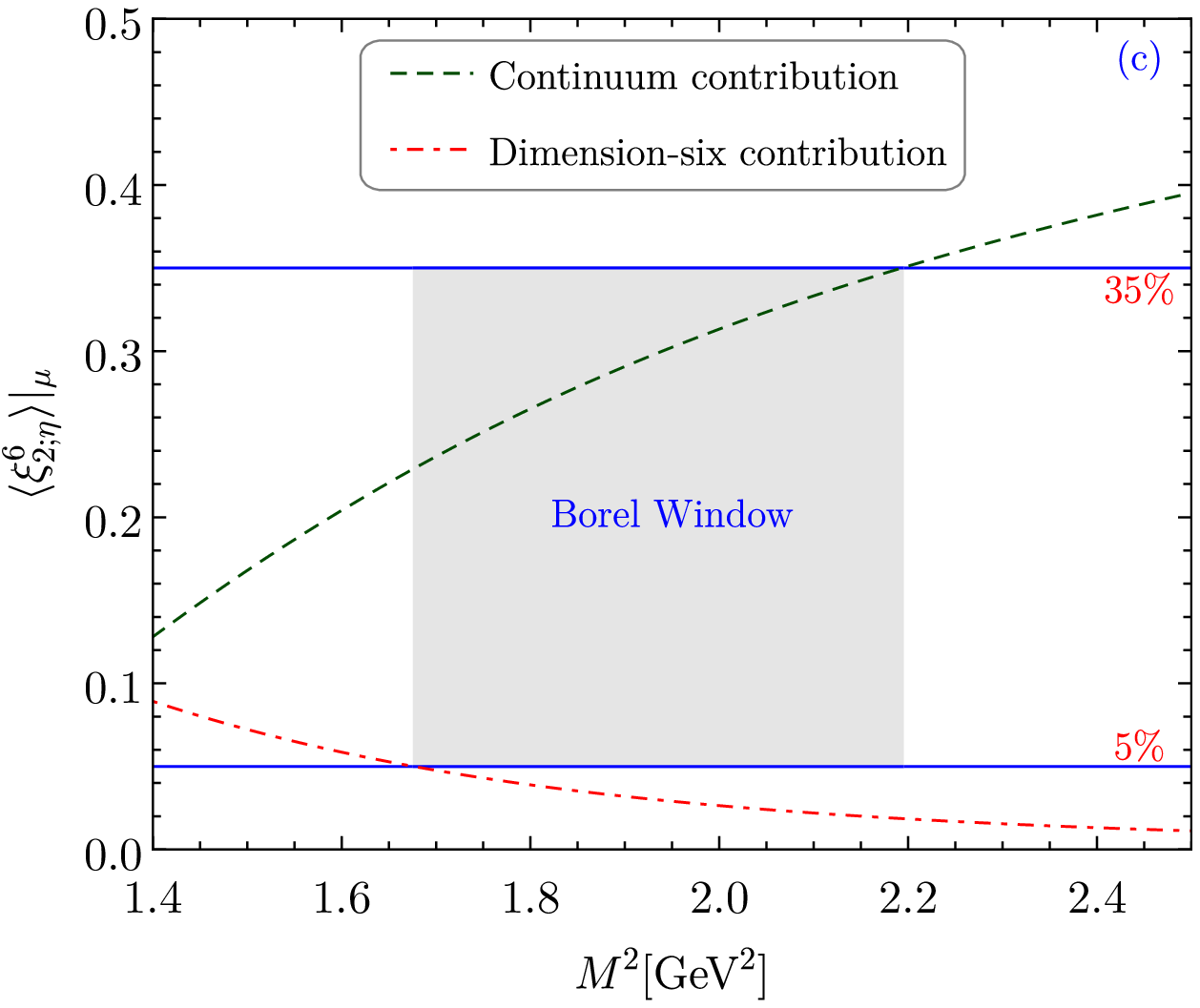}~\includegraphics[width=0.44\textwidth]{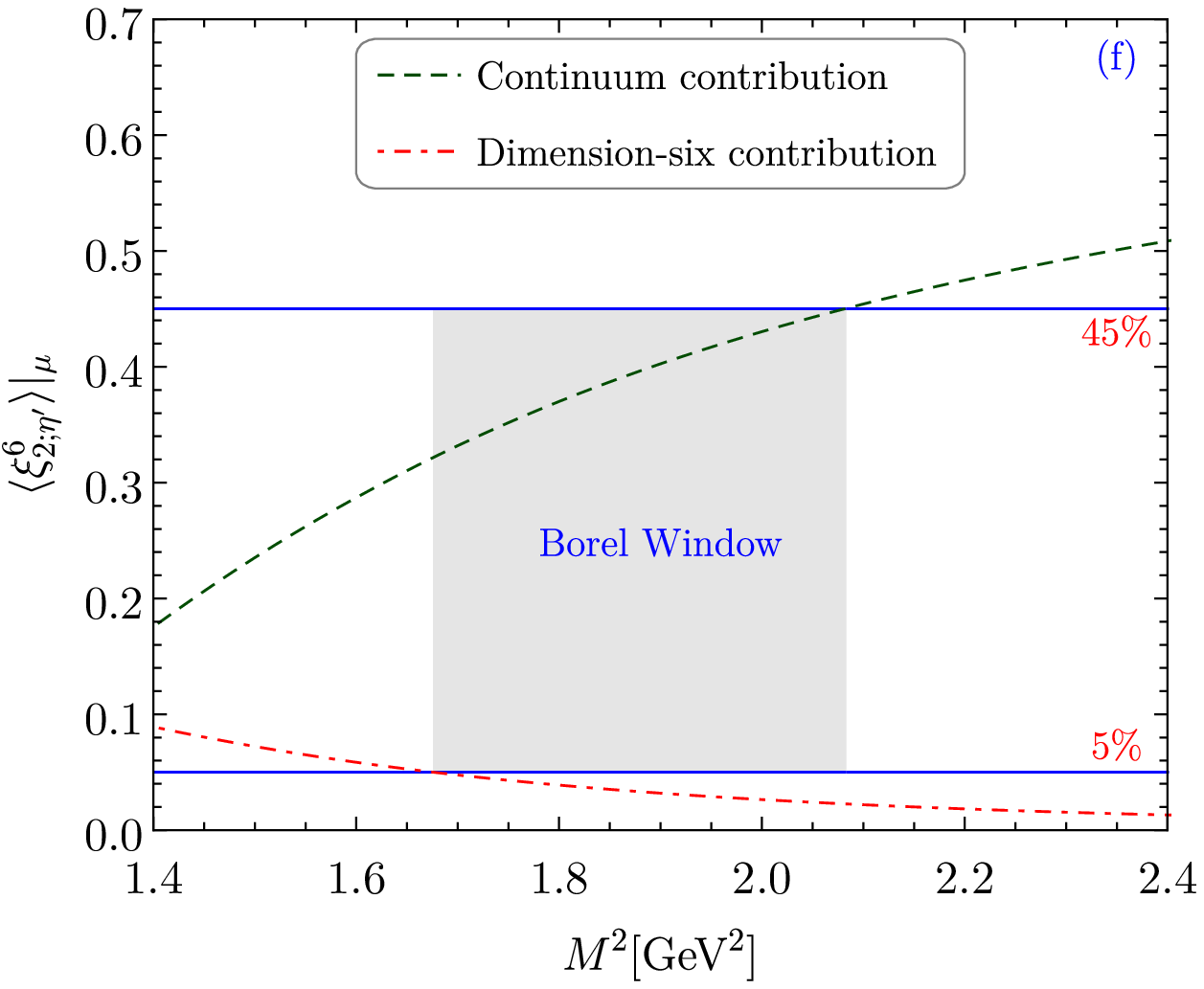}\\
\end{center}
\caption{Contributions from the continuum state and dimension-six condensates for the $\eta^{(\prime)}$-meson leading-twist LCDA moments $\langle {\xi^n_{2;\eta^{(\prime)} }}\rangle|_{\mu}$ versus the Borel parameter $M^2$, where all input parameters are set to be their central values. }
\label{fig:con}
\end{figure}

For the first four Gegenbauer moments $\langle\xi_{2;\eta^{(\prime)}}^n\rangle|_{\mu}$ with $n=(0,2,4,6)$, we list the allowable Borel region and their corresponding $\langle\xi_{2;\eta^{(\prime)}}^n\rangle|_{\mu}$ in Table~\ref{tab:m2}. When $n=(0,2,4,6)$, we have set the continuum contributions to be less than $20\%$, $25\%$, $30\%$, $35\%$ for $\langle\xi_{2;\eta}^n\rangle|_{\mu}$ and $30\%$, $35\%$, $40\%$, $45\%$ for $\langle\xi_{2;\eta'}^n\rangle|_{\mu}$, respectively, and the dimension-six condensates' contributions to be less than $5\%$ for all the order of $\langle\xi_{2;\eta^{(\prime)}}^n\rangle|_{\mu}$. Then, the determined continuum's and dimension-six condensates' contribution for $\langle\xi_{2;\eta}^n\rangle|_{\mu}$ and $\langle\xi_{2;\eta^{\prime}}^n\rangle|_{\mu}$ with $n=(2,4,6)$ are shown in the left and right panel of Fig.~\ref{fig:con}, respectively. In which, the shaded region stand for the Borel windows.

\begin{figure}[t]
\centering
\includegraphics[width=0.5\textwidth]{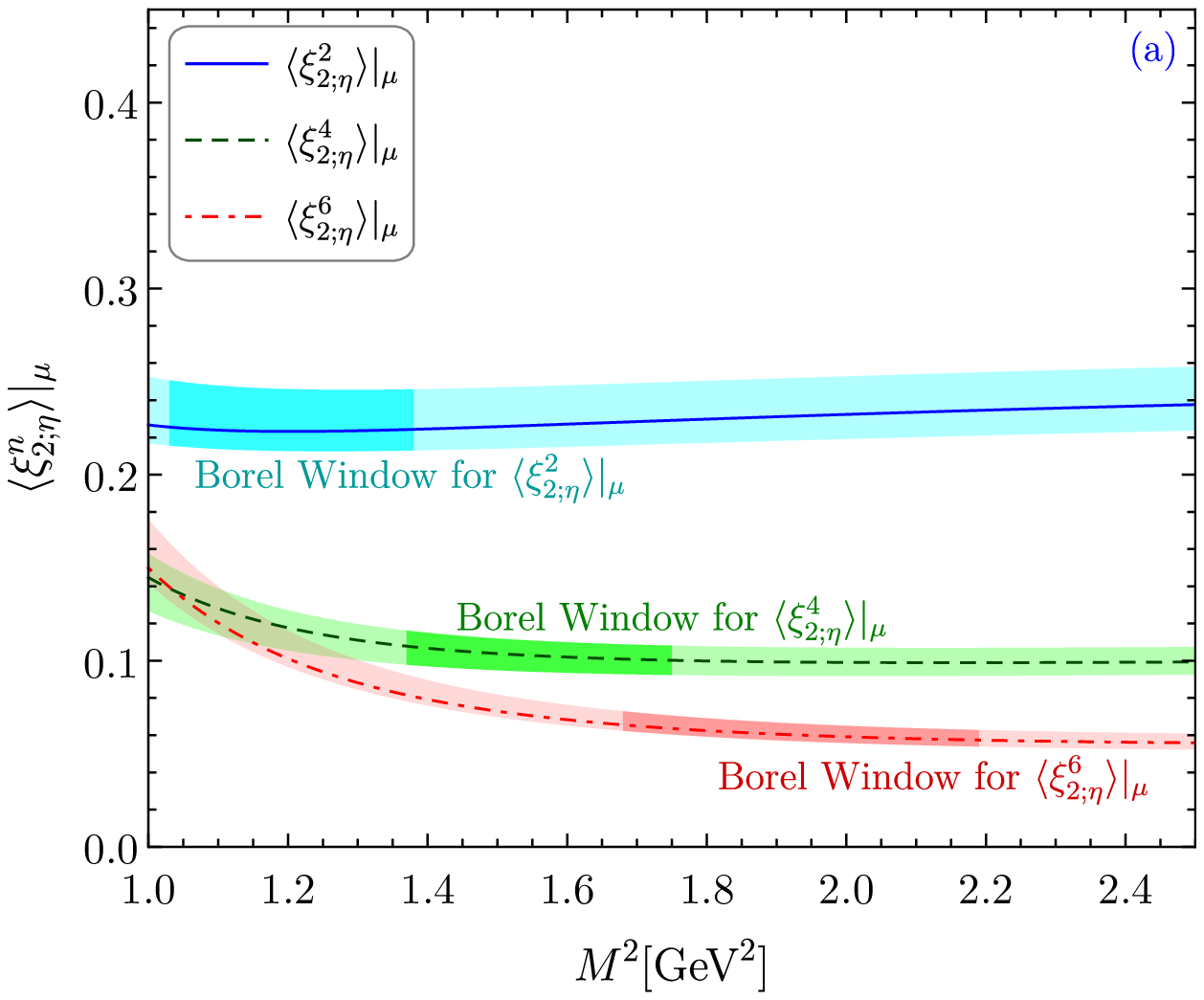}\includegraphics[width=0.511\textwidth]{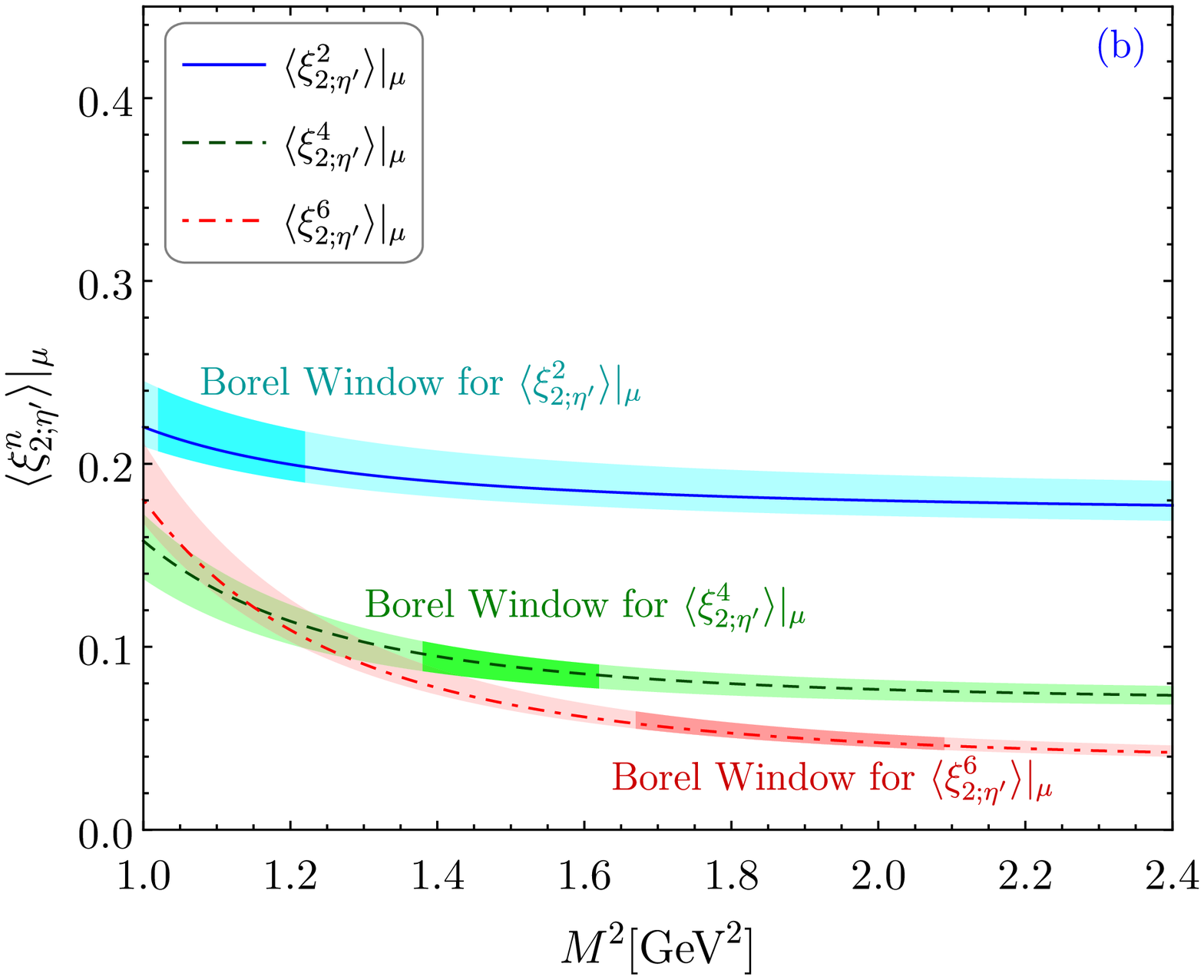}
\caption{The first three moments $\langle\xi^n_{2;\eta^{(\prime)}}\rangle|_{\mu}$ with $(n=2,4,6)$ versus the Borel parameter $M^2$. The darker shaded bands indicate the Borel windows for $\langle\xi^n_{2;\eta^{(\prime)}}\rangle|_{\mu}$, respectively.} \label{fig:xi}
\end{figure}

In order to provide a deeper insight into the flatness of the LCDA moments versus the Borel parameter $M^2$, we present the first three curves for the moments of $\eta^{(\prime)}$-meson twist-2 LCDA at the initial scale, i.e. $\langle \xi^n_{2;\eta^{(\prime)}} \rangle|_{\mu}$ with $n=(2,4,6)$ in Figure~\ref{fig:xi}. The determined Borel window are $M^2\in[1.0,2.5]~{\rm GeV}^2$ for $\eta$-meson and $M^2\in[1.0,2.4]~{\rm GeV}^2$ for $\eta'$-meson. It is noted that within this range, the moments $\langle \xi^2_{2;\eta^{(\prime)}} \rangle|_{\mu}$, $\langle \xi^4_{2;\eta^{(\prime)}} \rangle|_{\mu}$ and $\langle \xi^6_{2;\eta^{(\prime)}}\rangle|_{\mu}$ are almost flat, which vary less than $10\%$ for the total results in the Borel window.

By taking the squared average of all the uncertainty sources into consideration and making use of the relations between the Gegenbauer moments $a^n_{2;\eta^{(\prime)}}(\mu)$ and the LCDA moments $\langle\xi^n_{2;\eta^{(\prime)}}\rangle|_\mu$, i.e. Eq.~\eqref{xi}, we obtain the first three $a^n_{2;\eta^{(\prime)}}(\mu_0)$ and $\langle \xi^n_{2;\eta^{(\prime)}}\rangle|_{\mu_0}$ with $n = (2,4,6)$ for the leading-twist $\eta^{(\prime)}$-meson LCDA $\phi_{2;\eta^{(\prime)}}(u,\mu_0)$ predicted from BFTSR in Table~\ref{Tab:anxin}. The factorization scale is taken as the initial scale ${\mu _0} = 1~{\rm GeV}$. As a comparison, we also list the LCSR given in year 2013~\cite{Offen:2013nma}, the CLEO fit~\cite{Gronberg:1997fj}, the BABAR fit~\cite{BABAR:2011ad}, the ones given by Kroll~\cite{Kroll:2013iwa} and Ball~\cite{Ball:2004ye}, respectively. For the Ball's results, it is calculated by using the approximation $a_{2;\eta}^n(\mu) = a_{2;K}^n(\mu) = a_{2;\pi}^n(\mu)$. Our results for the second and fourth order $\eta$-meson LCDA's moments, e.g. $\langle\xi_{2;\eta}^2\rangle|_{\mu_0}$ and $\langle\xi_{2;\eta}^4\rangle|_{\mu_0}$ agree with the Ball's predictions within errors. But there still exist discrepancy for $a_{2;\eta}^4(\mu_0)$ with Ball's prediction. The main reason lies in the fourth order of equations between Gegenbauer and LCDA moments, i.e. Eq.~\eqref{xi} have large coefficients which will enlarge the small discrepancy of LCDA moments, even appears the opposite sign. At present, there are few studies on $\eta'$-meson's twist-2 LCDA.

\begin{table}[tb]
\renewcommand\arraystretch{1.2}
\begin{center}
\caption{First three Gegenbaner and LCDA moments $a^n_{2;\eta^{(\prime)}}(\mu_0)$ and $\langle \xi^n_{2;\eta^{(\prime)}}\rangle|_{\mu_0}$ with $n = (2,4,6)$ for the leading-twist $\eta^{(\prime)}$-meson LCDA $\phi_{2;\eta^{(\prime)}}(u,\mu_0)$, where the errors are squared averages of those from all the input parameters. Other theoretical predictions are also given as a comparison.} \label{Tab:anxin}
\begin{tabular}{lcccccccc}
\hline\hline
~~~~~~~~~~~~~~~~~~~~~~~~& &$\eta$-meson & &~~& & $\eta'$-meson& \\ \cline{2-4} \cline{6-8}
    &$\langle \xi^2_{2;\eta}\rangle|_{\mu_0}$
    & $\langle \xi^4_{2;\eta}\rangle|_{\mu_0}$
    & $\langle \xi^6_{2;\eta}\rangle|_{\mu_0}$
    && $\langle \xi^2_{2;\eta'}\rangle|_{\mu_0}$
    & $\langle \xi^4_{2;\eta'}\rangle|_{\mu_0}$
    & $\langle \xi^6_{2;\eta'}\rangle|_{\mu_0}$   \\ \hline
BFTSR (This work)  & $0.231_{-0.013}^{+0.010}$   & $0.109_{-0.007}^{+0.007}$  & $0.066_{-0.006}^{+0.006}$    && $0.211_{-0.017}^{+0.015}$   & $0.093_{-0.009}^{+0.009}$  & $0.054_{-0.008}^{+0.008}$    \\
CLEO fit~\cite{Gronberg:1997fj}   &    $0.176\pm0.010$    &-      &-     &&-      &-     &-    \\
BABAR fit~\cite{BABAR:2011ad}     &    $0.183\pm0.007 $   &-      &-     &&-      &-     &-   \\
P. Kroll~\cite{Kroll:2013iwa}     &    $0.183\pm0.007 $   &-      &-     &&-      &-     &-   \\
SR fit~\cite{Offen:2013nma}       &    $0.286\pm0.051$    &-      &-    &&-      &-     &-\\
P. Ball~\cite{Ball:2004ye}        &    0.239              & 0.110 &-    & &-      &-     &-\\
\hline\hline
~~~~~~~~~~~~~~~~~~~~~~~~& &$\eta$-meson & &~~& & $\eta'$-meson& \\ \cline{2-4} \cline{6-8}
   & $a^2_{2;\eta}(\mu_0)$
   & $a^4_{2;\eta}(\mu_0)$
   & $a^6_{2;\eta}(\mu_0)$
   && $a^2_{2;\eta'}(\mu_0)$
   & $a^4_{2;\eta'}(\mu_0)$
   & $a^6_{2;\eta'}(\mu_0)$  \\ \hline
BFTSR (This work)   & $0.090_{-0.037}^{+0.031}$   & $0.025_{-0.010}^{+0.003}$   & $0.033_{-0.058}^{+0.055}$
                    && $0.033_{-0.050}^{+0.042}$   & $-0.002_{-0.016}^{+0.007}$  & $0.043_{-0.072}^{+0.067}$  \\
CLEO fit~\cite{Gronberg:1997fj}   & $-0.07\pm0.03$      &-&-&&-&-&- \\
BABAR fit~\cite{BABAR:2011ad}     & $-0.05\pm0.02$    &-&-&&-&-&- \\
P. Kroll~\cite{Kroll:2013iwa}     & $-0.05\pm0.02 $        &-&-&&-&-&- \\
SR fit~\cite{Offen:2013nma}       & $0.25\pm0.15$        &-&-&&-&-&- \\
P. Ball~\cite{Ball:2004ye}        &  $0.115$ & $-0.015$   &-&&-&-&- \\ \hline\hline
\end{tabular}
\end{center}
\end{table}

\begin{figure}[t]
\centering
\includegraphics[width=0.5\textwidth]{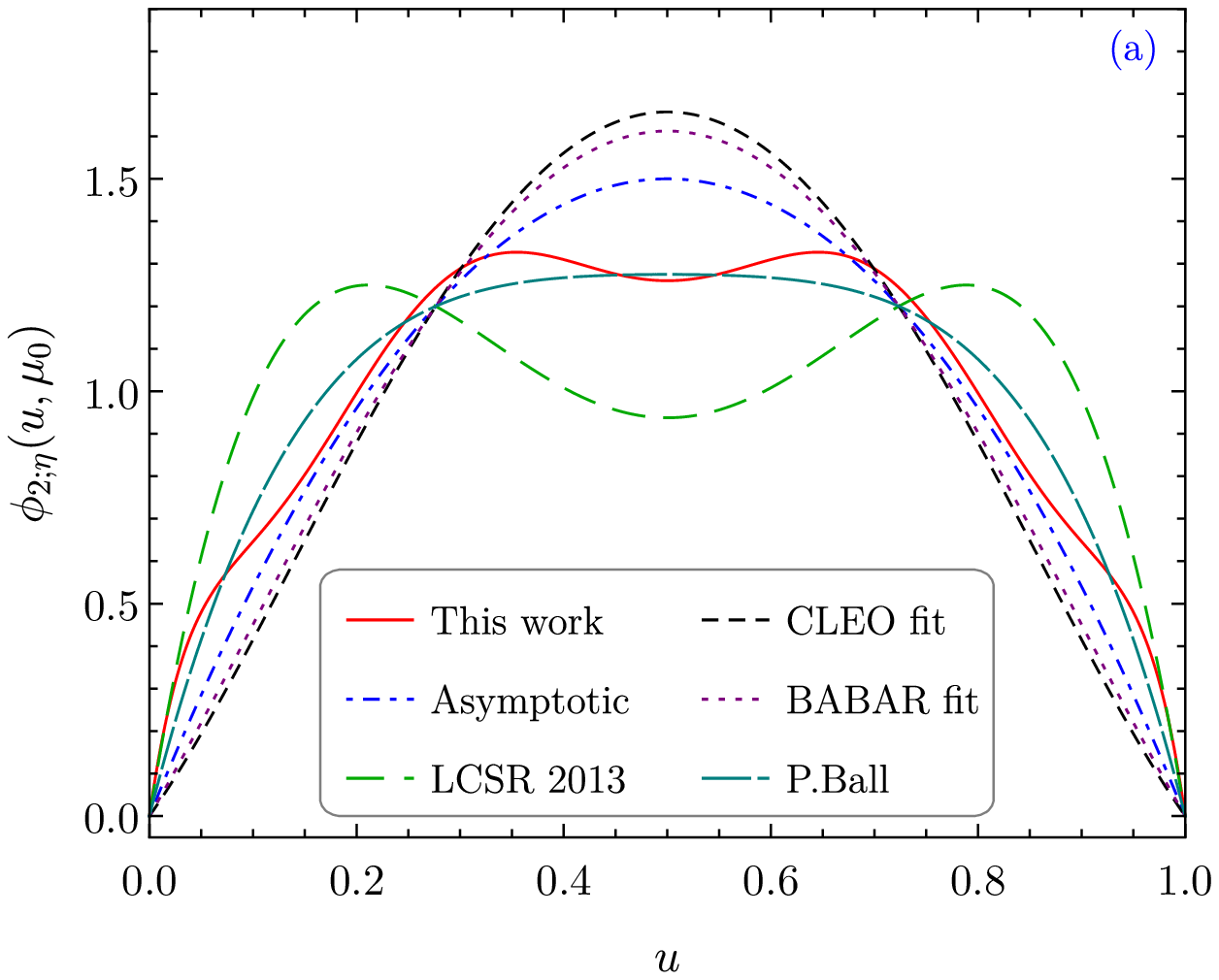}\includegraphics[width=0.5\textwidth]{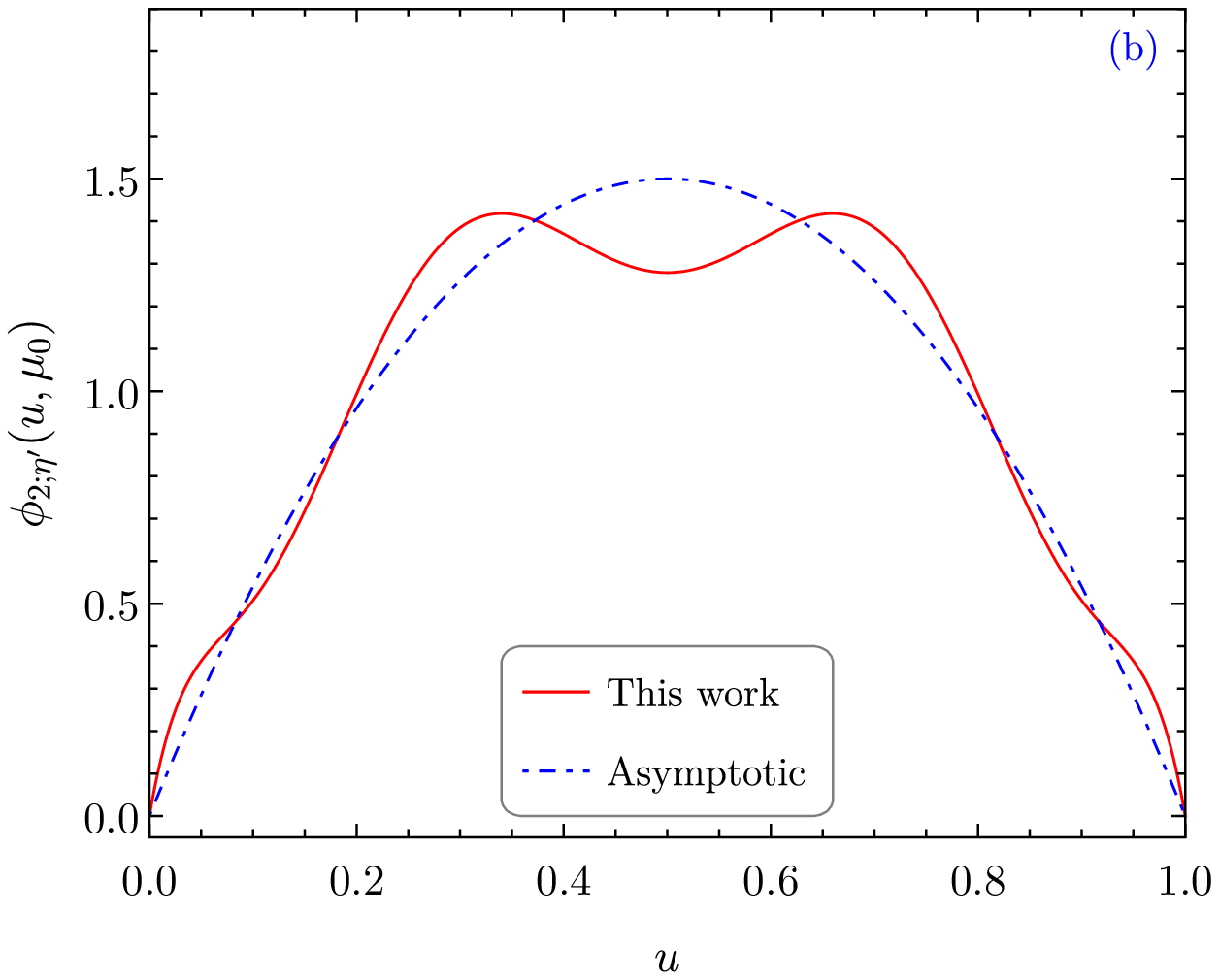}
\caption{The $\eta^{(\prime)}$-meson leading-twist LCDA $\phi_{2;\eta^{(\prime)}}(u,{\mu}_{0})$ predicted from the BFTSR. We make a comparison with the asymptotic form, the CLEO~\cite{Gronberg:1997fj}, SR fit~\cite{Offen:2013nma}, the BABAR~\cite{BABAR:2011ad}, and the predictions of Kroll~\cite{Kroll:2013iwa} and Ball~\cite{Ball:2004ye}.}
\label{fig:fi2}
\end{figure}

\begin{figure}[t]
\centering
\includegraphics[width=0.5\textwidth]{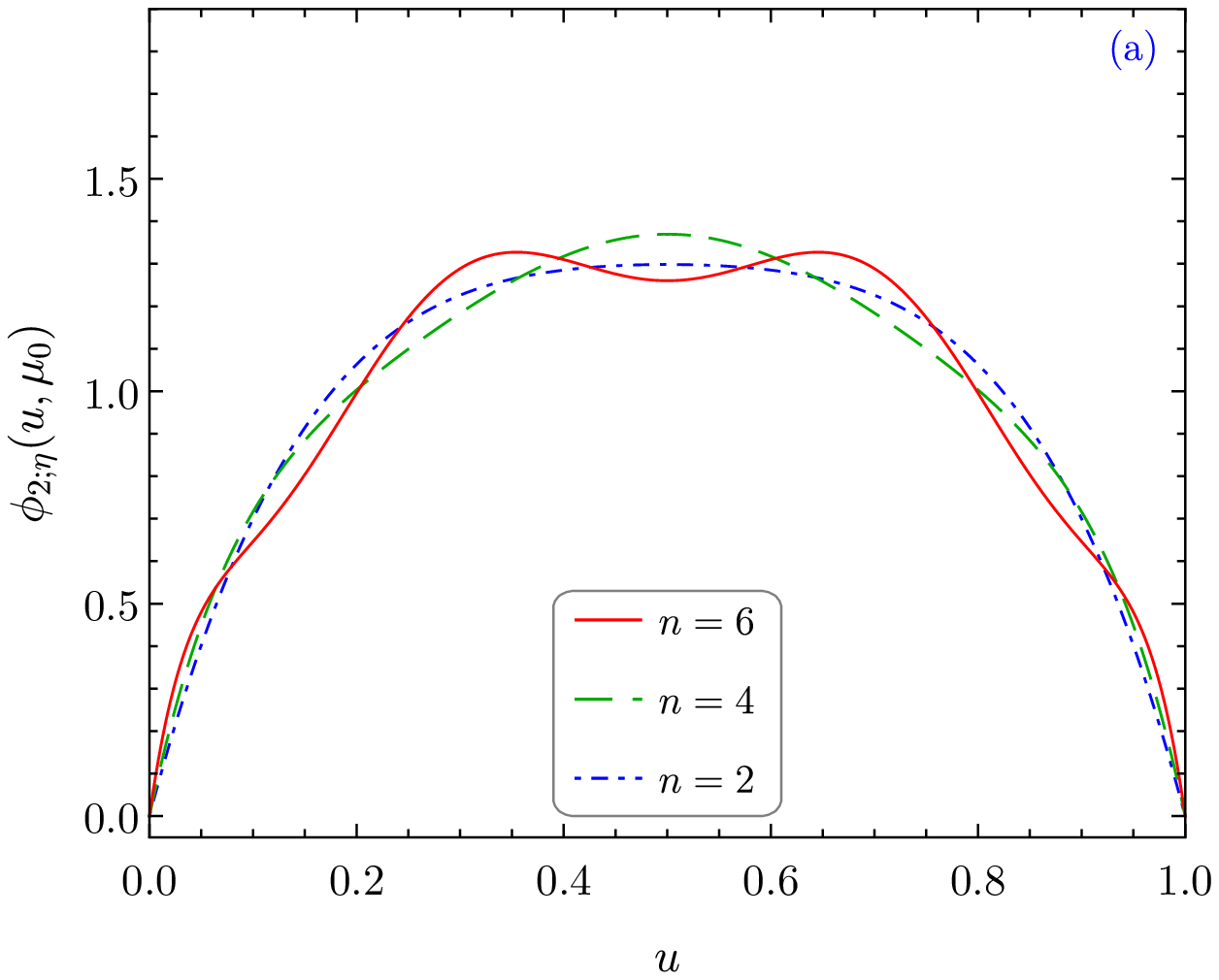}\includegraphics[width=0.5\textwidth]{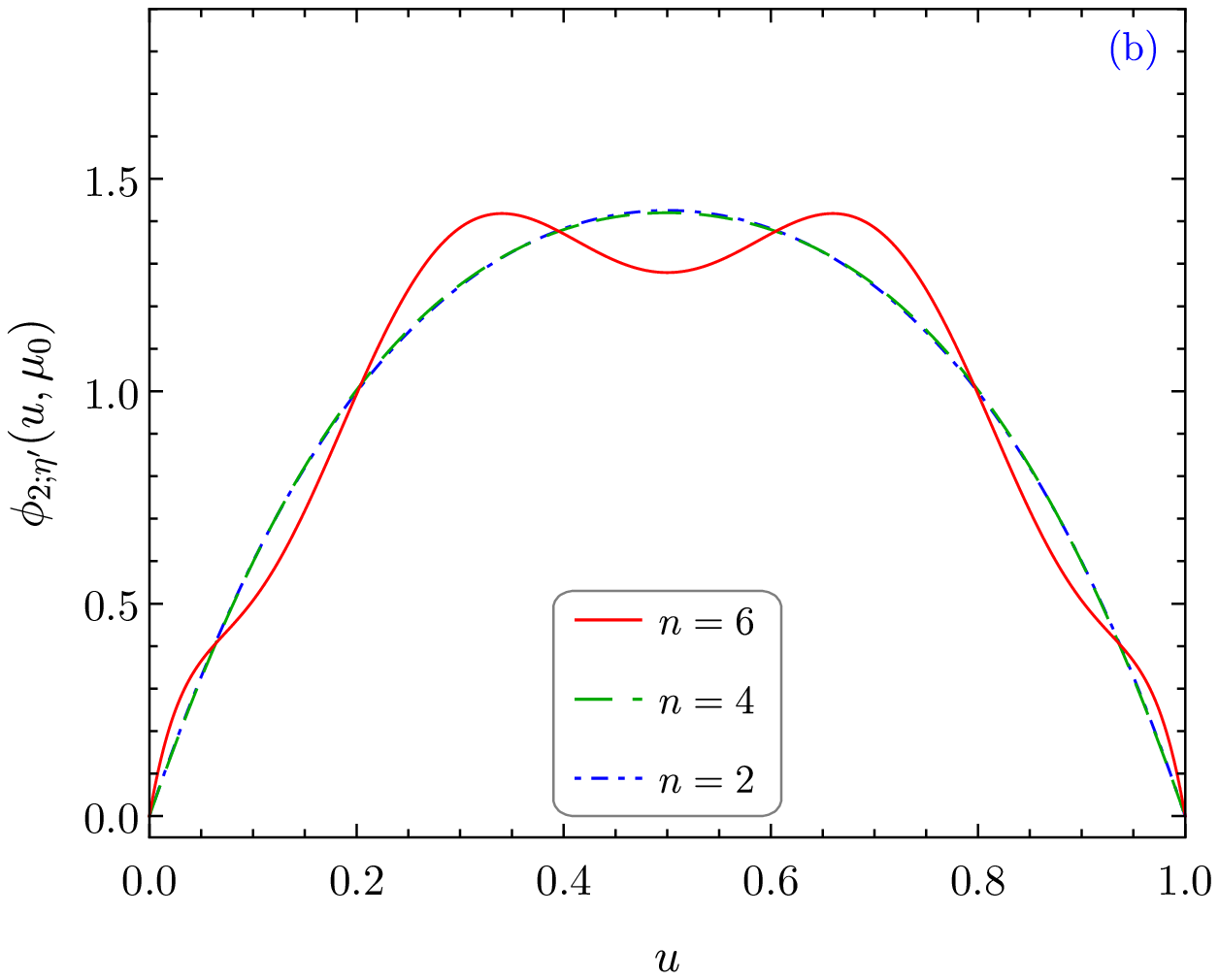}
\caption{The curves of $\eta$ and $\eta^{\prime}$-meson twist-2 LCDA with $n = (2,4,6)$ respectively.}
\label{fig:fi22}
\end{figure}

After considering the Gegenbauer moments $a_{2;\eta^{(\prime)}}^n(\mu)$ up to 6th-order into the conformal expansion of the Gegenbauer polynomial at initial scale, i.e. Eq.~\eqref{HPDA_CZ}, we present the $\eta$ and $\eta^{\prime}$-meson twist-2 LCDAs in Figure~\ref{fig:fi2}(a) and Figure~\ref{fig:fi2}(b) separately. For $\phi_{2;\eta}(u,\mu_0)$, we present the asymptotic form, the CLEO~\cite{Gronberg:1997fj}, the BABAR~\cite{BABAR:2011ad}, Kroll's prediction~\cite{Kroll:2013iwa}, the SR fit~\cite{Offen:2013nma} and the Ball's prediction~\cite{Ball:2004ye} as a comparison. Figure~\ref{fig:fi2}(a) shows that the LCSR 2013 and Ball's results prefer a double-peaked behavior. The reason lies in that they adopt the $\pi, K$-meson's LCDAs as those of $\eta$-meson LCDAs. Conversely, the CLEO~\cite{Gronberg:1997fj}, the BABAR~\cite{BABAR:2011ad} and the fitting results by Kroll indicate a single-peaked behavior. Our prediction tends to a double-peaked behavior. For $\phi_{2;\eta'}(u,\mu_0)$, we only exhibit the asymptotic form due to there are less results from references, which is shown in Figure~~\ref{fig:fi2}(b). Furthermore, in order to have a look at the evolution $\phi_{2;\eta^{(\prime)}}(u,\mu_0)$ with $n=(2,4,6)$, we present the different curves in Figure~\ref{fig:fi22}. If we take $n=(2,4)$, the behavior of LCDAs shall be closer to the asymptotic form. Furthermore, a small shake is observed when taking the 6th-order LCDA moment into consideration.

Other two-particle Fock state twist-3 and twist-4 LCDAs $\phi _{3;\eta^{(\prime)} }^p(u)$, $\phi _{3;\eta^{(\prime)} }^\sigma (u)$, ${\psi _{4;\eta^{(\prime)} }}(u)$ and ${\phi _{4;\eta^{(\prime)} }}(u)$ are defined as follows
\begin{align}
\phi_{3;\eta^{(\prime)}}^p(u) & = 1 + \Big(30\eta_3^{\eta^{(\prime)}}  - \frac52\rho_{\eta^{(\prime)}}^2 \Big) C_2^{1/2} (\xi) + \Big(-3\eta_3^{\eta^{(\prime)}} \omega_3^{\eta^{(\prime)}}
 -\frac{{27}}{{20}}\rho _{\eta^{(\prime)}} ^2 - \frac{81}{10}\rho_{\eta^{(\prime)}}^2a^2_{2;\eta^{(\prime)}} \Big)
\nonumber\\
&\times C_4^{1/2}(\xi),
\\
\phi_{3;\eta^{(\prime)} }^\sigma (u) & = 6u \bar u\Big(1 + 5\eta _3^{\eta^{(\prime)}}  - \frac{1}{2}\eta _3^{\eta^{(\prime)}} \omega _3^{\eta^{(\prime)}}  - \frac{7}{{20}}\rho _{\eta^{(\prime)}} ^2
- \frac{3}{5}\rho _{\eta^{(\prime)}} ^2
 a^2_{2;\eta^{(\prime)}}\Big)C_2^{3/2}(\xi),
\\
\psi_{4;\eta^{(\prime)}}(u) & = \frac52\varepsilon^2u^2\bar u^2 + \frac12\varepsilon \delta^2 \bigg [u\bar u(2+13u\bar u)+ 10u^3\ln u\ln\bar u (2-3u+\frac65u^2) + 10\,\bar u^3\, (2
\nonumber\\
&- 3\bar u+\frac{6}{5}{\bar u^2})\bigg],
\\
\phi_{4;\eta^{(\prime)}}(u) &= \frac{{10}}{3}{\delta^2}u \bar u(u - \bar u),
\end{align}
where, the values for the twist-3, 4 LCDAs parameters are taken from Refs.~\cite{Ball:2006wn, Huang:2001xb}. In order to run the hadronic parameters of the $\eta^{(\prime)}$-meson twist-2, 3, 4 LCDAs from the initial factorization scale to any other scale, especially for typical scale $\mu_{\rm IR}$, the renormalization group equation should be used, which has the form
\begin{align}
c_i(\mu_{\rm IR}) = {\cal L}^{\gamma_{c_i}/\beta_0}c_i(\mu_0) , \label{Ci}
\end{align}
where ${\cal L}=\alpha_s(\mu_{\rm IR})/\alpha_s (\mu_0)$, $\beta_0 = 11-2/3 n_f$, and the one-loop anomalous dimensions $\gamma_{c_i}$ can be seen in our previous work~\cite{Fu:2020uzy}. Taking the hadronic parameters at initial scale $\mu_0$ and using the renormalization function ~\eqref{Ci}, one can achieve the corresponding values at the typical scale $\mu_{\rm IR}$.

\begin{table}[t]
\renewcommand\arraystretch{1.2}
\centering
\caption{The TFFs $f_+^{\eta^{(\prime)}}(0)$ at the large recoil point $q^2=0$. As a comparison, we also present the predictions from various experimental and theoretical groups.}\label{Tab:fq0}
\begin{tabular}{l l l l}
\hline
~~~~~~~~~~~~~~~~~~~~~~~~~~~~~~~~~~~~&References  ~~~~~~~~~ ~~~~~~~~& $f_ +^{\eta} (0)$   ~~~~~~~~~~~~~~& $f_ + ^{\eta '}(0)$   \\      \hline
           &BESIII~\cite{Ablikim:2019rjz}         & $0.4576(70)$     & $0.490(51)$\\
Experimental results  &LQCD-I~\cite{Bali:2014pva}               & $0.542(13)$      & $0.404(25)$   \\
           &LQCD-II~\cite{Bali:2014pva}               & $0.564(11)$      & $0.437(18)$   \\
\hline
           &This work (LCSR)   & $0.476_{-0.036}^{+0.040}$     & $0.544_{-0.042}^{+0.046}$   \\
           &LFQM~\cite{Verma:2011yw}              & $0.76$            & -   \\
           &CQM~\cite{Melikhov:2000yu}             & $0.78$           & $0.78$   \\
           &CCQM~\cite{Soni:2018adu}               & $0.78(12)$       & $0.73(11)$   \\
\raisebox {2.0ex}[0pt]{Theoretical predictions}&CCQM~\cite{Ivanov:2019nqd}               & $0.49(7)$       & $0.59(9)$   \\
           &LCSR~2013~\cite{Offen:2013nma}               & $0.432(33)$    & $0.520(80)$ \\
           &LCSR~2015~\cite{Duplancic:2015zna}          & $0.495_{-0.030}^{+0.029}$    & $0.558_{-0.045}^{+0.047}$\\
            &QCD SR~\cite{Colangelo:2001cv}               & $0.50(4)$     & -   \\\hline
\end{tabular}
\end{table}

\begin{figure}[t]
\centering
\includegraphics[width=0.52\textwidth]{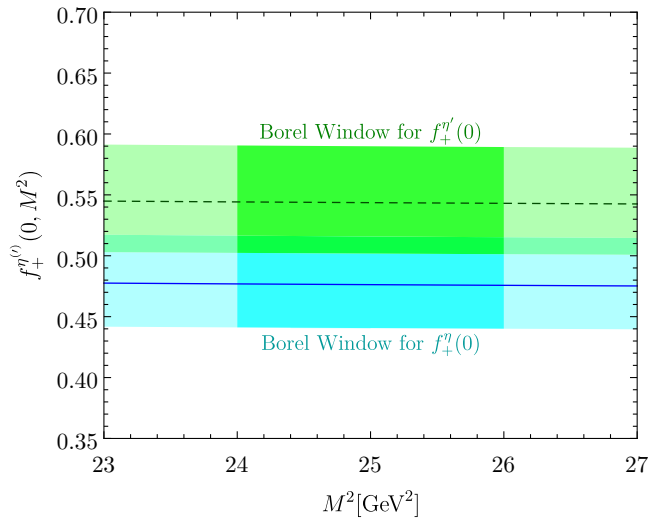}
\caption{The TFFs $f_+^{\eta^{(\prime)}}(0)$ versus the Borel parameters $M^2$, where the shaded band is induced by the variations of squared average of all input parameters.}
\label{Fig:fqM2}
\end{figure}

\subsection{TFFs and series expansion}

In order to determine the continuum threshold $s_0$ for the $D_s\to\eta^{(\prime)}$ TFFs within LCSR approach, i.e. Eq.~\eqref{Eq:fp}, one can follow the four criteria
\begin{itemize}
  \item The continuum contributions are less than $30\%$ of the total results;
  \item The contributions from the twist-4 LCDAs do not exceed $5\%$;
  \item We require the variations of the TFF within the Borel window be less than $10\%$;
  \item The continuum threshold $s_0$ should be closer to the squared mass of the first excited state of $D_s$-meson.
\end{itemize}

Based on the fourth term of the criteria, we take $s_0$ to be close to the squared mass of the excited state of $D_s$-meson $D_{s1}(2460)$, i.e. $s_0 = 6.1(3)~{\rm GeV^2}$. Furthermore, the Borel parameter is taken as $M^2 = 25 (1)~{\rm GeV^2}$. Furthermore, we can obtain the sum rule for $m_{D_s}$ by differentiating the form factors $f_+^{\eta^{(\prime)}}(q^2)f_{D_s}$ with respect to $- 1/M^2$~\cite{Fu:2014pba}. The resultant $m_{D_s}^{\rm LCSR} = 1.9653~{\rm GeV}$ agrees with the measured value $m_{D_s}^{\rm PDG} = 1.9685~{\rm GeV}$. In order to show the degree of stability of the TFFs versus the Borel parameter, we present the curve of TFFs in Figure~\ref{Fig:fqM2}, in which the shaded region shows the errors from all input parameters. The solid line with blue shaded band represents $f_{+}^{\eta}(0)$ and the dashed line with green band represents $f_{+}^{\eta'}(0)$. This figure shows that $f_+^{\eta^{(\prime)}}(0)$ changes less than $0.5\%$ within the range of $M^2\in [23,27]~{\rm GeV^2}$, which  satisfies the third term of the criteria. More definitely, we put the errors caused by different input parameters in the following,
\begin{align}
f_+^\eta(0) & = 0.476+ (_{-0.012}^{+0.011})_{s_0} + (_{-0.001}^{+0.001})_{M^2} + (_{-0.032}^{+0.036})_{m_c,f_{D_s}} + (_{-0.011}^{+0.011})_{f_\eta} + (_{-0.003}^{+0.004})_{a_{2;\eta}^2}
\nonumber\\
& + (_{-0.000}^{+0.000})_{a_{2;\eta }^4}+ (_{-0.000}^{+0.000})_{a_{2;\eta }^6}
\nonumber\\
& = 0.476_{-0.036}^{+0.040}
\\
\nonumber \\
f_+^{\eta'}(0)&= 0.544+(_{-0.016}^{+0.015})_{s_0} + (_{-0.001}^{+0.001})_{M^2} + (_{-0.037}^{+0.042})_{m_c,f_{D_s}} + (_{-0.010}^{+0.010})_{f_{\eta'}} + (_{-0.005}^{+0.006})_{a_{2;\eta'}^2}
\nonumber\\
&+(_{-0.001}^{+0.000})_{a_{2;\eta'}^4} + (_{-0.000}^{+0.000})_{a_{2;\eta '}^6}
\nonumber\\
& = 0.544_{-0.042}^{+0.046}
\end{align}
According to Eq.~\eqref{HPDA_CZ}, when the Gegenbauer moments are taken up to 2nd, 4th and 6th order levels, the central values of TFFs $f_{+}^{\eta^{(\prime)}}(0)$ are
\begin{align}
&f_+^{\eta}(0)|_{\rm 2nd} = 0.4745, && f_+^{\eta}(0)|_{\rm 4th} = 0.4755, &&f_+^{\eta}(0)|_{\rm 6th} = 0.4763, \nonumber\\
&f_+^{\eta'}(0)|_{\rm 2nd} = 0.5435, &&f_+^{\eta'}(0)|_{\rm 4th} = 0.5434, &&f_+^{\eta'}(0)|_{\rm 6th} = 0.5436. \nonumber
\end{align}
In comparing with the 2nd-order Gegenbauer moment's contribution, the 4th, 6th-order contributions shall be changed by about 0.211\%, 0.379\% for $f_+^{\eta}(0)$ and $-0.018\%$, 0.018\% for $f_+^{\eta'}(0)$, respectively. These ratios are really small, indicating the Gegenbauer series has good convergence over the moment expansion. Then, we list the TFFs for $D_s\to\eta^{(\prime)}$ at large recoil point, i.e. $f_+^{\eta^{(\prime)}}(0)$, in Table~\ref{Tab:fq0}, in which the uncertainties are from the squared average of all the mentioned error sources. As a comparison, we also present other theoretical and experimental predictions, such as the BESIII~\cite{Ablikim:2019rjz}, the LQCD~\cite{Bali:2014pva}, the LFQM~\cite{Verma:2011yw}, the CQM~\cite{Melikhov:2000yu}, the CCQM~\cite{Soni:2018adu, Ivanov:2019nqd}, the QCD SR~\cite{Colangelo:2001cv}, the LCSR at 2013 and 2015~\cite{Offen:2013nma, Duplancic:2015zna}, respectively. Our results agree with the BESIII, the CCQM, the LCSR, the LQCD within errors, but are lack of agreement with the Lattice QCD results.

The physically allowable ranges for the TFFs are $m_\ell^2 \le q^2\le (m_{D_s}-m_{\eta})^ 2 \approx 2~{\rm GeV^2}$ and $m_\ell^2 \le q^2\le (m_{D_s}-m_{\eta'})^ 2 \approx 1~{\rm GeV^2}$. Theoretically, the LCSRs approach for $D_s\to\eta^{(\prime)}$ TFFs are applicable in low and intermediate $q^2$-regions, i.e. $q^2\in[0,1.2]~{\rm GeV^2}$ of $\eta$-meson, $q^2\in[0,0.6]~{\rm GeV^2}$ of $\eta'$-meson. One can extrapolate it to whole $q^2$-regions via a rapidly $z(q^2, t)$ converging series expansion (SE)~\cite{Bharucha:2010im}:
\begin{align}
f^{\eta^{(\prime)}}_+(q^2)= \frac1{B(q^2)\phi_+^{\eta^{(\prime)}}(q^2)}\sum\limits_k^{K-1} \alpha_k z^k(q^2,t_0).
\end{align}

\begin{table}[t]
\centering
\renewcommand\arraystretch{1}
\caption{The fitting parameters $z(q^2,t_0^{\eta^{(\prime)}})$ of SSE for TFFs $f_ + ^{{\eta^{(\prime)}}}(q^2)$.}  \label{Tab:zTFFs}
\begin{tabular}{c ccc cccc}
\hline
~~~~~~~~~~~~~~~~~~~&$q^2({\rm GeV^2})$                            & 0                                & 0.4                             & 0.8                              & 1.2                                   & 1.6                                   & 2.0 \\
\raisebox {2.0ex}[0pt]{$\eta$-meson}  & $z(q^2,t_0^\eta)$                      & $0.048$                   & $0.032$                   & $0.014$                    & $-0.005$                       & $-0.025$                       & $-0.047$
\\ \hline
                                                            & $q^2({\rm GeV^2})$                            & 0                                & 0.2                             & 0.4                              & 0.6                                   & 0.8                                   & 1.0 \\
\raisebox {2.0ex}[0pt]{$\eta'$-meson} & $z(q^2,t_0^{\eta'})$                  & $0.016$                   & $0.004$                   & $-0.009$                    & $-0.022$                       & $-0.036$                       & $-0.051$                        \\ \hline
\end{tabular}
\end{table}

Normally, the parameter $K$ stands for the order of expansion. The $z(q^2, t)$ is the function:
\begin{align}
z(q^2, t)=\frac{\sqrt{t_+ - q^2} - \sqrt{t_+ -t}} {\sqrt{t_+ - q^2} +\sqrt{t_+ - t}},
\end{align}
where $t = (t_0^{\eta^{(\prime)}}, t_-^{\eta^{(\prime)}}, m_{D_s})$ with $t_\pm^{\eta^{(\prime)}} =(m_{D_s}\pm m_{\eta^{(\prime)}})^2$. Here, the $0\leq t_0^{\eta^{(\prime)}} \leq t_-^{\eta^{(\prime)}}$ is a free parameter which can be optimised to reduce the maximum value of $|z(q^2,t_0^{\eta^{(\prime)}})|$ in the physical TFFs range, $t_0^{\eta^{(\prime)}}|_{\rm opt.} = t_+^{\eta^{(\prime)}} (1-\sqrt{1-t_-^{\eta^{(\prime)}}/t_+^{\eta^{(\prime)}}})$. Here, we list the value of $z(q^2, t_0^{\eta^{(\prime)}})$ with different $q^2$ cases in Table~\ref{Tab:zTFFs}. The function $\phi_+^{\eta^{(\prime)}}(q^2)$ can be expressed as
\begin{eqnarray}
\phi_+^{\eta^{(\prime)}}(q^2) &=& \sqrt{\dfrac{\varsigma}{48\pi\chi_+^{\eta^{(\prime)}}(n)}} \dfrac{q^2 - t_+}{(t_+ - t_0)^{1/4}} \bigg(\dfrac{z(q^2, 0)}{-q^2}\bigg)^{(3+n)/2} \bigg(\dfrac{z(q^2, t_0)}{t_0 -q^2}\bigg)^{-1/2} \nonumber\\
&\times& \bigg(\dfrac{z(q^2, t_-)}{t_- -q^2}\bigg)^{-3/4}\bigg|_{n=2},
\end{eqnarray}
\begin{figure}[t]
\centering
\includegraphics[width=0.5\textwidth]{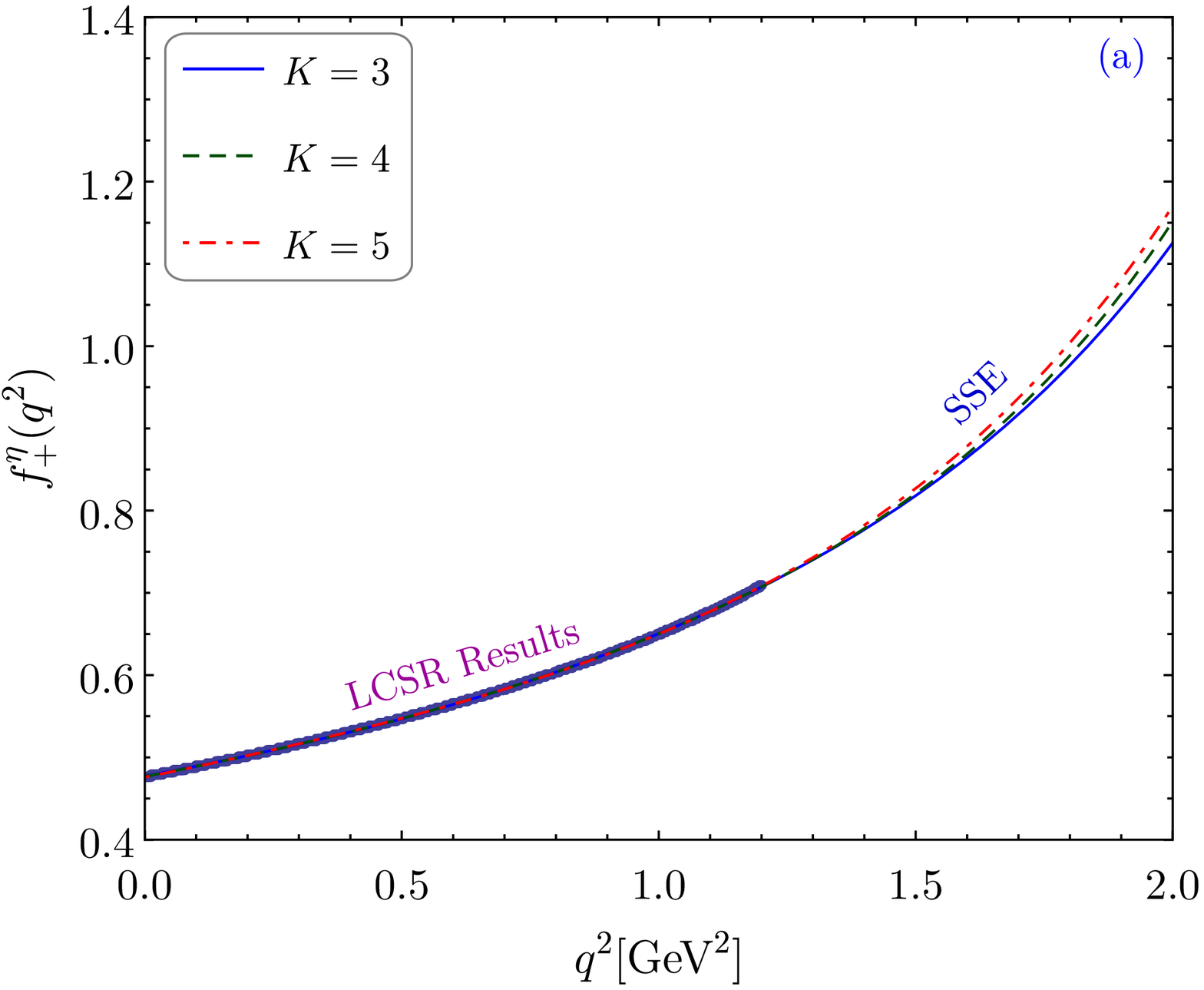}\includegraphics[width=0.5\textwidth]{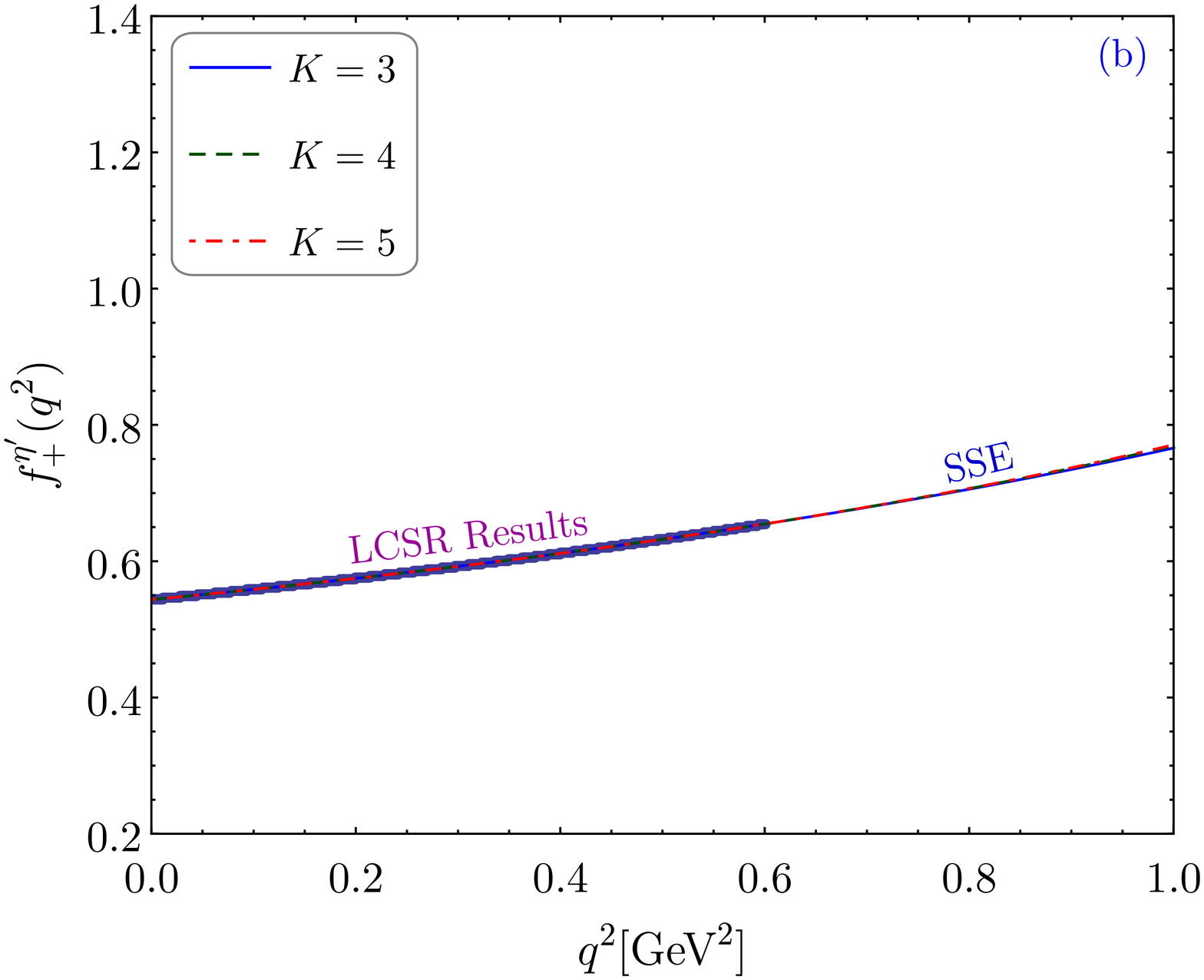}
\caption{The SSE for the TFFs $f^\eta_+(q^2)$ and $f^{\eta^{\prime}}_+(q^2)$ up to $K=3,4,5$ order.}
\label{Fig:TFFs_K345}
\end{figure}
where $\varsigma = 1 $ is an isospin-degeneracy factor for a given channel and $B(q^2) = \prod\limits_i z(q^2, m_{D_s^i}^2)$ stands for the Blaschke factor. The $m_{D_s^i}$ stands for the mass of each resonance state with $J^P = 0^-, 0^+, 1^-...$, respectively, which can be found in PDG~\cite{Zyla:2020zbs}. For the coefficients $\chi_+^{\eta^{(\prime)}}(n)$, it can be calculated by using QCD sum rules including perturbative LO and NLO results as well as the condensate contributions, which can be expressed as~\cite{Bharucha:2010im}
\begin{eqnarray}
\chi_+^{\eta^{(\prime)}} = \frac{3}{32\pi^2 m_c^2} \bigg[1+\frac{\alpha_s(m_c)C_F}{4\pi}\bigg(\frac{25}{6}+\frac{2\pi^2}{3} \bigg) \bigg]- \frac{\langle q\bar q\rangle}{m_c^5} - \frac{\langle \alpha_s G^2\rangle}{12\pi m_c^6} - \frac{\langle \bar qGq\rangle}{m_c^7}.
\end{eqnarray}
Furthermore, the coefficients $\alpha_k$ should satisfy the basic unitarity constraint,
\begin{eqnarray}
\sum\limits_k^{K-1} \alpha_k^2 <1
\end{eqnarray}

The simplified version of the series expansion (SSE) method is to replace the Blaschke factor $B(q^2)$ by a simple pole $P(q^2)$ to account for low-lying resonances, i.e.~\cite{Bourrely:2008za}
\begin{align}
f_+^{\eta^{(\prime)}}(q^2) =\frac1{P(t)}\sum\limits_k^{K-1} \beta_k z^k(q^2,t_0).
\end{align}
As for the SSE parameterization, imposing the unitarity bound and by comparing the SE and SSE parameterizations
\begin{align}
\alpha_i= \sum\limits_{k=0}^{\min [K-1,i]} \zeta_{i-k}\beta_k,~~~~~~~0\leq i \leq K-1, \label{Eq:abij}
\end{align}
we obtain the unitarity bound of the SSE
\begin{align}
\sum\limits_{j,k=0}^{\min [K-1]} C_{jk}\beta_j \beta_k\leq 1,
\end{align}
with the positive defined matrix
\begin{align}
C_{jk} = \sum\limits_{i=0}^{K-1-\max [j,k]} \zeta_i\zeta_{j-k}.
\end{align}

\begin{table}[t]
\renewcommand\arraystretch{1.2}
\caption{The fitting parameters for the central TFFs $f_ + ^{{\eta^{(\prime)}}(C)}( q^2)$, the upper TFFs $f_+^{{\eta^{(\prime)}}(U)}( q^2)$ and the lower TFFs $f_+^{{\eta^{(\prime)}}(L)}( q^2)$.}  \label{Tab:fit}
\begin{tabular}{l l l l l l l }
\hline
& $f_+^{\eta(C)}(q^2)$ ~& $f_+^{\eta(U)}(q^2)$   & $f_+^{\eta(L)}(q^2)$    & $f_+^{\eta^{\prime}(C)}(q^2)$   & $f_+^{\eta^{\prime}(U)}( q^2)$   & $f_+^{\eta^{\prime}(L)}( q^2)$  \\ \hline
$\alpha_0$                                          & $-0.0003$                   & $-0.0003$                   & $-0.0003$                    & $-0.00001$                       & $-0.00001$                       & $-0.00001$
\\
$\alpha_1$                                          & $-0.0015$                   & $-0.0017$                   & $-0.0015$                    & $-0.00017$                       & $-0.00018$                       & $-0.00016$
\\
$\alpha_2$                                          & $-0.0061$                   & $-0.0074$                   & $-0.0052$                    & $-0.00134$                       & $-0.00143$                       & $-0.00127$
\\
$\displaystyle\sum{\alpha_k^2}$  & $3.9 \times 10^{-5}$  & $5.8 \times 10^{-5}$  & $2.9 \times 10^{-5}$  & $1.8 \times 10^{-6}$      & $2.4 \times 10^{-6}$      & $1.6 \times 10^{-6}$
\\
\hline
$\beta_0$                                          & $0.512$                    & $0.555$                       & $0.476$                      & $0.563$                           & $0.611$                          & $0.519$
\\
$\beta_1$                                          & $-1.450$                     &  $-1.211$                   &  $-0.976$                 &  $-1.653$                         &  $-1.801$                        &  $-1.534$
\\
$\beta_2$                                          & $17.26$                    &  $12.24$                   &  $8.73$                  &  $28.95$                         &  $30.61$                                   &  $27.98$
\\
$\displaystyle\sum C_{i,j}\beta_i \beta_j$ & $0.020$                   &  $0.019$                      & $0.008$                     & $0.005$                              & $0.005$                            & $0.004$
\\
$\Delta$                                       & $0.075\%$                & $0.129\%$                    & $0.099\%$                 & $0.005\%$                          & $0.004\%$                         & $0.007\%$                            \\ \hline
\end{tabular}
\end{table}

Theoretically, the order $K$ can be taken up to infinite order. It has been proven that the higher order expansion shall give the same result with very small errors. We take $K=3,4,5$ as explicit examples, which are shown in Figure~\ref{Fig:TFFs_K345}. It is found that three SSE curves with $K=3,4,5$ for the TFFs $f_+^{\eta^{(\prime)}}(q^2)$ are almost coincide with each other. So, we will take $K=3$ to do our expansion in the following calculations, which also agrees with the choices of most of the theoretical groups. At the same time, the $\alpha_k$ of SE and $\beta_k$ of SSE should also satisfy the condition $\Delta<1\%$. From which, the parameter $\Delta$ is used to measure the quality of extrapolation, which is defined as
\begin{align}
\Delta = \dfrac{\sum_t|F_i(t) - F_i^{\rm fit}(t)|}{\sum_t|F_i(t)|} \times 100, \label{Eq:fit}
\end{align}
where $t\in[0,1/100,\cdots,100/100]\times 1.2~{\rm GeV}$ for $\eta$-meson, $t\in[0,1/100,\cdots,100/100]\times 0.6~{\rm GeV}$ for $\eta'$-meson. Numerical values of $\alpha_k$, $\beta_k$ with $k=(0,1,2)$ for central values, upper and lower limit of $f_+^{\eta^{(\prime)}}(q^2)$ are listed in Table~\ref{Tab:fit}. Here, the $\sum \alpha_k^2$ for SE unitarity bound and $\displaystyle\sum C_{i,j}\beta_i \beta_j$ for SSE are less than 1. The quality of extrapolation $\Delta$ are less than 0.13\%. Since there have good unitary bound and a small $\Delta$ value, the SSE results are in high agreement with our LCSR results.

\begin{figure}[t]
\centering
\includegraphics[width=0.5\textwidth]{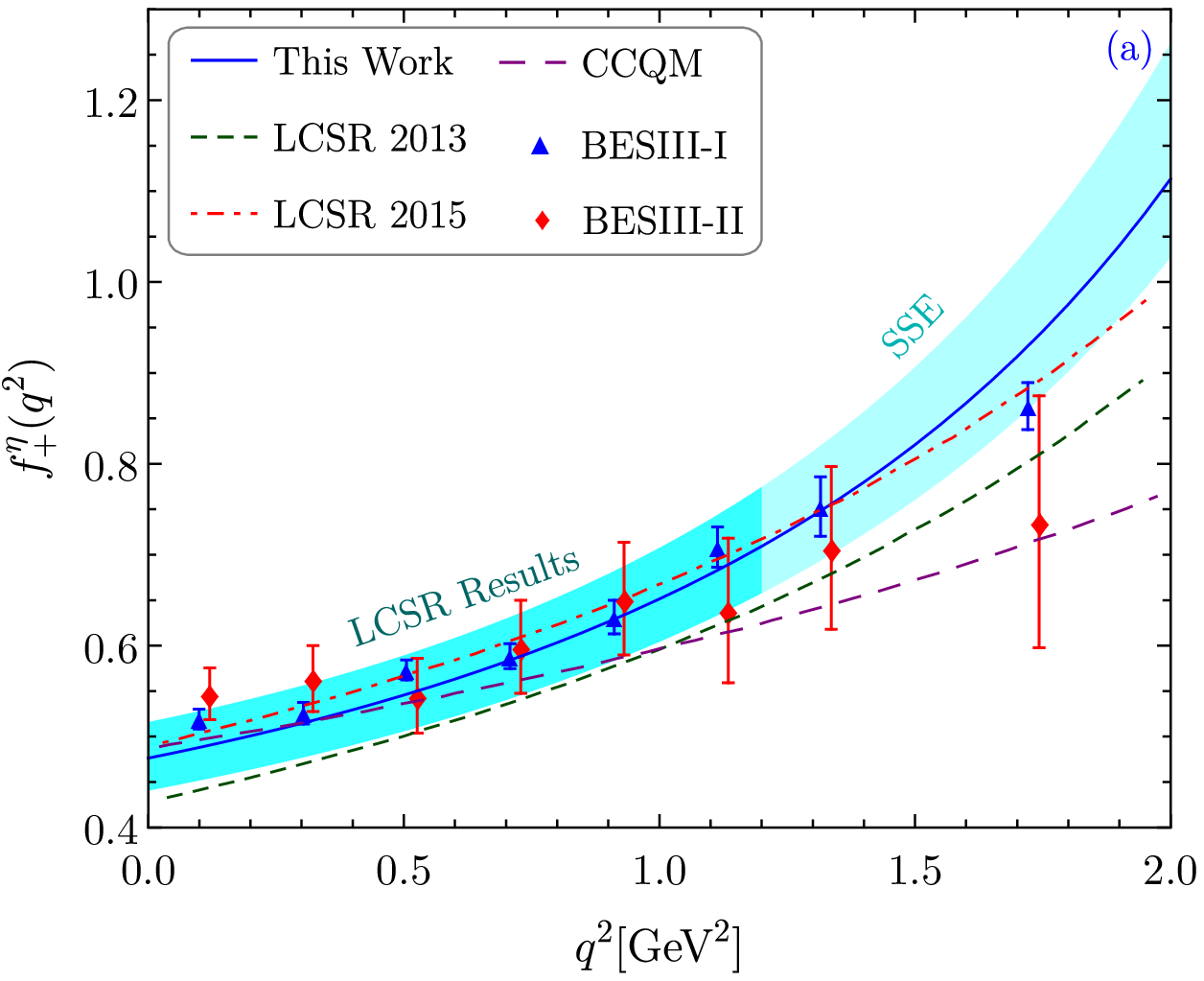}\includegraphics[width=0.5\textwidth]{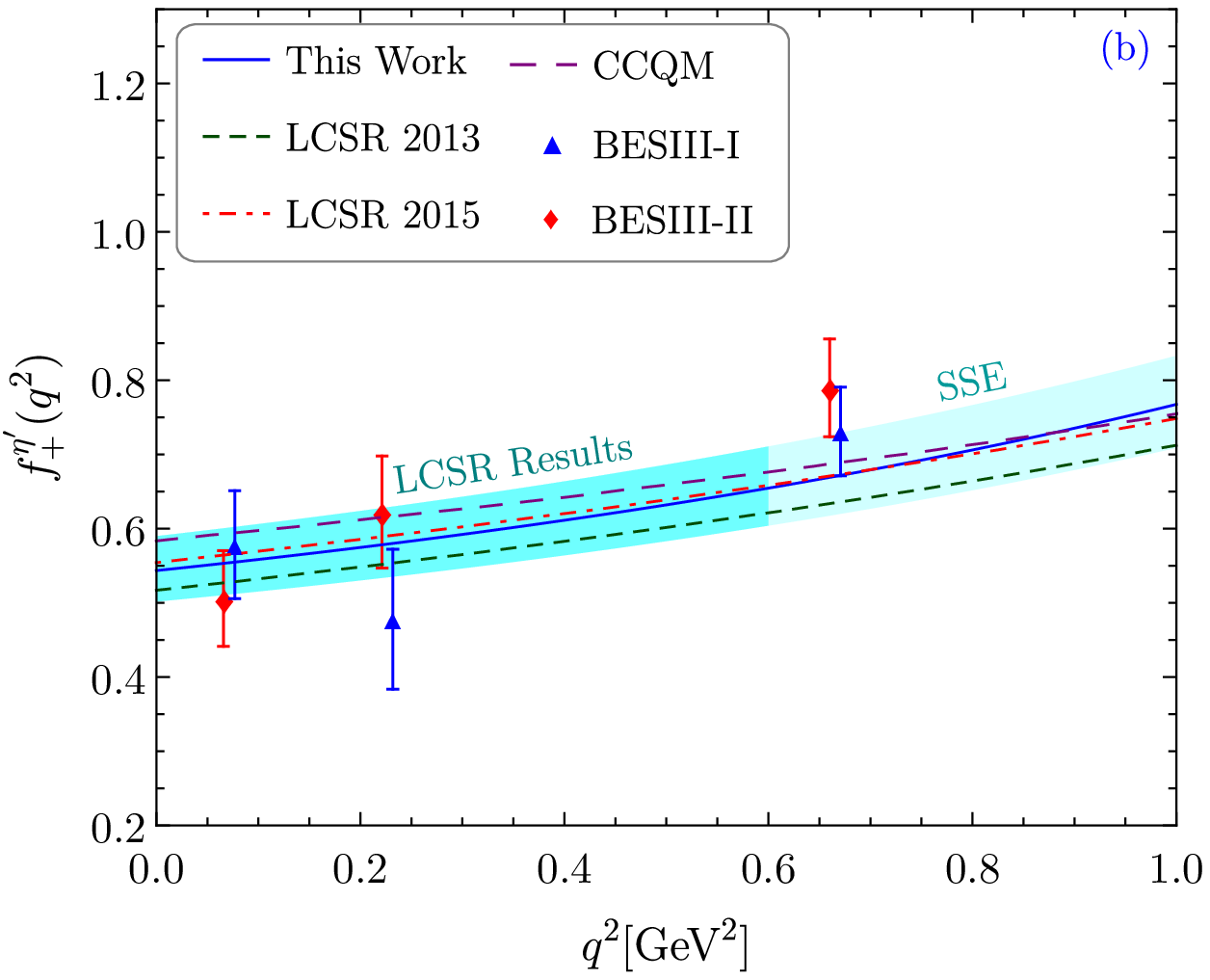}
\caption{The TFFs $f^{\eta^{(\prime)}}_+(q^2)$ together with its uncertainties. Results of the LCSR 2013~\cite{Offen:2013nma}, the LCSR 2015~\cite{Duplancic:2015zna}, the CCQM~\cite{Ivanov:2019nqd} and the BESIII~\cite{Ablikim:2019rjz} are also given as the comparison. }
\label{Fig:fq}
\end{figure}

The extrapolated TFFs in the whole $q^2$-region is shown in Figure~\ref{Fig:fq}, other theoretical and experimental results, such as those of the BESIII~\cite{Ablikim:2019rjz}, the LCSR 2013~\cite{Offen:2013nma}, the LCSR 2015~\cite{Duplancic:2015zna} and the CCQM~\cite{Ivanov:2019nqd} are present as a comparison. Two sets of BESIII are from the two different $\eta^{(\prime)}$ decay channels. For $\eta$-meson, the blue triangle stands for $\eta\to\gamma\gamma$ channel and the red diamond stands for $\eta\to\pi^0\pi^+\pi^-$ channel. For $\eta'$-meson the blue triangle stands for $\eta'\to\gamma\rho^0$ channel and red diamond stands for $\eta'\to\eta_{\gamma\gamma}\pi^+\pi^-$ channel. To compare with other theoretical and experimental groups, our results have the following characteristics,
\begin{itemize}
  \item Comparing with $f^\eta_+(q^2)$, the $f^{\eta'}_+(q^2)$ is more flat in the whole $q^2$ region.
  \item Our predictions of $f^\eta_+(q^2)$ are in good agreement with the recent BESIII predictions for the $\eta\to\gamma\gamma$ channel.
  \item In the LCSR $q^2$-region, our results have good agreement with the LCSR 2013 and 2015, the CCQM, and the two sets of BES-III predictions within errors.
  \item The SSE of $f^\eta_+(q^2)$ in the region of $q^2\in [1.2, 2.0]~{\rm GeV}^2$ have agreement with the LCSR in 2015 results within errors. However, our predictions are larger than the LCSR in 2013 and CCQM predictions due to the different $\eta$-meson distribution amplitudes or different method.
\end{itemize}

\begin{figure}[t]
\centering
\includegraphics[width=0.5\textwidth]{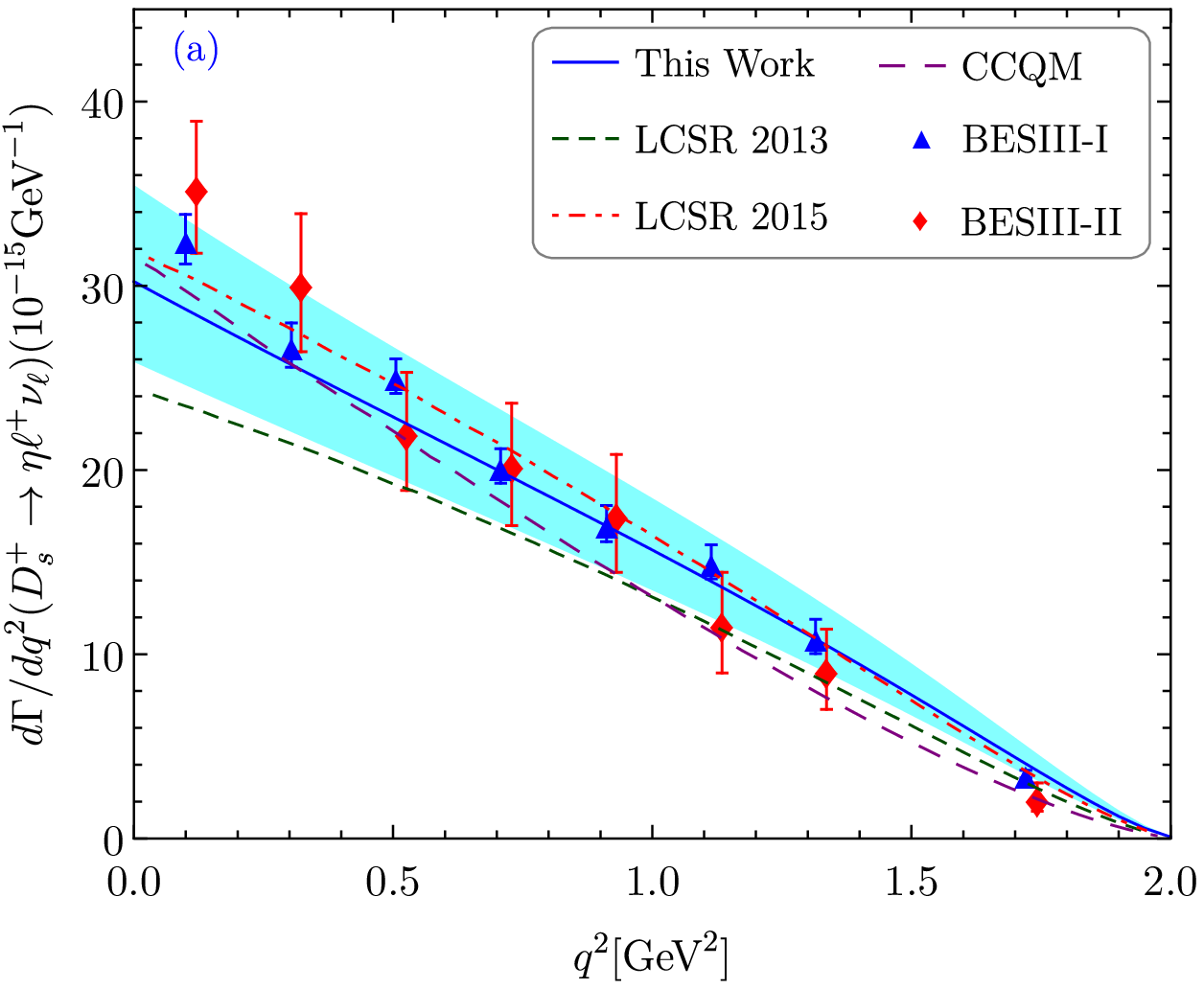}\includegraphics[width=0.5\textwidth]{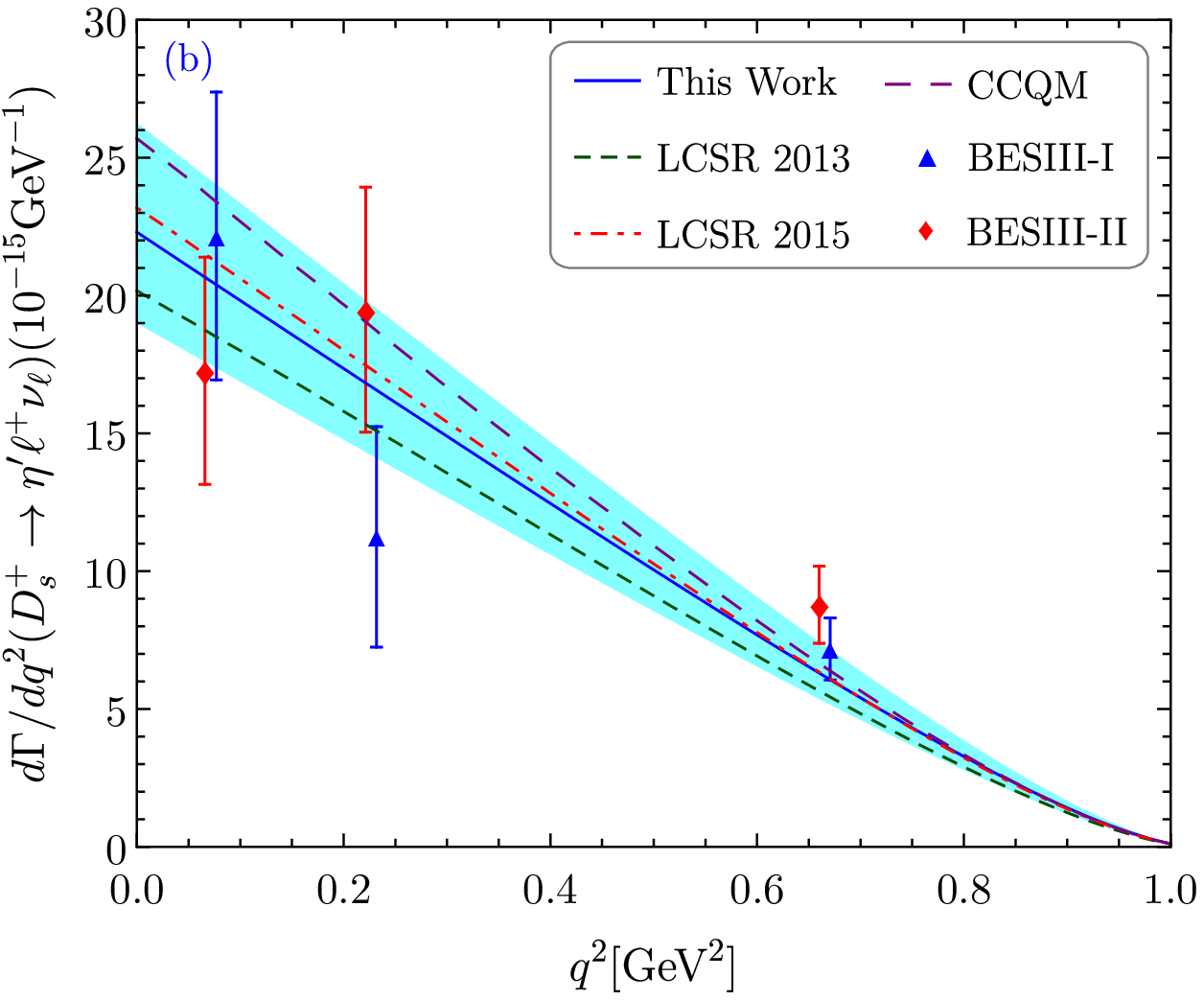}
\caption{Decay width for the $D_s^+\to \eta^{(\prime)}\ell ^+\nu_\ell (\ell =e,\mu )$ versus $q^2$ within uncertainties. The LCSR in 2013~\cite{Offen:2013nma} and 2015~\cite{Duplancic:2015zna}, the CCQM~\cite{Ivanov:2019nqd}, and two sets of BESIII collaboration~\cite{Ablikim:2019rjz} predictions are also present as a comparison.}
\label{Fig:dGamma}
\end{figure}

\subsection{Decay widthes and branching fractions for the semileptonic decay $D_s^+ \to \eta^{(\prime)} \ell^+ \nu_\ell$}

One can get the differential decay widths by using the formula ~\eqref{Eq:dGamma}. For the CKM matrix element $|V_{cs}|$, we take it to be the average value of leptonic and semileptonic decay $c\to s$ processes coming from PDG~\cite{Narison:2014ska}, i.e. $|V_{cs}| = 0.987\pm0.011$. After taking the derived $D_s\to{\eta^{(\prime)}}$ TFFs into the decay widths, we present the differential decay widths in Figure~\ref{Fig:dGamma}. As a comparison, we also give the LCSR in 2013~\cite{Offen:2013nma} and 2015~\cite{Duplancic:2015zna}, the CCQM~\cite{Ivanov:2019nqd}, and two sets of BESIII collaboration~\cite{Ablikim:2019rjz} predictions. The LCSR and the CCQM results are calculated by applying their TFFs into the width formula. Those figures show that our prediction for $D_s^+\to\eta\ell^+\nu_\ell$ is in agreement with LCSR 2015, CCQM, the BESIII-I and the BESIII-II results within errors, and the $D_s^+\to\eta'\ell^+\nu_\ell$ agrees with the LCSR 2013 and 2015, the CCQM, the BESIII-I and BESIII-II predictions within errors. All the results are convergence to zero at the small recoil region $q^2_{\max} = (m_{D_s}-m_{\eta^{(\prime)}})^2$, which indicates that our results are reasonable.

\begin{table}[t]
\renewcommand\arraystretch{1.2}
\centering
\caption{Branching factions of $D_s^+\to \eta^{(\prime)} \ell^+ \nu_\ell$ with $\ell =e$ and $\mu$ (in unit $10^{-2}$). The errors are squared averages of all the mentioned error sources. As a comparison, we also present the predictions for various methods.}\label{tab:BF}
\begin{tabular}{l l l l l l}
\hline\hline
~~~~~~~~~~~~~~~~~~~~~~~~~~~~~~~~~~~~~~ &Mode~~~~~~~~~~~~~~~~~~~~~~~~~~           & ${\cal B}(D_s^+ \to\eta e^+\nu_e)$~~~~~~~~~~~~~~~~  & ${\cal B}(D_s^ + \to\eta \mu^+ \nu_\mu)$
\\\hline
&BESIII~\cite{Ablikim:2019rjz,Ablikim:2017omq}    & $2.323\pm 0.063\pm 0.063$           & $2.42\pm 0.46\pm 0.11$
\\
&CLEO~\cite{Hietala:2015jqa}                      & $2.28\pm 0.14\pm 0.19$              &-                                         \\
\raisebox {2.0ex}[0pt]{Experimental results}&CLEO~\cite{Yelton:2009aa}                        & $2.48\pm 0.29\pm 0.13$              &-                                         \\
&PDG~\cite{Zyla:2020zbs}                          & $2.32\pm 0.08$                      & $2.4\pm 0.5$
\\
\hline
&This work (LCSR)                                 & $2.346_{-0.331}^{+0.418}$           & $2.320_{-0.327}^{+0.413}$                \\
&LFQM~\cite{Cheng:2017pcq}                        & $2.26\pm 0.21$                      & $2.22\pm 0.20$                           \\
&CCQM~\cite{Ivanov:2019nqd}                       & $2.24$                              & $2.18$                                   \\
Theoretical predictions &LCSR~\cite{Offen:2013nma}                        & $2.00\pm 0.32$                      &-                                         \\
&LCSR~\cite{Duplancic:2015zna}                    & $2.40\pm 0.28$                      &-                                         \\
&QCD SR-I~\cite{Colangelo:2001cv}                 & $2.6\pm 0.7$                        &-                                         \\
&QCD SR-II~\cite{Colangelo:2001cv}                & $2.3\pm 0.4$                        &-                                         \\
\hline\hline

&Mode                                             & ${\cal B}(D_s^+ \to \eta' e^+\nu_e)$  & ${\cal B}(D_s^ + \to\eta'\mu^+\nu_\mu)$\\  \hline
&BESIII~\cite{Ablikim:2019rjz,Ablikim:2017omq}    & $0.824\pm 0.073\pm 0.027$             & $1.06\pm 0.54\pm 0.07$
\\
&CLEO~\cite{Hietala:2015jqa}                      & $0.68\pm 0.15\pm 0.06$                &-                                       \\
\raisebox {2.0ex}[0pt]{Experimental results}&CLEO~\cite{Yelton:2009aa}                        & $0.91\pm 0.33\pm 0.05$                &-                                       \\
&PDG~\cite{Zyla:2020zbs}                          & $0.80\pm 0.07$                        & $1.1\pm 0.5$
\\
\hline
&This work (LCSR)                                 & $0.792_{-0.118}^{+0.141}$             & $0.773_{-0.115}^{+0.138}$
\\
&LFQM~\cite{Cheng:2017pcq}                        & $0.89\pm 0.09$                        & $0.85\pm 0.08$
\\
&CCQM~\cite{Ivanov:2019nqd}                       & $0.83$                                & $0.79$                                 \\
Theoretical predictions &LCSR~\cite{Offen:2013nma}                        & $0.75\pm 0.23$                        &-                                       \\
&LCSR~\cite{Duplancic:2015zna}                    & $0.79\pm 0.14$                        &-                                       \\
&QCD SR-I~\cite{Colangelo:2001cv}                 & $0.89\pm 0.34$                        &-                                       \\
&QCD SR-II~\cite{Colangelo:2001cv}                & $1.0\pm 0.2$                          &-                                       \\
\hline\hline
\end{tabular}
\end{table}

After integrating over the whole $m_\ell^2\leq q^2\leq  (m_{D_s}-m_{\eta^{(\prime)}})^2$ region for the differential decay widths, we obtain the total decay widths for $D_s^+ \to \eta^{(\prime)} \ell^+ \nu_\ell$ with two different channel $\Gamma(D_s^+\to{\eta^{(\prime)}} e^+\nu_e)$ and $\Gamma(D_s^+\to{\eta^{(\prime)}}\mu^+\nu_\mu)$, i.e.,
\begin{eqnarray}
&& \Gamma (D_s^{+}\to \eta e^+ \nu_e)        = 30.634_{-4.323}^{+5.453} \times 10^{-15}~{\rm GeV} , \nonumber\\
&& \Gamma (D_s^{+}\to\eta\mu^+ \nu_\mu)  = 30.298_{-4.275}^{+5.395} \times 10^{-15}~{\rm GeV} , \nonumber\\
&& \Gamma (D_s^{+}\to \eta' e^+ \nu_e)       = 10.345_{-1.534}^{+1.844} \times 10^{-15}~{\rm GeV} , \nonumber\\
&& \Gamma (D_s^{+}\to\eta'\mu^+ \nu_\mu) = 10.096_{-1.498}^{+1.800}   \times 10^{-15}~{\rm GeV} . \label{Eq:GammaNumerical}
\end{eqnarray}

Then, by using the lifetime of the initial state $D_s^+$-meson, $\tau_{D_s^+}=(0.504\pm 0.007)~ {\rm ps}$~\cite{Zyla:2020zbs}, the branching fractions for the two different semileptonic decay channels $D_s\to{\eta^{(\prime)}}{{\ell}^{+}}\nu_\ell$ with $\ell = (e,\mu)$ can be obtained, which are presented in Table~\ref{tab:BF}. Here, we also listed the BESIII~\cite{Ablikim:2019rjz,Ablikim:2017omq}, the PDG~\cite{Zyla:2020zbs}, the CLEO~\cite{Yelton:2009aa} for the experimental results, and the CCQM~\cite{Ivanov:2019nqd}, the LFQM~\cite{Cheng:2017pcq}, the QCDSR-I, II~\cite{Colangelo:2001cv}, the LCSR~\cite{Duplancic:2015zna, Offen:2013nma} for theoretical predictions. Our results are closer to the BESIII, PDG, CLEO results, all of which are within $1\sigma$ uncertainties.

\begin{figure}[t]
\centering
\includegraphics[width=0.52\textwidth]{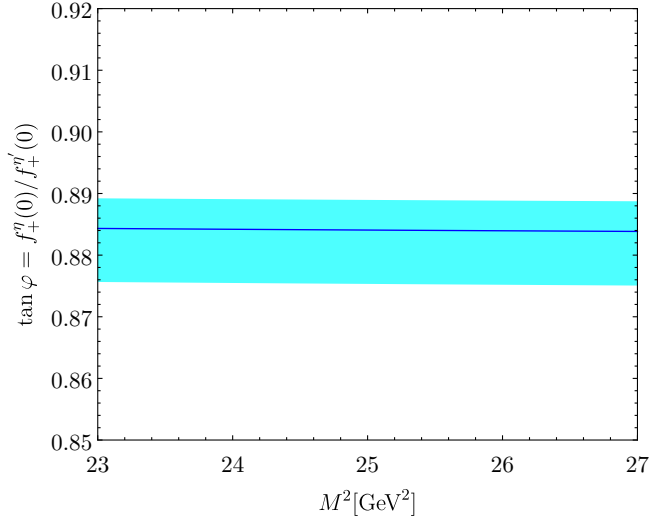}
\caption{$f^{\eta}_{+}(0)/f^{\eta'}_{+}(0)$ as a function of the Borel parameter $M^2$, where the shaded band is induced by the variations of squared average of all input parameters}
\label{Fig:BT}
\end{figure}

\begin{table}[t]
\renewcommand\arraystretch{1.2}
\centering
\caption{The $\varphi$ with respect to mixing angle, ${\cal R}_{\eta'/\eta}^\ell$ for different models and experimental values. As a comparison, we also present the experimental and theoretical predictions.}\label{tab:RDs}
\begin{tabular}{l l l}
\hline
Mode~~~~~~~~~~~~~~~~~~~~~~~~~~~~~~~~~~~~& Angle $\varphi$ ~~~~~~~~~~~~~~~~~~& ${\cal R}_{\eta'/\eta}^\ell$ \\ \hline
This work~(LCSR) with $\ell = e$     & ${41.2^ \circ }_{ - 0.06}^{ + 0.05}$      & $0.338_{ - 0.051}^{ + 0.057}$   \\
This work~(LCSR) with $\ell = \mu$     & ${41.2^ \circ }_{- 0.06}^{ + 0.05}$     & $0.333_{-0.058}^{+0.050}$   \\
CLEO~\cite{Brandenburg:1995qq}   &-        & $0.35\pm 0.09\pm 0.07$   \\
BESIII~\cite{Ablikim:2016rqq}    &-        & $0.40\pm 0.14\pm 0.02$  \\
LFQM~\cite{Wei:2009nc}   & ${39}^\circ$    & $0.39$  \\
LCSR~\cite{Duplancic:2015zna}     & ${41.8}^\circ$      & $0.33\pm 0.07$   \\
QCD SR~\cite{Colangelo:2001cv}      & ${40}^\circ$      & $0.44\pm 0.01$     \\
KLOE~\cite{Ambrosino:2006gk}     & ${41.4}^\circ$      &-        \\
LCSR-I~\cite{Azizi:2010zj}     & ${39.7}^\circ$        & $0.32\pm 0.02$    \\
LCSR-II~\cite{Azizi:2010zj}     & ${41.5}^\circ$        & $0.27\pm 0.01$    \\
\hline
\end{tabular}
\end{table}

After substituting the corresponding terms of Eq.~\eqref{bt}, one can get the mixing angle of $\eta-\eta'$. The mixing angle $\tan \varphi$ with the parameter of Borel parameter $M^2$ are shown in Figure~\ref{Fig:BT}. In the whole Borel parameter $M^2$ region, the mixing angle is changed slightly, which also indicates that there have a stable Borel window for $\tan\varphi$. Numerical results for the single mixing angle $\varphi$ are present in Table~\ref{tab:RDs}. Our result is closer to the KLOE~\cite{Ambrosino:2006gk} and the LCSR predictions~\cite{Duplancic:2015zna, Azizi:2010zj}, which is slightly larger than other LFQM~\cite{Wei:2009nc} and QCD SR~\cite{Colangelo:2001cv} predictions.

Furthermore, it is useful to study the ratio for the different decay channel ${\cal R}_{\eta'/\eta}^\ell$ related to the mixing angle, which has the basic definition
\begin{eqnarray}
{\cal R}_{\eta'/\eta}^\ell = \frac{{\cal B}(D_s \to \eta'\ell^+ \nu_\ell)}{{\cal B}(D_s \to \eta \ell^+ \nu_\ell)}.
\end{eqnarray}
Numerical results together with different experimental and theoretical predictions are given in Table~\ref{tab:RDs}. Our results are in agreement with the CLEO~\cite{Brandenburg:1995qq} and the BESIII~\cite{Ablikim:2016rqq} predictions, and the LCSR predictions~\cite{Duplancic:2015zna,Azizi:2010zj}. This can be considered as a good test of the correctness of the considered internal structure for the $D_s$-meson as well as the mixing angle between $\eta$ and $\eta'$ states.

\section{Summary}\label{sec:summary}

In this paper, we have calculated the moments $\langle\xi_{2;\eta^{(\prime)}}^n\rangle|_\mu$ of $\eta^{(\prime)}$-meson LCDA with $n=(2,4,6)$ up to NLO correction and completely dimension-six condensates within BFTSR, which are shown in Eq.~\eqref{xi2}. Due to the $\eta^{(\prime)}$-meson should be considered as $s\bar s$ component in $D_s\to\eta^{(\prime)}$ decay processes, we have also taken the $I_{m_s^2}(n,M^2)$ corrections into consideration, i.e. Eq.~\eqref{ms}, the detailed terms of OPE are listed in the Appendix~\ref{sec:appendixB}.

Then, we have sought a reasonable continuum threshold $s_0$ and stable Borel windows for  $\langle\xi_{2;\eta^{(\prime)}}^n\rangle|_\mu$ with $n=(2,4,6)$ by using the traditional three criteria for the SVZ sum rules, which are present in Figs.~\ref{fig:con}, \ref{fig:xi} and Table~\ref{tab:m2}. By using the expression between two different Gegenbauer and LCDA moments, i.e. Eq.~\eqref{xi}, we have presented the first three Gegenbaner and LCDA moments $a^n_{2;\eta^{(\prime)}}(\mu_0)$ and $\langle \xi^n_{2;\eta^{(\prime)}}\rangle|_{\mu_0}$ with $n = (2,4,6)$ for the $\phi_{2;\eta^{(\prime)}}(u,\mu_0)$ within errors in Table~\ref{Tab:anxin}. Our results are in agreement with the CLEO fit and the BABAR fit predictions. Meanwhile, we have exhibited the curves of $\phi_{2;\eta^{(\prime)}}(u,\mu_0)$ of our prediction with $n=(2,4,6)$ and compare with others.

Furthermore, the TFFs $f^{\eta^{(\prime)}}_+(q^2)$ have been given in Eq.~\eqref{Eq:fp} up to NLO QCD corrections for twist-2, 3 LCDA contributions. The TFFs at large recoil region have been presented in Table~\ref{Tab:fq0} with respect to other theoretical predictions. After extrapolating it to the whole physical $q^2$-region via simplified series expansion, we have shown the behavior of TFFs in Figure~\ref{Fig:fq}. The differential/total decay widths and branching fractions in this work have also been given. Our results are in agreement with the BESIII and the PDG average value within errors. Finally, we have presented the mixing angle $\tan\varphi$ and ratio for different decay channel ${\cal R}_{\eta'/\eta}^\ell$, which agree with theoretical and experimental results within errors. Thus, the QCDSR within BFTSR can be considered a good approach in dealing with the heavy-to-light semileptonic processes, and we hope more data can be achieved in the near future for more precise studies.

\section{Acknowledgments}

We are grateful to Dr. Xu-Chang Zheng for helpful discussions and valuable suggestions. Hai-Bing Fu would like to thank the Institute of Theoretical Physics in Chongqing University (CQUITP) for kind hospitality. This work was supported in part by the National Natural Science Foundation of China under Grant No.11765007, No.11947406, No.11625520, and No.12047564, the Project of Guizhou Provincial Department of Science and Technology under Grant No.KY[2019]1171, and No.ZK[2021]024, the Project of Guizhou Provincial Department of Education under Grant No.KY[2021]030 and No.KY[2021]003, the China Postdoctoral Science Foundation under Grant Nos.2019TQ0329, 2020M670476, the Chongqing Graduate Research and Innovation Foundation under Grant No.ydstd1912, the Fundamental Research Funds for the Central Universities under Grant No.2020CQJQY-Z003, and the Project of Guizhou Minzu University under Grant No. GZMU[2019]YB19.

\appendix

\section{Quark propagators and vertex for background field theory framework} \label{sec:appendixA}

\subsection{Quark propagators}\label{sec:appendix_I1}

Quarks field and its propagators satisfy the equation Eq.~\eqref{qu} in the BFT, slove to Eq.~\eqref{qu}, we have the following form
\begin{align}
S_F(x,0)=(i\DSS D + m){\rm \cal D}(x,0),
\end{align}
where ${\rm \cal D}(x,0)$ satisfies the equation
\begin{align}
(\partial^2 - {\cal P}_\mu \partial^\mu  - {\cal Q} + m^2 ){\cal D}(x,0) = \delta^4(x),
\end{align}
with ${\cal P}_\mu=2i {\cal A}_\mu$, ${\cal Q}=\gamma^\nu \gamma^\mu [{\cal A}_\nu (x) {\cal A}_\mu (x) + i\partial _\nu {\cal A}_\mu(x)]$. After using above equation up here we can get quark propagator $S_F(x,0)$ subexpression
\begin{align}
S_F(x,0) = S_F^0(x,0) + S_F^2(x,0) + S_F^3(x,0) +\sum\limits_{i=1}^2 S_F^{4(i)} (x,0) + \sum\limits_{i=1}^3 S_F^{5(i)} (x,0) + \sum\limits_{i=1}^5 S_F^{6(i)}(x,0).
\end{align}
In order to have a clear look at the detailed quark propagator $S_F(x,0)$ separately, we list the expression of $S_F^d(x,0)$ with $d\leq 6$ as follows,

\begin{figure}
\includegraphics[width=0.9\textwidth]{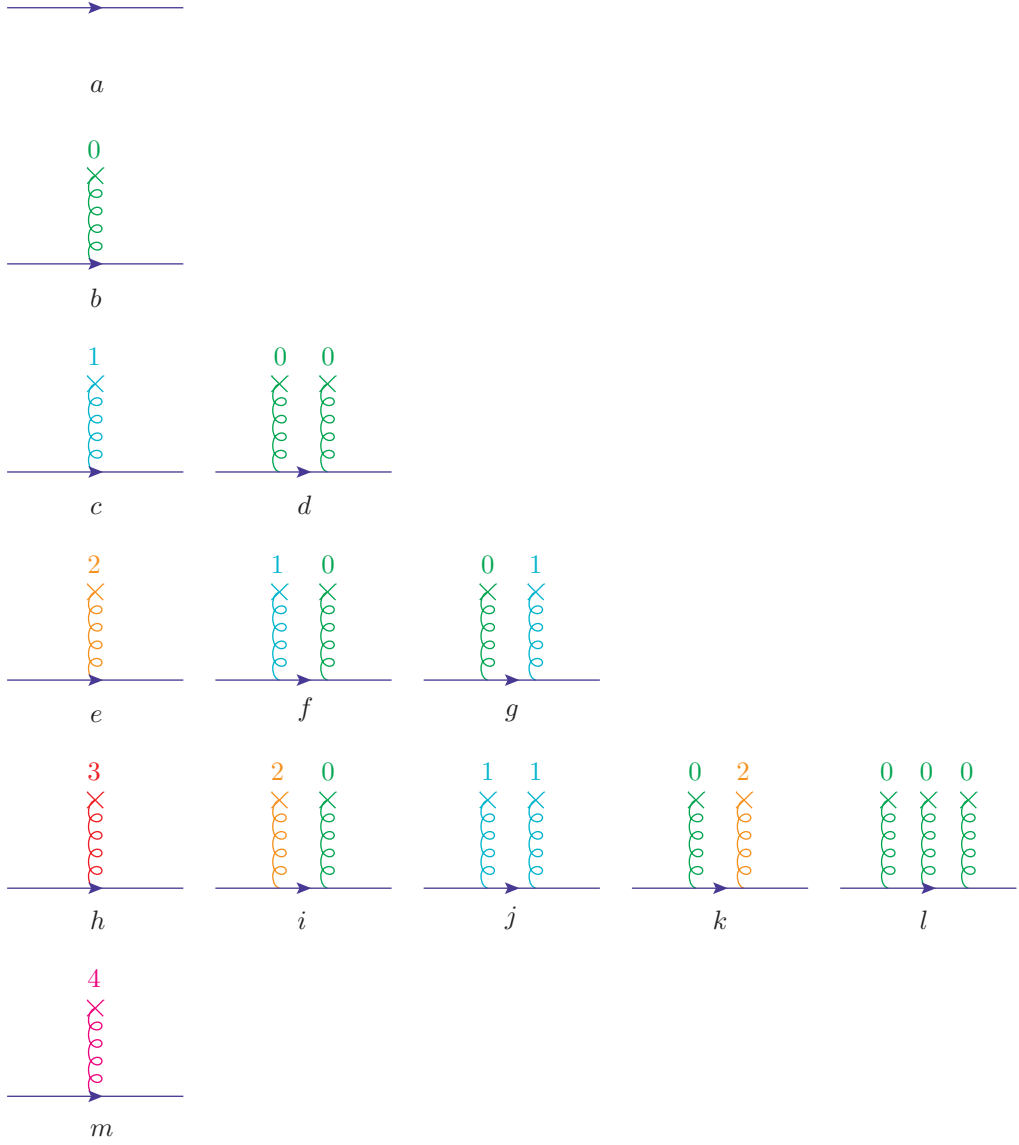}
\caption{Feynman diagrams for the quark propagator within the background field theory which shall results in operators up to dimension-six. The symbol ``$\times$'' attached to the gluon line indicates the tensor of the local gluon background field, in which ``$n$" stands for $n$-th order covariant derivative.} \label{propagator}
\end{figure}

\begin{align}
&S_F^0(x,0)=i\int\frac{d^4p}{(2\pi)^4} e^{-ip\cdot x}\bigg[-\frac{m+\DS p} {m^2-p^2}\bigg],
\label{eq:propagator_0}
\\[1ex]
&S_F^2(x,0)=i\int\frac{d^4p}{(2\pi)^4} e^{-ip\cdot x} \bigg[ -\frac i2\frac{\gamma^\mu (m-\DS p) \gamma^\nu }{(m^2-p^2)^2} G_{\mu\nu}\bigg],
\label{eq:propagator_2}
\\[1ex]
&S_F^3(x,0)=i\int\frac{d^4p}{(2\pi)^4} e^{-ip\cdot x}\bigg\{-\frac23\bigg[\frac{( \gamma^\mu  p^\rho + \gamma^\rho p^\mu)(m-\DS p)}{(m^2-p^2)^3}-\frac{g^{\mu \rho }}{(m^2-p^2)^2}\bigg] \gamma^\nu  G_{\mu\nu;\rho}\bigg\},
\label{eq:propagator_3}
\\[1ex]
&S_F^{4(1)}(x,0)=i\int\frac{d^4p}{(2\pi)^4} e^{-ip\cdot x} \bigg\{\frac14 \bigg[\frac{\gamma^\mu (m-\DS p)}{(m^2-p^2)^3}-\frac{ 2p^\mu}{(m^2-p^2)^3}\bigg]\gamma^\nu \gamma^\rho \gamma^\sigma +\frac{1}{2}~\bigg[\frac{(m+\DS p) \gamma^\mu}{(m^2-p^2)^3}~g^{\nu\sigma}
\nonumber
\\[1ex]
&\hspace{1.7cm} + 4\frac{\gamma^\mu (m-\DS p)}{(m^2-p^2)^4} p^\nu p^\sigma \bigg]\gamma^\rho \bigg\}G_{\mu\nu}G_{\rho\sigma},
\label{eq:propagator_41}
\\[1ex]
&S_F^{4(2)}(x,0) =i\int\frac{d^4p}{(2\pi)^4} e^{-ip\cdot x}\bigg\{~\frac i4 \bigg[\frac{{g^{\{\mu \rho }}\gamma ^{\sigma \}}(m-\DS p)}{(m^2-p^2)^3}-\frac{2g^{\{\mu\rho}p^{\sigma \}}}{(m^2-p^2)^3}+4\frac{{\gamma ^{\{\mu }} p^{\rho } p^{\sigma \}}(m-\DS p)}{(m^2-p^2)^4}\bigg]\gamma^\nu
\nonumber
\\[1ex]
&\hspace{1.7cm} \times   G_{\mu \nu ;\rho \sigma }\bigg\},
\label{eq:propagator_42}
\\[1ex]
&S_F^{5(1)}(x,0) =i\int \frac{d^4p}{(2\pi)^4} e^{-ip\cdot x}\bigg\{\!-\frac{i}{3}\bigg[\bigg(3\frac{\gamma^\mu (m-\DS p)\gamma ^\nu }{(m^2-p^2)^4}(p^\lambda \gamma^\rho + p^\rho \gamma^\lambda )-\frac{\gamma^\nu ({g^{\mu \lambda }}\gamma ^\rho + g^{\mu \rho} \gamma^\lambda )}{(m^2- p^2)^3}\bigg)
\nonumber
\\[1ex]
&\hspace{1.7cm} \cdot~ \gamma^\sigma + 4~\bigg(\frac{{\gamma^\mu }(m-\DS p)}{(m^2-p^2)^4}{g^{\{\nu \sigma }}{p^{\lambda \}}}+2\frac{p^\mu {g^{\{\nu \sigma }}p^{\lambda \}}}{(m^2-p^2)^4}+6~\frac{\gamma^\mu (m-\DS p)}{(m^2-p^2)^5}p^\nu p^\sigma  p^\lambda \bigg) \gamma^\rho \bigg]~G_{\mu \nu }
\nonumber
\\[1ex]
&\hspace{1.7cm} \times G_{\rho \sigma ;\lambda }\bigg\},
\label{eq:propagator_51}
\\[1ex]
&S_F^{5(2)}(x,0)=i\int{\frac{d^4p}{(2\pi)^4}}e^{-ip\cdot x}\bigg\{\frac{2i}{3}\bigg[\bigg(\frac{g^{\mu \lambda }}{(m^2-p^2)^3} +6\frac{p^\mu p^\lambda }{(m^2-p^2)^4}\bigg)\gamma^\nu \gamma ^\rho \gamma^\sigma -2~\bigg(\frac{\gamma^\mu (m-\DS p)}{(m^2-p^2)^4}
\nonumber
\\[1ex]
&\hspace{1.7cm}\times {g^{\{\nu \sigma}}{p^{\lambda \}}}+2\frac{p^\mu {g^{\{\nu \sigma }}p^{\lambda \}}}{(m^2-p^2)^4}+6\frac{\gamma^\mu (m-\DS p)}{(m^2-p^2)^5} p^\nu p^\sigma p^\lambda \bigg)\gamma^\rho \bigg]G_{\mu \nu ;\lambda }G_{\rho \sigma }\bigg\},
\label{eq:propagator_52}
\\[1ex]
&S_F^{5(3)}(x,0) =i\int{\frac{d^4p}{(2\pi)^4}}e^{-ip\cdot x}~\frac4{15}\bigg[\frac{{g^{\{\rho \sigma }}p^\lambda \gamma^{\mu\}}(m-\DS p)}{(m^2-p^2)^4}-\frac{2{g^{\{\rho \sigma }}p^\lambda p^{\mu \}}}{(m^2-p^2)^4}+6\frac{{\gamma^{\{\mu }}p^\rho p^\sigma p^{\lambda \}}(m-\DS p)}{(m^2-p^2)^5}
\nonumber
\\[1ex]
&\hspace{1.7cm} - \frac{g^{(\mu \nu \sigma \lambda )}}{(m^2 - p^2)^3}\bigg]\gamma^\nu G_{\mu \nu ;\rho \sigma \lambda },
\label{eq:propagator_53}
\\[1ex]
&S_F^{6(1)}(x,0)=i\int{\frac{d^4p}{(2\pi)^4}}e^{-ip\cdot x}\frac{i}{8}\bigg\{\bigg[\frac{\gamma^{\mu}(m-\DS p)}{(m^2-p^2)^4}-\!4\frac{p^\mu }{(m^2-p^2)^4}\bigg]\gamma^\nu \gamma^\rho \gamma^\sigma \gamma^\lambda \gamma^\tau \!+\!2\bigg[\frac{3 \gamma^\mu (m-\DS p)}{(m^2-p^2)^4}
\nonumber
\\[1ex]
&\hspace{1.7cm} \times g^{\sigma \tau } +16\frac{\gamma^\mu (m-\DS p)}{(m^2-p^2)^5}p^\sigma p^\tau \!-\!4\frac{g^{\mu \sigma} p^\tau + g^{\mu \tau } p^\sigma }{(m^2-p^2)^4}\bigg]\gamma^\nu \gamma^\rho \gamma^\lambda \bigg\}G_{\mu \nu }G_{\rho \sigma }G_{\lambda \tau },
\label{eq:propagator_61}
\\[1ex]
&S_F^{6(2)}(x,0)=i\int{\frac{d^4p}{(2\pi)^4}}e^{-ip\cdot x}\bigg(-\frac{1}{8}\bigg)\bigg\{\bigg[3\frac{\gamma ^{\mu }(m-\DS p)\gamma^\nu }{(m^2-p^2)^4}{g^{\{\lambda \tau }}\gamma^{\rho \}}+16\frac{\gamma^\mu (m-\DS p) \gamma^\nu }{(m^2-p^2)^5}~ \gamma^{\{\rho}p^\lambda p^{\tau \} }
\nonumber
\\[1ex]
&\hspace{1.7cm}-4\frac{\gamma^\nu }{(m^2-p^2)^4}{g^{\mu \{\lambda }}p^{\tau }\gamma^{\rho \}}\bigg]{\gamma^\sigma }+4\bigg[\frac{m+\DS p}{(m^2-p^2)^4}~g^{(\nu \sigma \tau \lambda )}+6\frac{m+\DS p}{(m^2-p^2)^5}~\,{g^{\{\nu \sigma }}p^\tau p^{\lambda \}}
\nonumber
\\[1ex]
&\hspace{1.7cm} +48\frac{m+\DS p}{(m^2-p^2)^6}p^\nu p^\sigma p^\tau p^\lambda \bigg]\gamma^\mu \gamma^\rho \bigg\}G_{\mu \nu }G_{\rho\sigma;\lambda\tau}, \label{eq:propagator_62}
\\[1ex]
&S_F^{6(3)}(x,0)=i\int{\frac{d^4p}{(2\pi)^4}}e^{-ip\cdot x}\bigg(-\frac{2}{9}\bigg)\bigg\{3\bigg[\bigg(2\frac{{\gamma^{\mu }}(m-\DS p)}{(m^2-p^2)^5}p^\lambda p^\tau -\frac{{g^{\{\mu \lambda }}p^{\tau \}}}{(m^2-p^2)^4}-\frac{4p^\mu p^\lambda p^\tau }{(m^2-p^2)^5}\bigg)
\nonumber
\\[1ex]
&\hspace{1.7cm} \cdot~\gamma^\nu \gamma^\rho+(\mu \leftrightarrow \lambda )+(\rho \leftrightarrow \tau )+(\mu \leftrightarrow \lambda ,\rho \leftrightarrow \tau )\bigg]\gamma^\sigma +4\,\bigg[\frac{m+\DS p}{(m^2-p^2)^4}g^{(\nu \lambda \sigma \tau )}+6
\nonumber
\\[1ex]
&\hspace{1.7cm} \times\frac{m+\DS p}{(m^2-p^2)^5}g^{\{\nu \lambda}{p^\sigma }p^{\tau \}}+48\frac{m+\DS p}{(m^2-p^2)^6}p^\nu p^\lambda p^\sigma p^\tau \bigg]\gamma^\mu \gamma^\rho \bigg\}G_{\mu \nu ;\lambda }G_{\rho \sigma ;\tau },
\label{eq:propagator_63}
\\[1ex]
&S_F^{6(4)}(x,0)=i\int{\frac{d^4p}{(2\pi)^4}}e^{-ip\cdot x}~\bigg(-\frac12\bigg)\bigg\{\bigg[\frac{m+\DS p}{(m^2-p^2)^4}g^{\{ \nu \tau \lambda \sigma \} }+6\frac{m+\DS p}{(m^2-p^2)^5}{g^{\{\nu \tau }} p^\lambda p^{\sigma \}}+48
\nonumber
\\[1ex]
&\hspace{1.7cm}\times \frac{(m+\DS p)p^\nu p^\tau p^\lambda p^\sigma}{(m^2-p^2)^6} \bigg]\gamma^\mu \gamma^\rho -3\bigg[\frac{{g^{\{\mu \lambda }}p^{\tau \}}}{(m^2-p^2)^4}+\frac{8 p^\mu  p^\lambda p^\tau }{(m^2-p^2)^5}\bigg]\gamma^\nu \gamma^\rho \gamma^\sigma \bigg\}G_{\mu \nu ;\lambda \tau }G_{\rho \sigma },
\label{eq:propagator_64}
\\[1ex]
&S_F^{6(5)}(x,0)=i\int{\frac{d^4p}{(2\pi)^4}}e^{-ip\cdot x}\bigg\{-\frac{i}{18}~\bigg[\frac{{g^{[\rho \sigma \lambda \tau }}{\gamma ^{\mu ]}}(m-\DS p)}{(m^2-p^2)^4}~-4~\frac{{g^{\{\rho \sigma \lambda \tau }}p^{\mu \}}}{(m^2-p^2)^4}+6~g^{\{\rho\sigma} p^\lambda p^\tau \gamma^{\mu\}}
\nonumber
\\[1ex]
&\hspace{1.7cm}\times \frac{m-\DS p}{(m^2-p^2)^5} -12\frac{{g^{\{\rho \sigma }}p^\lambda p^\tau p^{\mu \}}}{(m^2-p^2)^5}+48\frac{{p^{\{\rho }}p^\sigma p^\lambda p^\tau p^{\mu \}}(m-\DS p)}{(m^2-p^2)^6}\bigg]\gamma^\nu G_{\mu \nu ;\rho \sigma \lambda \tau }\bigg\}.
\label{eq:propagator_65}
\end{align}

The Feynman diagrams for the quark propagators Eqs.~\eqref{eq:propagator_0}-\eqref{eq:propagator_65} that with various gauge invariant tensors are shown in Figure \ref{propagator}, where thirteen figures, i.e. Figure~\ref{propagator}a, $...$, \ref{propagator}m, correspond to $S_F^{0}(x,0)$, $...$, $S_F^{6(5)}(x,0)$, respectively. The symbol ``$\times$'' attached to the gluon line indicates the tensor of the local gluon background field with ``$n$" stands for $n$-th order covariant derivative. For quark propagator from $x$ to $0$, the following relation can be used,
\begin{align}
S_F(0,x|{\cal A}) = C S_F^T(x,0| - {\cal A}^T){C^{ - 1}}.
\end{align}
The detailed expression are
\begin{align}
&S_F^0(0,x)=i\int\frac{d^4p}{(2\pi)^4} e^{-ip\cdot x}\bigg[-\frac{m-\DS p} {m^2-p^2}\bigg],
\label{eq:propagatorx_0}
\\[1ex]
&S_F^2(0,x)=i\int\frac{d^4p}{(2\pi)^4} e^{-ip\cdot x} \bigg[ \frac i2\frac{\gamma^\nu (m+\DS p) \gamma^\mu }{(m^2-p^2)^2} G_{\mu\nu}\bigg],
\label{eq:propagatorx_2}
\\[1ex]
&S_F^3(0,x)=i\int\frac{d^4p}{(2\pi)^4} e^{-ip\cdot x}\bigg\{\frac23 \gamma^\nu \bigg[\frac{(m+\DS p)( \gamma^\mu  p^\rho + \gamma^\rho p^\mu)}{(m^2-p^2)^3}+\frac{g^{\mu \rho }}{(m^2-p^2)^2}\bigg] G_{\mu\nu;\rho}\bigg\},
\label{eq:propagatorx_3}
\\[1ex]
&S_F^{4(1)}(0,x)=i\int\frac{d^4p}{(2\pi)^4} e^{-ip\cdot x} \bigg\{\frac14\gamma^\sigma \gamma^\rho \gamma^\nu \bigg[\frac{ (m+\DS p)\gamma^\mu}{(m^2-p^2)^3}+\frac{ 2p^\mu}{(m^2-p^2)^3}\bigg] +\frac{1}{2}\gamma^\rho\bigg[\frac{ \gamma^\mu(m-\DS p)}{(m^2-p^2)^3}~g^{\nu\sigma}
\nonumber
\\[1ex]
&\hspace{1.7cm} + 4\frac{ (m+\DS p)\gamma^\mu}{(m^2-p^2)^4} p^\nu p^\sigma \bigg] \bigg\}G_{\rho\sigma}G_{\mu\nu},
\label{eq:propagatorx_41}
\\[1ex]
&S_F^{4(2)}(0,x) =i\int\frac{d^4p}{(2\pi)^4} e^{-ip\cdot x}\bigg\{-\frac i4 \gamma^\nu\bigg[\frac{(m+\DS p){g^{\{\mu \rho }}\gamma ^{\sigma \}}}{(m^2-p^2)^3}+\frac{2g^{\{\mu\rho}p^{\sigma \}}}{(m^2-p^2)^3}+4\frac{(m+\DS p){\gamma ^{\{\mu }} p^{\rho } p^{\sigma \}}}{(m^2-p^2)^4}\bigg]
\nonumber
\\[1ex]
&\hspace{1.7cm} \times   G_{\mu \nu ;\rho \sigma }\bigg\},
\label{eq:propagatorx_42}
\\[1ex]
&S_F^{5(1)}(0,x) =i\int \frac{d^4p}{(2\pi)^4} e^{-ip\cdot x}\bigg\{\!-\frac{i}{3}\bigg[\gamma^\sigma\bigg(-3(p^\lambda \gamma^\rho + p^\rho \gamma^\lambda )\frac{\gamma ^\nu (m+\DS p)\gamma^\mu }{(m^2-p^2)^4}-\frac{({g^{\mu \lambda }}\gamma ^\rho + g^{\mu \rho} \gamma^\lambda )}{(m^2- p^2)^3}
\nonumber
\\[1ex]
&\hspace{1.7cm} ~\cdot ~ \gamma^\nu \bigg)  + 4\gamma^\rho\bigg(\frac{{\gamma^\mu }(m-\DS p)}{(m^2-p^2)^4}{g^{\{\nu \sigma }}{p^{\lambda \}}}+2\frac{p^\mu {g^{\{\nu \sigma }}p^{\lambda \}}}{(m^2-p^2)^4}+6\frac{\gamma^\mu (m-\DS p)}{(m^2-p^2)^5}p^\nu p^\sigma  p^\lambda \bigg)  \bigg]G_{\rho \sigma ;\lambda }
\nonumber
\\[1ex]
&\hspace{1.7cm} \times G_{\mu \nu }\bigg\},
\label{eq:propagatorx_51}
\\[1ex]
&S_F^{5(2)}(0,x)=i\int{\frac{d^4p}{(2\pi)^4}}e^{-ip\cdot x}\frac{2i}{3}\bigg[\gamma^\sigma \gamma ^\rho\gamma^\nu\bigg(\frac{g^{\mu \lambda }}{(m^2-p^2)^3} +6\frac{p^\mu p^\lambda }{(m^2-p^2)^4}\bigg)\!-2\gamma^\rho\bigg(\!\!-\frac{ (m+\DS p)\gamma^\mu}{(m^2-p^2)^4}
\nonumber
\\[1ex]
&\hspace{1.7cm}\times {g^{\{\nu \sigma}}{p^{\lambda \}}}-2\frac{p^\mu {g^{\{\nu \sigma }}p^{\lambda \}}}{(m^2-p^2)^4}-6\frac{ (m+\DS p)\gamma^\mu}{(m^2-p^2)^5} p^\nu p^\sigma p^\lambda \bigg) \bigg]G_{\rho \sigma }G_{\mu \nu ;\lambda },
\label{eq:propagatorx_52}
\\[1ex]
&S_F^{5(3)}(0,x) =i\int{\frac{d^4p}{(2\pi)^4}}e^{-ip\cdot x}~\frac4{15}~\gamma^\nu~\bigg[-\frac{(m+\DS p){g^{\{\rho \sigma }}p^\lambda \gamma^{\mu\}}}{(m^2-p^2)^4}~-~\frac{2{g^{\{\rho \sigma }}p^\lambda p^{\mu \}}}{(m^2-p^2)^4}-6~\frac{(m-\DS p)}{(m^2-p^2)^5}
\nonumber
\\[1ex]
&\hspace{1.7cm}~\cdot~{\gamma^{\{\mu }}p^\rho p^\sigma p^{\lambda \}} - \frac{g^{(\mu \nu \sigma \lambda )}}{(m^2 - p^2)^3}\bigg] G_{\mu \nu ;\rho \sigma \lambda },
\label{eq:propagatorx_53}
\\[1ex]
&S_F^{6(1)}(0,x)=i\int{\frac{d^4p}{(2\pi)^4}}e^{-ip\cdot x}\Big(-\frac{i}{8}\Big)\bigg\{ \gamma^\tau\gamma^\lambda\gamma^\sigma \gamma^\rho\gamma^\nu  \bigg[\frac{(m+\DS p)\gamma^{\mu}}{(m^2-p^2)^4}+4\frac{p^\mu }{(m^2-p^2)^4}\bigg] +2\gamma^\lambda\gamma^\rho\gamma^\nu
\nonumber
\\[1ex]
&\hspace{1.7cm} \times \bigg[\frac{3 (m+\DS p) \gamma^\mu}{(m^2-p^2)^4}g^{\sigma \tau } +16\frac{(m+\DS p)\gamma^\mu }{(m^2-p^2)^5}p^\sigma p^\tau \!+\!4\frac{g^{\mu \sigma} p^\tau + g^{\mu \tau } p^\sigma }{(m^2-p^2)^4}\bigg] \bigg\}G_{\lambda \tau }G_{\rho \sigma }G_{\mu \nu },
\label{eq:propagatorx_61}
\\[1ex]
&S_F^{6(2)}(0,x)=i\int{\frac{d^4p}{(2\pi)^4}}e^{-ip\cdot x}\bigg(-\frac{1}{8}\bigg)\bigg\{\gamma^\sigma \!\bigg[3{g^{\{\lambda \tau }}\gamma^{\rho \}}\frac{\gamma^\nu(m+\DS p)\gamma ^\mu }{(m^2-p^2)^4}+16\gamma^{\{\rho}p^\lambda p^{\tau \} }\frac{\gamma^\nu (m+\DS p) \gamma^\mu }{(m^2-p^2)^5}
\nonumber
\\[1ex]
&\hspace{1.7cm}+4\frac{{g^{\mu \{\lambda }}p^{\tau }\gamma^{\rho \}}\gamma^\nu }{(m^2-p^2)^4}\bigg]+4~\gamma^\rho\gamma^\mu~ \bigg[\frac{m-\DS p}{(m^2-p^2)^4}~g^{(\nu \sigma \tau \lambda )}~+6\frac{m-\DS p}{(m^2-p^2)^5}~{g^{\{\nu \sigma }}p^\tau p^{\lambda \}}+48
\nonumber
\\[1ex]
&\hspace{1.7cm} \times\frac{m-\DS p}{(m^2-p^2)^6}p^\nu p^\sigma p^\tau p^\lambda \bigg] \bigg\}G_{\rho\sigma;\lambda\tau}G_{\mu \nu },
\label{eq:propagatorx_62}
\\[1ex]
&S_F^{6(3)}(0,x)=i\int{\frac{d^4p}{(2\pi)^4}}e^{-ip\cdot x}\bigg(-\frac{2}{9}\bigg)\bigg\{~3\gamma^\sigma\bigg[~\gamma^\rho\gamma^\nu \bigg(2\frac{(m+\DS p){\gamma^{\mu }}}{(m^2-p^2)^5}p^\lambda p^\tau +\frac{{g^{\{\mu \lambda }}p^{\tau \}}}{(m^2-p^2)^4}\,4p^\mu p^\lambda p^\tau
\nonumber
\\[1ex]
&\hspace{1.7cm}\times\frac{1}{(m^2-p^2)^5}\bigg)  +(\mu \leftrightarrow \lambda )+(\rho \leftrightarrow \tau )+(\mu \leftrightarrow \lambda ,\rho \leftrightarrow \tau )\bigg] +4 ~\gamma^\rho \gamma^\mu ~\bigg[~\frac{m-\DS p}{(m^2-p^2)^4}
\nonumber
\\[1ex]
&\hspace{1.7cm} \times g^{(\nu \lambda \sigma \tau )}~+~6\frac{m-\DS p}{(m^2-p^2)^5}g^{\{\nu \lambda}{p^\sigma }p^{\tau \}}~+~48~\frac{m-\DS p}{(m^2-p^2)^6}p^\nu p^\lambda p^\sigma p^\tau \bigg]~ \bigg\}~G_{\rho \sigma ;\tau }G_{\mu \nu ;\lambda },
\label{eq:propagatorx_63}
\\[1ex]
&S_F^{6(4)}(0,x)=i\int{\frac{d^4p}{(2\pi)^4}}~e^{-ip\cdot x}~\bigg(-\frac12\bigg) \bigg\{ \gamma^\rho\gamma^\mu~\bigg[~
\frac{m-\DS p}{(m^2-p^2)^4}g^{\{ \nu \tau \lambda \sigma \} }+6\frac{m-\DS p}{(m^2-p^2)^5}{g^{\{\nu \tau }} p^\lambda p^{\sigma \}}
\nonumber
\\[1ex]
&\hspace{1.7cm}+48\, \frac{(m-\DS p)p^\nu p^\tau p^\lambda p^\sigma}{(m^2-p^2)^6} ~\bigg] +3~\gamma^\sigma\gamma^\rho\gamma^\nu ~ \bigg[\frac{{g^{\{\mu \lambda }}p^{\tau \}}}{(m^2-p^2)^4}+\frac{8 p^\mu  p^\lambda p^\tau }{(m^2-p^2)^5}\bigg] ~\bigg\}~G_{\rho \sigma }G_{\mu \nu ;\lambda \tau },
\label{eq:propagatorx_64}
\\[1ex]
&S_F^{6(5)}(0,x)=i\int{\frac{d^4p}{(2\pi)^4}}e^{-ip\cdot x}\bigg\{-\frac{i}{18}~\gamma^\nu\bigg[\frac{{(m+\DS p)}{g^{[\rho \sigma \lambda \tau }}\gamma ^{\mu ]}}{(m^2-p^2)^4}~+4~\frac{{g^{\{\rho \sigma \lambda \tau }}p^{\mu \}}}{(m^2-p^2)^4}+6~\frac{(m+\DS p) }{(m^2-p^2)^5}
\nonumber
\\[1ex]
&\hspace{1.7cm}~\cdot~g^{\{\rho\sigma} p^\lambda p^\tau \gamma^{\mu\}} +12\frac{{g^{\{\rho \sigma }}p^\lambda p^\tau p^{\mu \}}}{(m^2-p^2)^5}-48\frac{{p^{\{\rho }}p^\sigma p^\lambda p^\tau p^{\mu \}}(m+\DS p)}{(m^2-p^2)^6}\bigg] G_{\mu \nu ;\rho \sigma \lambda \tau }\bigg\}.
\label{eq:propagatorx_65}
\end{align}
where
\begin{align}
&{g^{[\mu \nu \rho \sigma }}p^{\lambda ]} =g^{(\mu \nu \rho \sigma )} {p^\lambda} +g^{(\lambda \nu \rho \sigma)} p^\mu +g^{(\mu \lambda \rho \sigma )} {p^\nu }+g^{(\mu \nu \lambda \sigma )}{p^\rho }+g^{(\mu \nu \rho \lambda )}{p^\sigma},
\\
&{g^{\{\mu \nu }} p^\rho p^\sigma p^{\lambda \}}=g^{\mu\nu}p^\rho p^\sigma p^\lambda +g^{\mu \rho}p^\nu p^\sigma p^\lambda +g^{\mu \sigma }p^\rho p^\nu p^\lambda + g^{\mu \lambda} p^\rho p^\sigma p^\nu +g^{\nu \rho }p^\mu p^\sigma p^\lambda
\nonumber
\\
&\hspace{3cm} +g^{\nu \sigma }p^\rho p^\mu p^\lambda + g^{\nu \lambda } p^\rho  p^\sigma p^\mu +g^{\rho \sigma }p^\mu p^\nu p^\lambda +g^{\rho \lambda }p^\mu p^\sigma  p^\nu +g^{\sigma \lambda } p^\rho  p^\mu p^\nu,
\\
&{g^{\{\mu \rho }}{\gamma ^{\sigma \}}} =g^{\mu \rho }\gamma^\sigma +g^{\mu \sigma } \gamma^\rho + g^{\rho \sigma }\gamma ^\mu,
\\
&{g^{[\rho \sigma \lambda \tau }} \gamma ^{\mu ]} =\gamma ^\mu g^{(\rho \sigma \lambda \tau )}+(\mu \leftrightarrow \rho )+(\mu \leftrightarrow \sigma )+(\mu \leftrightarrow \lambda )+(\mu \leftrightarrow \tau ),
\\
&{g^{\{\rho \sigma }}p^\lambda \gamma ^{\mu \}} =\gamma^\mu g^{\{\rho \sigma }p^{\lambda \}}+(\mu \leftrightarrow \rho )+(\mu \leftrightarrow \sigma )+(\mu \leftrightarrow \lambda ),
\\
&{g^{\{\rho \sigma }}p^\lambda p^\tau \gamma ^{\mu \}} =\gamma^\mu {g^{\{\rho \sigma }}p^{\lambda }p^{\tau \}}+(\mu \leftrightarrow \rho )+(\mu \leftrightarrow \sigma )+(\mu \leftrightarrow \lambda )+(\mu \leftrightarrow \tau ),
\\
&{\gamma ^{\{\mu }}p^\rho p^{\sigma \}} =\gamma^\mu p^\rho p^\sigma +\gamma^\rho p^\mu p^\sigma +\gamma^\sigma p^\mu p^\rho,
\\
&{\gamma ^{\{\mu }}p^\rho p^\sigma p^{\lambda \}}=\gamma^\mu p^\rho p^\sigma p^\lambda +(\mu \leftrightarrow \rho )+(\mu \leftrightarrow \sigma )+(\mu \leftrightarrow \lambda ),
\\
&{\gamma ^{\{\mu }}p^\rho p^\sigma p^\lambda p^{\tau \}}=\gamma^\mu p^\rho p^\sigma p^\lambda p^\tau +(\mu \leftrightarrow \rho )+(\mu \leftrightarrow \sigma )+(\mu \leftrightarrow \lambda )+(\mu \leftrightarrow \tau ),
\\
&{g^{\mu \{\lambda }}p^\tau \gamma^{\rho \}}=g^{\mu\lambda} p^\tau  \gamma^\rho +g^{\mu \tau }p^\lambda \gamma^\rho +g^{\mu\rho} p^\tau \gamma^\lambda + g^{\mu \tau } p^\rho \gamma^\lambda + g^{\mu\rho} p^\lambda \gamma^\tau +g^{\mu \lambda }p^\rho \gamma^\tau.
\end{align}
In these expressions, the mass term is retained explicitly, which application not only in the case of light quarks, but also in the case of heavy quarks.

\begin{figure}[t]
\includegraphics[width=0.95\textwidth]{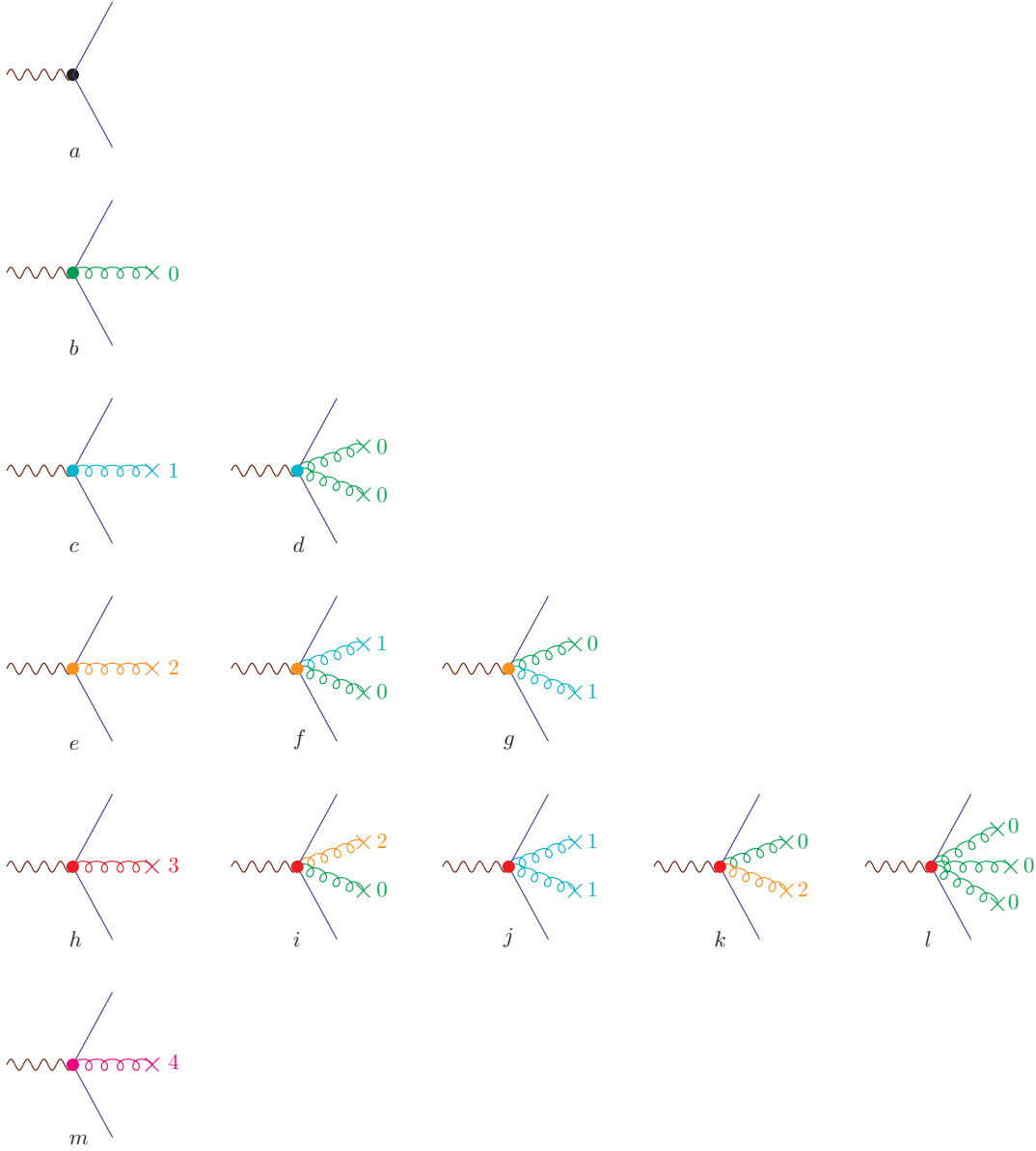}
\caption{Feynman diagrams for the vertex operator $\Gamma (z\cdot \tensor{D})^n$ under the background field theory up to dimension-six. The symbol ``$\times$'' attached to the gluon line indicates the tensor of the local gluon background field, in which ``$n$" stands for $n$-th order covariant derivative.}
\label{vertex}
\end{figure}

\subsection{Vertex operator} \label{sec:appendix_I2}

The vertex operator $(z\cdot \tensor D)^n$ is encountered when we calculate the distributed amplitude of moment with the SVZ sum rule. Here it is only the dimension $d \le 6$ terms are retained, vertex operator $(z \cdot \tensor D)^n$ in dimension-six order, the results are as follows
\begin{align}
(z \cdot \tensor D)_0^n &= (z\cdot\tensor\partial)^n,
\\
(z \cdot \tensor D)_2^n &= - i(z\cdot\tensor\partial)^{n-1} \underline x^\mu z^\nu G_{\mu \nu },
\\
(z \cdot \tensor D)_3^n &= -\frac{2i}{3}(z \cdot  \tensor \partial  )^{n-1} \underline x^\mu \underline x^\rho z^\nu G_{\mu \nu;\rho },
\\
(z \cdot \tensor D)_{4(1)}^n &= -\frac{n(n-1)}{2}(z \cdot  \tensor \partial  )^{n-2} \underline x^\mu \underline x^\rho z^\nu z^\sigma G_{\mu \nu }G_{\rho \sigma },
\\
(z \cdot \tensor D)_{4(2)}^n &= \bigg[ -\frac{i}{4} n(z \cdot  \tensor \partial  )^{n-2} \underline x^{\mu }\underline x^{\rho }\underline x^{\sigma }-\frac{i}{12}(n-1)(n-2)(z \cdot  \tensor \partial  )^{n-3}\underline x^\mu z^\rho z^\sigma z^\nu \bigg]G_{\mu \nu ;\rho \sigma },
\\
(z \cdot \tensor D)_{5(1)}^n &= -\frac{n(n-1)}{3}(z \cdot\tensor\partial )^{n-2} \underline x^\mu \underline x^\rho \underline x^\sigma \underline x^\lambda z^\nu z^\sigma G_{\mu \nu } G_{\rho \sigma ;\lambda},
\\
(z \cdot \tensor D)_{5(2)}^n &= -\frac{n(n-1)}{3}(z \cdot  \tensor \partial  )^{n-2} \underline x^\mu \underline x^\rho \underline x^\sigma \underline x^\lambda z^\nu z^\sigma G_{\mu \nu ;\lambda} G_{\rho \sigma },
\\
(z \cdot \tensor D)_{5(3)}^n &= \bigg[-\frac{i}{15} n(z \cdot \tensor\partial)^{n-1} \underline x^{\mu }\underline x^\rho \underline x^\sigma \underline x^\lambda z^\nu -\frac{i}{45}n(n-1)(n-2)(z \cdot  \tensor \partial  )^{n-2}\underline x^\mu (\underline x^\rho z^\sigma  z^\lambda
\nonumber\\
&+ \underline x^\sigma z^\rho z^\lambda +\underline x^\lambda z^\sigma z^\rho )z^\nu \bigg]G_{\mu \nu ;\rho \sigma \lambda},
\\
(z \cdot \tensor D)_{6(1)}^n &= \frac i6 n(n-1)(n-2)(z \cdot \tensor\partial  )^{n-3}\underline x^\mu \underline x^\rho \underline x^\lambda z^\nu z^\sigma z^\tau G_{\mu\nu} G_{\rho\sigma} G_{\lambda \tau},
\\
(z \cdot \tensor D)_{6(2)}^n &= \bigg[-\frac{n(n-1)}{8}(z \cdot  \tensor \partial  )^{n-2}\underline x^\mu \underline x^\rho \underline x^\lambda \underline x^\tau z^\nu z^\sigma -\frac{1}{12}n(n-1)(n-2)(n-3)(z \cdot \tensor\partial)^{n-4}
\nonumber\\
& \times\underline x^{\mu }\underline x^\rho z^\lambda z^\tau z^\nu z^\sigma \bigg]G_{\mu \nu } G_{\rho \sigma ;\lambda \tau},
\\
\ (z \cdot \tensor D)_{6(3)}^n &= -\frac29 n(n-1)(z \cdot  \tensor\partial  )^{n-2} \underline x^\mu \underline x^\rho \underline x^\lambda \underline x^\tau z^\nu z^\sigma G_{\mu \nu ;\lambda} G_{\rho \sigma; \tau},
\nonumber\\
(z \cdot \tensor D)_{6(4)}^n &= -\frac{n(n-1)}8 (z \cdot \tensor\partial  )^{n-2} \underline x^\mu \underline x^\rho \underline x^\lambda \underline x^\tau z^\nu z^\sigma G_{\mu \nu ;\lambda \tau} G_{\rho \sigma },
\\
(z\cdot\tensor D)_{6(5)}^n &= \bigg[-\frac i{72}\times (z \cdot  \tensor \partial  )^{n-1}\underline x^\mu \underline x^\rho \underline x^\lambda \underline x^\tau z^{\nu }-\frac i{216}n(n-1)(n-2)(z \cdot  \tensor \partial )^{n-3}\underline x^\mu \bigg(z^\rho z^\sigma \underline x^\lambda \underline x^\tau
\nonumber\\
&+ z^\rho \underline x^\sigma z^\lambda \underline x^\tau + z^\rho \underline x^\sigma \underline x^\tau z^\tau +\underline x^\rho z^\sigma z^\lambda \underline x ^\tau +\underline x^\rho z^\sigma \underline x^\lambda z^\tau +\underline x^\rho \underline x^\sigma z^\lambda z^\tau \bigg)z^\nu -\frac i{360} n(n-1)
\nonumber\\
&\times(n-2)(n-3)(n-4)(z \cdot \tensor\partial)^{n-5} \underline x^\mu z^\rho z^\sigma z^\lambda z^\tau z^\nu \bigg]G_{\mu\nu ;\lambda\tau\rho\sigma }.
\end{align}

where the subscript $k(m)$ with $k=(1,...,6)$ stands for the dimension of the operator, in which $(m)$ stands for the $m$-th type of the operator with same dimension. For example, there is two type of dimension four operators, three type of dimension five operators and five type of dimension-six operators. Figure~\ref{vertex} represents the Feynman diagrams for the vertex operator $\Gamma (z\cdot \tensor{D})^n$ under the background field theory, where thirteen figures, i.e. Figs. \ref{vertex}a, $...$, \ref{vertex}m, correspond to $\Gamma(z\cdot \tensor{D})^n_1, ..., \Gamma(z\cdot \tensor{D})^n_{6(5)}$, respectively. The symbol ``$\times$'' attached to the gluon line indicates the tensor of the local gluon background field with ``$n$" stands for the $n$-th order covariant derivative.

\subsection{Vacuum matrix elements}  \label{sec:appendix_I3}

In addition to quark propagators $S_F(x,0)$ and vertex operators $(z\cdot \tensor D)^n$, one may also encounter vacuum condensation in the process of calculation, which are the basic input parameter. The specific expression for vacuum matrix element in the $D$-dimension space have the following forms,
\begin{eqnarray}
&& \langle 0| G^a_{\mu\nu}G^b_{\rho\sigma}|0\rangle
= \frac{1}{96} \delta^{ab}\langle G^2\rangle [g_{\mu\rho}g_{\nu\sigma}-(\sigma \leftrightarrow \rho)],
\\
&&{\rm Tr} \langle 0| G_{\mu\nu}G_{\rho\sigma}|0\rangle
= \frac{\pi}{6}\langle\alpha_sG^2\rangle[g_{\mu\rho}g_{\nu\sigma}-(\sigma \leftrightarrow \rho)],
\\
&&
\langle 0| G^a_{\mu\nu;\lambda} G^b_{\rho\sigma;\tau}|0\rangle
= -\frac{1}{648} \sum_{u,d,s} \delta^{ab} \langle g_s\bar\psi\psi\rangle ^2
[ 2 g_{\lambda\tau} (g_{\mu\rho}g_{\nu\sigma} - g_{\mu\sigma}g_{\nu\rho})
\nonumber\\
&& \hspace{3cm} + g_{\lambda\rho} (g_{\tau\mu}g_{\sigma\nu} - g_{\tau\nu}g_{\sigma\mu})- g_{\lambda\sigma} (g_{\tau\mu}g_{\rho\nu} - g_{\tau\nu}g_{\rho\mu}) ],
\\
&& {\rm Tr} \langle 0| G_{\mu\nu;\lambda} G_{\rho\sigma;\tau}|0\rangle
= -\frac{1}{162} \sum_{u,d,s} \langle  g_s^2\bar{\psi}\psi\rangle ^2
[ 2 g_{\lambda\tau} (g_{\mu\rho}g_{\nu\sigma} - g_{\mu\sigma}g_{\nu\rho})
\nonumber\\
&& \hspace{3cm} + g_{\lambda\rho} (g_{\tau\mu}g_{\sigma\nu} - g_{\tau\nu}g_{\sigma\mu})- g_{\lambda\sigma} (g_{\tau\mu}g_{\rho\nu} - g_{\tau\nu}g_{\rho\mu}) ],
\\
&& \langle 0| G^a_{\mu\nu} G^b_{\rho\sigma} G^c_{\lambda\tau}|0\rangle
= \frac{1 }{576} f^{abc}\langle G^aG^bG^c\rangle \{ [ g_{\mu\rho} (g_{\nu\lambda}g_{\sigma\tau} - g_{\nu\tau}g_{\sigma\lambda})
\nonumber\\
&& \hspace{3cm} - g_{\mu\sigma} (g_{\nu\lambda}g_{\rho\tau} - g_{\nu\tau}g_{\rho\lambda}) ] - (\mu\leftrightarrow\nu)  \},
\\
&& {\rm Tr} \langle 0| G_{\mu\nu} G_{\rho\sigma} G_{\lambda\tau}|0\rangle
= \frac{i}{96} \langle g_s^3fG^3\rangle
\{ [ g_{\mu\rho} (g_{\nu\lambda}g_{\sigma\tau} - g_{\nu\tau}g_{\sigma\lambda})
\nonumber\\
&& \hspace{3cm} - g_{\mu\sigma} (g_{\nu\lambda}g_{\rho\tau} - g_{\nu\tau}g_{\rho\lambda}) ] - (\mu\leftrightarrow\nu)  \},
\\
&& \langle 0| G^a_{\mu\nu} G^b_{\rho\sigma;\lambda\tau}|0\rangle
= \delta^{ab} \bigg\{ \bigg( - \frac1{1296} \sum_{u,d,s} \langle g_s\bar{\psi}\psi\rangle ^2 - \frac1{384} \langle  g_s f G^3 \rangle  \bigg)
\nonumber
\\
&& \hspace{3cm}
\times [ 2 g_{\lambda\tau}(g_{\mu\sigma}g_{\nu\rho} - g_{\mu\rho}g_{\nu\sigma}) + g_{\rho\tau} (g_{\mu\sigma}g_{\nu\lambda} - g_{\mu\lambda}g_{\nu\sigma})
\nonumber\\
&& \hspace{3cm}
- g_{\sigma\tau} (g_{\mu\lambda}g_{\nu\rho} - g_{\mu\rho}g_{\nu\lambda}) ]
\nonumber\\
&& \hspace{3cm}
+ \bigg( -\frac1{1296}\sum_{u,d,s} \langle g_s\bar{\psi}\psi\rangle ^2 + \frac1{384}  \langle  g_s f G^3 \rangle  \bigg)
\nonumber\\
&& \hspace{3cm}
\times[ g_{\mu\tau} (g_{\rho\nu}g_{\sigma\lambda} - g_{\rho\lambda}g_{\nu\sigma}) +(\mu\leftrightarrow\nu) ] \bigg\},
\\
&&
{\rm Tr} \langle 0| G_{\mu\nu} G_{\rho\sigma;\lambda\tau}|0\rangle = 4\bigg( -\frac1{1296} g_s^2 \sum_{u,d,s} \langle g_s\bar{\psi}\psi\rangle ^2 - \frac1{384}  \langle  g_s^3 f G^3 \rangle  \bigg)
\nonumber\\
&& \hspace{3cm}
\times [ 2 g_{\lambda\tau} (g_{\mu\sigma}g_{\nu\rho} - g_{\mu\rho}g_{\nu\sigma}) + g_{\rho\tau} (g_{\mu\sigma}g_{\nu\lambda} - g_{\mu\lambda}g_{\nu\sigma})
\nonumber\\
&& \hspace{3cm}
- g_{\sigma\tau} (g_{\mu\lambda}g_{\nu\rho} - g_{\mu\rho}g_{\nu\lambda}) ]
\nonumber\\
&& \hspace{3cm}
+ 4\bigg( -\frac1{1296} g_s^2 \sum_{u,d,s} \langle g_s\bar{\psi}\psi\rangle ^2 + \frac1{384} \langle  g_s^3 f G^3 \rangle  \bigg)
\nonumber\\
&& \hspace{3cm}
\times [ g_{\mu\tau} (g_{\rho\nu}g_{\sigma\lambda} - g_{\rho\lambda}g_{\nu\sigma}) +(\mu\leftrightarrow\nu) ],
\\
&&
\langle 0|\bar\psi_\alpha^a(x)\psi_\beta^b(y)|0\rangle = \delta^{ab}\bigg\{\langle\bar\psi\psi\rangle\bigg[\frac1{12}g_{\beta\alpha}+\frac{i m_s}{48}(\DS x-\DS y)_{\beta\alpha}-\frac{m_s^2}{96}(x-y)^2g_{\beta\alpha}
\nonumber\\
&& \hspace{3cm}
-\frac{im_s^3}{576}(x-y)^2(\DS x-\DS y)_{\beta\alpha}\bigg] +g_s\langle\bar\psi TG\psi\rangle\bigg[\frac1{192}(x-y)^2g_{\beta\alpha}
\nonumber\\
&& \hspace{3cm}
+\frac{im_s}{864}(x-y)^2(\DS x-\DS y)_{\beta\alpha}\bigg]\bigg\}
\label{eq:condensate_qqxy}
\\
&&
\langle 0|\bar\psi_\alpha^a(x)\psi_\beta^b(y) G^A_{\mu\nu}|0\rangle = \frac1{192} \langle\bar\psi TG\psi\rangle (\sigma_{\mu\nu})_{\beta\alpha} (T^A)^{ba}
+\bigg\{-\frac1{864}g_s\langle\bar\psi\psi\rangle^2
\nonumber\\
&& \hspace{3cm}
\times (g_{\mu\rho}\gamma_\nu - g_{\nu\rho}\gamma_\mu)_{\beta\alpha} (x+y)^\rho
+i(x-y)^\rho\bigg[\frac1{864}g_s\langle\bar\psi\psi\rangle^2
\nonumber\\
&& \hspace{3cm}
+\frac{m_s}{384}\langle\bar\psi TG\psi\rangle \bigg] (\epsilon_{\rho\mu\nu\sigma}\gamma_5\gamma^\sigma)_{\beta\alpha}\bigg\} (T^A)^{ba}
\label{eq:condensate_qqg2xy}
\\
&&\langle 0|\bar\psi_\alpha^a(0)\psi_\beta^b(0) G^A_{\mu\nu;\rho}|0\rangle =\frac1{432}g_s\langle\bar\psi\psi\rangle^2 (g_{\mu\rho}\gamma_\nu - g_{\nu\rho}\gamma_\mu)_{\beta\alpha} (T^A)^{ba}
\label{eq:condensate_qqg300}
\end{eqnarray}
the symbol ``${\rm Tr}$'' stands for tracing to the colour matrixes. Obviously,
\begin{eqnarray}
\langle 0| G^B_{\rho\sigma;\lambda\tau} G^A_{\mu\nu}|0\rangle  = \langle 0| G^A_{\mu\nu} G^B_{\rho\sigma;\lambda\tau}|0\rangle .
\end{eqnarray}
As a final remark, by making use of the equation
\begin{equation}
[ \widetilde{D}_\mu, \widetilde{D}_\nu ]^{AB} = -g_s f^{ABC} G^C_{\mu\nu},
\end{equation}
we can obtain a useful relation
\begin{equation}
\langle g_s^3 fG^3\rangle  = \frac{8}{27}  \sum_{u,d,s} \langle g_s^2\bar{\psi}\psi\rangle ^2. \label{relgluqua}
\end{equation}
which, can also be referred to our previous work~\cite{Zhong:2014jla}.

\section{Detailed expressions for $I_{m_s^0}(n,M^2)$ and $I_{m_s^2}(n,M^2)$ in BFTSR for $\langle\xi_{2;\eta^{(\prime)}}^n\rangle|_\mu$}\label{sec:appendixB}

According to the Feynman rule, we can obtain the following detailed expressions within  $I_{m_s^0}(n,M^2)$ and $I_{m_s^2}(n,M^2)$ for the final BFTSR, i.e. Eq.~\eqref{xi2}. Some of the terms have the symmetric  structure, we can combined these terms together. Here, we only list the key expression for the last step of the BFTSR. Firstly, we give the $I_{m_s^0}(n,M^2)$ terms which are expressed as,
\begin{align}
&\hat{\cal B}_{M^2} I_{000} = \frac{3}{8\pi^2} \frac{1+(-1)^n}{(n+1)(n+3)},\label{eq:ms0_I000}
\\[1ex]
&\hat{\cal B}_{M^2} I_{006+600} = ~\frac{g_s^2 \sum \langle g_s\bar{q}q\rangle ^2}{M^6}~ [1+(-1)^n]~~ \frac{-17}{324\pi^2}~ \bigg\{~ \left[ (2n+1)~ \left( -\ln \frac{M^2}{\mu^2} \right) ~- (n+2) \right]
\nonumber\\[1ex]
&\hspace{2.1cm} + \theta(n-2)~ \bigg[~ \frac{1}{2}~(2n+1)~ \bigg( \psi\Big(\frac{n+1}2\Big) ~-~ \psi\Big(\frac n2 \Big) ~+~ \ln4 \bigg) ~- \frac{(n+1)^2}{n} \bigg] ~\bigg\},
\\[1ex]
&\hat{\cal B}_{M^2} I_{024+420} = \frac{g_s^2 \sum \langle g_s\bar{q}q\rangle ^2}{ M^6 } [1+(-1)^n] \frac{-1}{162\pi^2} \theta(n-2) \bigg\{ n \bigg( -\ln \frac{M^2}{\mu^2} \bigg) + \bigg[ \frac{n}{2} \bigg( \psi\Big(\frac{n+1}2\Big)
\nonumber\\[1ex]
&\hspace{2.1cm}  - \psi\Big(\frac n2 \Big) + \ln4 \bigg) - 1 \bigg] \bigg\},
\\[1ex]
&\hat{\cal B}_{M^2} I_{033+330} = \frac{g_s^2 \sum \langle g_s\bar{q}q\rangle ^2}{M^6}~ [1+(-1)^n]~\frac{2}{243\pi^2} ~\theta(n-2)~\bigg\{~n~\left( -\ln \frac{M^2}{\mu^2} ~-~ \frac{1}{2} \right) ~+ \bigg[ \frac{n}{2}
\nonumber\\[1ex]
&\hspace{2.1cm}  \times \left( \psi\Big(\frac{n+1}2\Big) - \psi\Big(\frac n2\Big) + \ln4 \right) - 1 \bigg] \bigg\},
\\[1ex]
&\hat{\cal B}_{M^2} I_{202+040} = \frac{\langle \alpha_sG^2\rangle }{M^4} \frac{1 + n\theta(n-2)}{24\pi(n+1)} [1+(-1)^n],
\\[1ex]
&\hat{\cal B}_{M^2} I_{060} = \frac{16 g_s^2 \sum \langle g_s\bar{q}q\rangle ^2 - 81 \langle g_s^3fG^3\rangle }{M^6} \frac{n}{7776\pi^2} [1+(-1)^n]\, \theta(n-2),
\\[1ex]
&\hat{\cal B}_{M^2} I_{204+402}= \frac{g_s^2 \sum \langle g_s\bar{q}q\rangle ^2}{ M^6 } [1+(-1)^n]\, \frac{1}{324\pi^2} \bigg\{ \left( -\ln \frac{M^2}{\mu^2} \right) +  \theta(n-2) ~\bigg[ ~\frac{1}{2} \bigg( \psi\Big(\frac{n+1}2\Big)
\nonumber\\[1ex]
&\hspace{2.1cm}  - \psi\Big(\frac n2 \Big) + \ln4 \bigg) - \frac{1}{n} \bigg] \bigg\},
\\[1ex]
&\hat{\cal B}_{M^2} I_{303(1)+303(3)}= ~\frac{g_s^2 \sum \langle g_s\bar{q}q\rangle ^2}{ M^6 } ~[1+(-1)^n]~~\frac{-2}{243\pi^2}~ \bigg\{ ~\bigg(~ -\ln \frac{M^2}{\mu^2}~ \bigg) ~+ \theta(n-2)~ \bigg[ ~\frac12
\nonumber\\[1ex]
&\hspace{2.8cm}\times \bigg( \psi\Big(\frac{n+1}2\Big) - \psi\Big(\frac n2\Big) + \ln4 \bigg) - \frac{1}{n} \bigg] \bigg\},
\\[1ex]
&\hat{\cal B}_{M^2} I_{303(2)} = \frac{g_s^2 \sum \langle g_s\bar{q}q\rangle ^2}{M^6 }[1+(-1)^n] \frac{1}{162\pi^2 }~ \bigg\{ \bigg( -\ln \frac{M^2}{\mu^2} + \frac12 \bigg) + \frac{ \theta(n-2)}2\bigg( \,\psi\Big(\frac{n+1}2\Big)
\nonumber\\[1ex]
&\hspace{1.8cm} - \psi\Big(\frac n2 \Big) + \ln4 \bigg) \bigg\}.\label{eq:ms0_I3032}
\end{align}
In which, the subscript $a,b,c,m$ in $ \hat{\cal B}_{M^2} I_{abc(m)}$ for Eqs.~\eqref{eq:ms0_I000}-\eqref{eq:ms0_I3032} have the meaning that, $a,b,c$ stand for $a,b,c$th-order of the propagator, vertex and propagator, while $m$ means the $m$th terms of $I^{m_s^2}_{abc}$. Then, we give the expression for $I_{m_s^2}(n,M^2)$ terms separately,
\begin{align}
&\hat{\cal B}_{M^2} I_{004+400}^{m_s^2} = \frac{\langle\alpha_sG^2\rangle}{M^6} m_s^2 \frac{-1}{12\pi} [1+(-1)^n] \bigg\{~\bigg[~(2n+1) \Big(-\ln\frac{M^2}{\mu^2}+\psi(n+1) + 2\gamma_E \Big) - n
\nonumber
\\[1ex]
&\hspace{2.1cm}-2~\bigg]~+~ \theta(n-2) \bigg[ ~\frac{2n+1}{2} \bigg( \psi\Big( \frac{n+1}{2} \Big) - \psi \Big( \frac{n}{2} \Big) + \ln 4 \bigg) - \frac{(n+1)^2}{n} \bigg] \bigg\},  \label{eq:I004400}
\\[1ex]
&\hat{\cal B}_{M^2} I_{006(1)+600(1)}^{m_s^2} = \frac{\langle g_s^2\bar{q}q\rangle^2}{M^8} m_s^2 \frac{17}{324\pi^2}[ 1 + (-1)^n ] \bigg\{ \delta^{n0} + 2n \Big( -\ln\frac{M^2}{\mu^2} + \psi(n+1) + 2\gamma_E \Big)
\nonumber
\\[1ex]
&\hspace{2.8cm} - n-3+ \theta(n-2)~\left[~ n \left( \psi\Big( \frac{n+1}{2}\Big) ~-~ \psi\Big( \frac{n}{2}\Big) ~+~ \ln 4 \right) - \left( n + 1 \right)\, \right] ~\bigg\},
\\[1ex]
&\hat{\cal B}_{M^2} I_{006(2)+600(2)}^{m_s^2} = \frac{8\langle g_s^2\bar{q}q\rangle^2-3\langle g_s^3fG^3\rangle}{M^8} m_s^2 \frac{1}{144\pi^2} [1+(-1)^n]\,\bigg\{ \frac12~ \delta^{n0} - \bigg[ \,2n^2 \bigg( -\ln\frac{M^2}{\mu^2}
\nonumber
\\[1ex]
&\hspace{2.8cm} + \psi(n+1)+ 2\gamma_E \bigg) -\frac{7n^2 + 19n + 2}{4} \bigg] \bigg\}  +\bigg\{\theta(n-3) \bigg[ -n^2 \bigg(  \,\psi \Big( \frac{n+1}{2} \Big)
\nonumber
\\[1ex]
&\hspace{2.8cm} - \psi \Big( \frac{n}{2} \Big)+ \frac 14( 5n^2+5n+2)\bigg]\bigg\},
\\[1ex]
&\hat{\cal B}_{M^2} I_{024(1)+420(1)}^{m_s^2} = \frac{\langle g_s^2\bar{q}q\rangle^2}{M^8} m_s^2 \frac{n}{162\pi^2} [1+(-1)^n]\, \bigg\{\theta(n-1)\bigg(-\ln\frac{M^2}{\mu^2}+\psi(n+1)+2\gamma_E
\nonumber
\\[1ex]
&\hspace{2.8cm} -\frac{1}{n}\bigg)+ \frac12\theta(n-2) \left[\psi\Big(\frac{n+1}2\Big) - \psi\Big(\frac n2\Big) + \ln 4 \right] \bigg\},
\\[1ex]
&\hat{\cal B}_{M^2} I_{024(2)+420(2)}^{m_s^2} = \frac{8\langle g_s^2\bar{q}q\rangle^2 - 27\langle g_s^3fG^3\rangle}{M^8} ~m_s^2~ \frac{-n}{3888\pi^2} ~[1+(-1)^n]\,~\bigg\{~ \theta(n-1)~\bigg[ (2n-1)
\nonumber
\\[1ex]
&\hspace{2.8cm}\times \bigg( -\ln\frac{M^2}{\mu^2}+\psi(n+1) + 2\gamma_E ~\bigg) - n - 3 ~+ \frac{1}{n}~\bigg]+~\theta(n-3)~\bigg[ ~\frac{2n-1}{2}
\nonumber
\\[1ex]
&\hspace{2.8cm}\times \bigg(\psi\Big(\frac{n+1}2\Big) - \psi\Big(\frac n2\Big)+ \ln 4 \bigg)- ( n + 1 )\bigg]\bigg\},
\\[1ex]
&\hat{\cal B}_{M^2} I_{033(1)+330(1)}^{m_s^2} = \frac{\langle g_s^2\bar{q}q\rangle^2}{M^8} m_s^2 \frac{-2n}{243\pi^2} [1+(-1)^n]\, \!\bigg\{\! \theta(n-1) \Big[ (2n-1) \bigg( \!\!\! -\ln\frac{M^2}{\mu^2}+ \psi(n+1)
\nonumber
\\[1ex]
&\hspace{2.8cm} + 2\gamma_E \Big)- n - 3 + \frac{1}{n} \bigg] + \theta(n-3) \bigg[ ~\frac{2n-1}{2}~ \bigg(\psi\Big(\frac{n+1}2\Big)- \psi\Big(\frac n2\Big) + \ln 4 \bigg)
\nonumber
\\[1ex]
&\hspace{2.8cm} - (n+1) \bigg] \bigg\},
\\[1ex]
&\hat{\cal B}_{M^2} I_{033(2,1)+330(2,1)}^{m_s^2} = \frac{\langle g_s^2\bar{q}q\rangle^2}{M^8} m_s^2 \frac{-[1+(-1)^n] 2n^2 \theta(n-1)}{243\pi^2},
\\[1ex]
&\hat{\cal B}_{M^2} I_{033(2,2)+330(2,2)}^{m_s^2} =\, \frac{\langle g_s^2\bar{q}q\rangle^2}{M^8} ~m_s^2 ~\frac{2n}{243\pi^2}~ [1+(-1)^n]\,~ \bigg\{~ \theta(n-1)~ \bigg[~ (2n-1) ~\bigg( -\ln\frac{M^2}{\mu^2}
\nonumber
\\[1ex]
&\hspace{3.3cm}+ \psi(n+1)~+2\gamma_E \bigg) + n - 3 + \frac{1}{n} ~\bigg] + \theta(n-3) ~\bigg[ ~\frac{2n-1}{2} \bigg(~\psi\Big(\frac{n+1}2\Big)
\nonumber
\\[1ex]
&\hspace{3.3cm} - \psi\Big(\frac n2\Big)+ \ln 4 \bigg)-( n + 1) \bigg] \bigg \},
\\[1ex]
&\hat{\cal B}_{M^2} I_{033(3,1)+330(3,1)}^{m_s^2}= \frac{\langle g_s^2\bar{q}q\rangle^2}{M^8} m_s^2 \frac{\left[ 1 ~+~ (-1)^n \right] n \theta(n-1)}{243\pi^2},
\\[1ex]
&\hat{\cal B}_{M^2} I_{033(3,2)+330(3,2)}^{m_s^2}= \frac{\langle g_s^2\bar{q}q\rangle^2}{M^8}~ m_s^2 ~\frac{-2n}{243\pi^2} ~[1+(-1)^n]\,~ \bigg\{ ~\theta(n-1)~ \bigg( -\ln\frac{M^2}{\mu^2} + \psi(n+1)
\nonumber\\
&\hspace{3.3cm}+ 2\gamma_E - \frac{1}{n}\bigg)+ \frac{1}{2}\theta(n-2) \left[\psi\Big(\frac{n+1}2\Big)- \psi\Big(\frac n2\Big) + \ln 4 \right] \bigg\},
\\[1ex]
&\hat{\cal B}_{M^2} I_{040(1)+040(2)}^{m_s^2} = \frac{\langle\alpha_sG^2\rangle}{(m^2)^3} m_s^2 \frac{-[1+(-1)^n]\, n \theta(n-2)}{12\pi},
\\
&\hat{\cal B}_{M^2} I_{060(1)+060(2)}^{m_s^2} = \frac{16\langle g_s^2\bar{q}q\rangle^2 - 81\langle g_s^3fG^3\rangle}{M^8} m_s^2 \frac{-[1+(-1)^n]\, n \theta(n-2)}{3888\pi^2},
\\[1ex]
&\hat{\cal B}_{M^2} I_{202(1)+202(2)}^{m_s^2} = \frac{\langle\alpha_sG^2\rangle}{(m^2)^3} m_s^2 \frac{1}{12\pi} [1+(-1)^n]\, \bigg\{ \!\!\left( -\ln\frac{M^2}{\mu^2}+ \psi(n+1) + 2\gamma_E\! \right)+\theta(n-1)
\nonumber\\[1ex]
&\hspace{2.8cm}\times \frac{1}{2}\bigg[\psi\Big(\frac{n+1}2\Big) - \psi\Big(\frac n2\Big) + \ln 4 - \frac{2}{n} \bigg] \bigg\},
\\[1ex]
&\hat{\cal B}_{M^2} I_{204(1)}^{m_s^2} = \frac{\langle g_s^2\bar{q}q\rangle^2}{M^8} m_s^2 \frac{1}{324\pi^2} [1+(-1)^n]\, \left\{ - \left( -\ln\frac{M^2}{\mu^2} + \psi(n+1)+ 2\gamma_E \right) \,\,+ \frac{\theta(n-1)}{n} \right. \nonumber\\
&\hspace{1.78cm}- \left. \theta(n-2) \frac{1}{2} \left[\psi\Big(\frac{n+1}2\Big)~ -~ \psi\Big(\frac n2\Big) + \ln 4 \right] \right\},
\\[1ex]
&\hat{\cal B}_{M^2} I_{402(1)}^{m_s^2} = \frac{\langle g_s^2\bar{q}q\rangle^2}{M^8} m_s^2 \frac{1}{324\pi^2} [1+(-1)^n]\, \left\{ - \left( -\ln\frac{M^2}{\mu^2} + \psi(n+1) + 2\gamma_E \right) \,\,+ \frac{\theta(n-1)}{n} \right.
\nonumber\\[1ex]
&\hspace{1.78cm}- \left. \theta(n-2) \frac{1}{2} \left[\psi\Big(\frac{n+1}2\Big) - \psi\Big(\frac n2\Big)~ + ~\ln 4 \right] \right\},
\\[1ex]
&\hat{\cal B}_{M^2} I_{204(2)}^{m_s^2} = \frac{\langle g_s^2\bar{q}q\rangle^2}{M^8} m_s^2 \frac{1}{243\pi^2} [1+(-1)^n]\, \left\{ - \left( -\ln\frac{M^2}{\mu^2} + \psi(n+1) + 2\gamma_E \right) \,\,+ \frac{\theta(n-1)}{n} \right.
\nonumber\\[1ex]
&\hspace{1.78cm}- \left. \theta(n-2) \frac{1}{2} \left[\psi\Big(\frac{n+1}2\Big) - \psi\Big(\frac n2\Big) + \ln 4 \right] \right\},
\\[1ex]
&\hat{\cal B}_{M^2} I_{402(2)}^{m_s^2}= \frac{\langle g_s^2\bar{q}q\rangle^2}{M^8} m_s^2 \frac{1}{243\pi^2} [1+(-1)^n]\, \left\{ - \left( -\ln\frac{M^2}{\mu^2} + \psi(n+1) + 2\gamma_E \right) \,\,+ \frac{\theta(n-1)}{n} \right.
\nonumber\\[1ex]
&\hspace{1.78cm}- \left. \theta(n-2) \frac{1}{2} \left[\psi\Big(\frac{n+1}2\Big) - \psi\Big(\frac n2\Big) + \ln 4 \right] \right\},
\\[1ex]
&\hat{\cal B}_{M^2} I_{204(3)+402(3)}^{m_s^2} = \frac{4\langle g_s^2\bar{q}q\rangle^2 - 27\langle g_s^3fG^3\rangle}{M^8} ~m_s^2 ~\frac{1}{3888\pi^2} ~[1+(-1)^n]\,~ \bigg\{ \delta^{n0} + 2n \, \bigg( -\ln\frac{M^2}{\mu^2}
\nonumber\\[1ex]
&\hspace{2.8cm} + \psi(n+1)+ 2\gamma_E \bigg)-n - 3  +  \theta(n-2)~ \bigg[ ~n\, \bigg(\psi\Big(\frac{n+1}2\Big) - \psi\Big(\frac n2\Big) + \ln 4 \bigg)
\nonumber\\[1ex]
&\hspace{2.8cm} - (n+1) \bigg] \bigg\},
\\[1ex]
&\hat{\cal B}_{M^2} I_{303(1)+303(2)}^{m_s^2} = \frac{\langle g_s^2\bar{q}q\rangle^2}{M^8} m_s^2 \frac{-5}{486\pi^2} [1+(-1)^n]\, \bigg\{ \delta^{n0} + 2n \left(\! -\ln\frac{M^2}{\mu^2} + \psi(n+1) + 2\gamma_E \right)
\nonumber\\[1ex]
&\hspace{2.8cm}- n - 3 + \theta(n-2) \left[ n \left(\psi\Big(\frac{n+1}2\Big) - \psi\Big(\frac n2\Big) + \ln 4 \right) - \left( n + 1 \right) \right] \bigg\},
\\[1ex]
&\hat{\cal B}_{M^2} I_{303(3)+303(4)}^{m_s^2} = \frac{\langle g_s^2\bar{q}q\rangle^2}{M^8} m_s^2 \frac{-1}{81\pi^2} [1+(-1)^n]\, \bigg \{ \delta^{n0} + 2n \left( -\ln\frac{M^2}{\mu^2} ~+ \psi(n+1) + 2\gamma_E \right)
 \nonumber\\[1ex]
&\hspace{2.8cm}- n- 3 +\theta(n-2) \left[ n \left(\psi\Big(\frac{n+1}2\Big) - \psi\Big(\frac n2\Big) + \ln 4 \right) - \left( n + 1 \right) \right] \bigg\},
\\[1ex]
&\hat{\cal B}_{M^2} I_{303(5)}^{m_s^2} = \frac{\langle g_s^2\bar{q}q\rangle^2}{M^8} m_s^2 \frac{-2}{243\pi^2} [1+(-1)^n]\, \left\{ - \left( -\ln\frac{M^2}{\mu^2} + \psi(n+1) + 2\gamma_E \right) ~+ \frac{\theta(n-1)}{n} \right.
\nonumber\\[1ex]
&\hspace{1.78cm}- \left. \theta(n-2) \frac{1}{2} \left[\psi\Big(\frac{n+1}2\Big) - \psi\Big(\frac n2\Big) + \ln 4 \right] \right\},
\\[1ex]
&\hat{\cal B}_{M^2} I_{303(6)}^{m_s^2} = \frac{\langle g_s^2\bar{q}q\rangle^2}{M^8} m_s^2 \frac{-2}{243\pi^2} [1+(-1)^n]\, \left\{ - \left( -\ln\frac{M^2}{\mu^2} + \psi(n+1) + 2\gamma_E \right) +~ \frac{\theta(n-1)}{n} \right.
\nonumber\\[1ex]
&\hspace{1.78cm}- \left. \theta(n-2) \frac{1}{2} \left[\psi\Big(\frac{n+1}2\Big) - \psi\Big(\frac n2\Big) + \ln 4 \right] \right\},
\\[1ex]
&\hat{\cal B}_{M^2} I_{303(7)}^{m_s^2} = \frac{\langle g_s^2\bar{q}q\rangle^2}{M^8} m_s^2 ~\frac{-7}{162\pi^2} [1+(-1)^n]\, \left\{ - \left( -\ln\frac{M^2}{\mu^2} + \psi(n+1) + 2\gamma_E \right) + \frac{\theta(n-1)}{n} \right.
\nonumber\\[1ex]
&\hspace{1.78cm}- \left. \theta(n-2) \frac{1}{2} \left[\psi\Big(\frac{n+1}2\Big) - \psi\Big(\frac n2\Big) + \ln 4 \right] \right\},
\\[1ex]
&\hat{\cal B}_{M^2} I_{303(8,1)+303(9,1)}^{m_s^2} = \frac{\langle g_s^2\bar{q}q\rangle^2}{M^8} m_s^2 \frac{4}{162\pi^2} \left[ 1+(-1)^n \right] \bigg\{ \!\!-2 \delta^{n0}\bigg ( -\ln \frac{M^2}{\mu^2} +\psi(n+1) + 2\gamma_E
\nonumber
\\[1ex]
&\hspace{3.3cm} + 1 \bigg)+ 2 \delta^{n1} \left(\!\!-\ln \frac{M^2}{\mu^2} + \psi(n+1) + 2\gamma_E - 2 \right)\!-\!\theta(n-2)\frac{n^2+4n-2}{n}
\nonumber
\\[1ex]
&\hspace{3.3cm}+ \theta(n-3) \bigg[ (n-1) \left(\psi\Big(\frac{n+1}2\Big) - \psi\Big(\frac n2\Big) + \ln 4 \right) -n \bigg] \bigg\},
\\[1ex]
&\hat{\cal B}_{M^2} I_{303(8,2)+303(9,2)}^{m_s^2} = \frac{\langle g_s^2\bar{q}q\rangle^2}{M^8} m_s^2 \frac{1}{162\pi^2} [1+(-1)^n]\, \left( 2n-2 + 2\delta^{n1} \right). \label{eq:I3038292}
\end{align}
Here the subscript $a,b,c,m,n$ in $\hat{\cal B}_{M^2} I^{m_s^2}_{abc(m,n)}$ in the expression within BFTSR, i.e. Eqs.~\eqref{eq:I004400}-\eqref{eq:I3038292} have following meaning. $a,b,c$ stand for the $a,b,c$th-order of the propagator, vertex and propagator, while $m$ means the $m$th terms of $I^{m_s^2}_{abc}$ and $n$ means the $n$th subterm of $I^{m_s^2}_{abc(m)}$. Meanwhile, for the perturbative part we have
\begin{align}
{\rm Im} I_{\rm pert.}(s) = \frac{3v^{n + 1}}{8\pi (n + 1)(n + 3)}\bigg\{ [1 + {( - 1)^n}](n + 1)\frac{1 - {v^2}}{2} + [1 + {( - 1)^n}]\bigg\}.
\end{align}
Considering $n$ is even, then $1+(-1)^n=2$, $\delta^{n1}=0$, $\theta(n-1)=\theta(n-2)$, $\theta(n-3) = \theta(n-4)$. Add up all the above expressions we obtain the moment of distributed amplitude Eq.~\eqref{xi2}.

\section{Matrix elements of $\eta^{(\prime)}$-meson twist-2, 3, 4 LCDAs}\label{sec:appendixC}

The expression of twist-2, 3, 4 LCDAs matrix elements used in the OPE calculation of LCSR have the following forms~\cite{Fu:2020uzy, Ball:2006wn}
\begin{align}
&\langle \eta^{(\prime)}(p)|\bar s(x)\gamma_\mu \gamma_5 s(0)|0\rangle = - ip_\mu f_{\eta^{(\prime)}}~\int_0^1 du e^{iup \cdot x}[\phi_{2;\eta^{(\prime)}}(u,\mu)+ x^2 \psi_{4;\eta^{(\prime)}}(u)] + f_{\eta^{(\prime)}} \left(x_\mu - \frac{x^2 p_\mu} {p \cdot x} \right)
\nonumber\\
& \hspace{5.5cm} \times\int_0^1 du e^{iup \cdot x} \phi_{4;\eta^{(\prime)}}(u),
\label{Eq:DAeta1}
\\
&\langle \eta^{(\prime)}(p)|\bar s(x)i\gamma_5s(0)|0\rangle  = \frac{m_{\eta^{(\prime)}}^2 f_{\eta^{(\prime)}}}{2m_s}\!\int_0^1 du e^{iup \cdot x}\phi_{3;\eta^{(\prime)}}^p(u),
\label{Eq:DAeta2}
\\
&\langle \eta^{(\prime)}(p)|\bar s(x)\sigma_{\mu \nu}\gamma_5 s(0)|0\rangle  = i(p_\mu x_\nu  - p_\nu x_\mu)~\frac{m_{\eta^{(\prime)}}^2 f_{\eta^{(\prime)}}}{12m_s} \int_0^1 du e^{iup\cdot x} \phi_{3;\eta^{(\prime)}}^\sigma (u),
\label{Eq:DAeta3}
\\
&\langle \eta^{(\prime)}(p)|\bar s(x)\sigma_{\mu\nu}\gamma_5 g_s G_{\alpha\beta}(vx)s(0)|0\rangle  = i\frac{m_{\eta^{(\prime)}}^2 f_{\eta^{(\prime)}}}{2m_s}\left(
p_\alpha p_\mu g_{\nu\beta}^\bot - p_\alpha p_\nu g_{\mu\beta}^\bot  - (\alpha\leftrightarrow \beta) \right)\Phi_{3;\eta^{(\prime)}}(v,p \cdot x)
\label{Eq:DAeta4}
\\[2ex]
& \langle \eta^{(\prime)}(p)|\bar s(x)\gamma_\mu \gamma_5 g_s G_{\alpha\beta}(vx)s(0)|0\rangle  = {f_{\eta^{(\prime)}} }\bigg[ \frac{{{p_\mu }}}{{p\cdot x}}({p_\alpha }{x_\beta } - {p_\beta }{x_\alpha }){\Phi _{4;\eta^{(\prime)} }}(v,p \cdot x) + ({p_\beta }g_{\alpha \mu }^ \bot  - {p_\alpha }g_{\beta \mu }^ \bot )
\nonumber\\
& \hspace{5.5cm}\times{\Psi _{4;\eta^{(\prime)} }}(v,p \cdot x)\bigg],
\label{Eq:DAeta5}
\\[1ex]
&\langle \eta^{(\prime)} (p)|\bar s(x)\gamma_\mu g_s \widetilde G_{\alpha\beta}(vx) s(0)|0\rangle = i{f_{\eta^{(\prime)}} }\bigg[ \frac{{{p_\mu }}}{{p\cdot x}}({p_\alpha }{x_\beta } - {p_\beta }{x_\alpha }){{\tilde \Phi }_{4;\eta^{(\prime)} }}(v,p \cdot x) + ({p_\beta }g_{\alpha \mu }^ \bot  - {p_\alpha }g_{\beta \mu }^ \bot )
\nonumber\\
& \hspace{5.5cm}\times \tilde\Psi_{4;\eta^{(\prime)}}(v,p \cdot x)\bigg].
\label{Eq:DAeta6}
\end{align}
In which, we have set
\begin{align}
&g_{\mu\nu}^\bot = g_{\mu\nu} - \frac{p_\mu x_\nu+ p_\nu x_\mu}{p\cdot x};
\nonumber\\
&K(v,p\cdot x) = \int_0^1 d {\alpha _1}d{\alpha _2}d{\alpha _3}\delta (1 - \alpha_1 - \alpha_2 - \alpha_3) e^{-i(\alpha_2 - \alpha_1 + v\alpha_3) p \cdot x}K(\alpha_i).
\end{align}

\end{document}